\documentclass[11pt,a4paper]{article}
\usepackage{longtable}
\usepackage{amsmath}
\usepackage{amssymb}
\usepackage{graphicx}
\usepackage{bm}
\usepackage{footmisc}
\usepackage{afterpage}
\usepackage{float}
\usepackage{lscape}
\usepackage{longtable}
\usepackage{float}
\usepackage{slashed}
\usepackage{morefloats}
\usepackage{afterpage}

\newcommand*{\no}{\noindent}
\newcommand*{\bea}{\begin{eqnarray}}
\newcommand*{\eea}{\end{eqnarray}}
\newcommand*{\be}{\begin{equation}}
\newcommand*{\ee}{\end{equation}}
\newcommand*{\pd}{\partial}
\newcommand*{\pdm}{\pd_{\mu}}

\newcommand*{\pref}[1]{(\ref{#1})}

\newcommand*{\nn}{\nonumber}

\newcommand*{\tr}{\mathrm{tr}}

\newcommand{\bma}{\begin{pmatrix}}
\newcommand{\ema}{\end{pmatrix}}

\usepackage[square,comma,sort&compress,numbers]{natbib}

\title{The quenched SU(2) adjoint scalar propagator in minimal Landau gauge}

\author{Axel Maas\\
Institute of Physics, NAWI Graz, University of Graz,\\
Universit\"atsplatz 5, A-8010 Graz, Austria}

\begin{document}

\maketitle

\begin{abstract}

It is a long-standing question whether the confinement of matter fields in QCD has an imprint in the (gauge-dependent) correlation functions, especially the propagators. In particular in the quenched case a fundamental difference could be expected between adjoint and fundamental matter. In a preceding investigation the propagator of a fundamental scalar has been studied, showing no obvious sign of confinement. Here, complementary, the adjoint scalar propagator is investigated over a wide range of parameters in the minimal Landau gauge using lattice gauge theory. This study is performed in two, three, and four dimensions in quenched SU(2) Yang-Mills theory, both in momentum space and position space. No conclusive difference between both cases is found.

\end{abstract}

\section{Introduction}

The confinement\footnote{Confinement is here understood, if not noted otherwise, in the sense that a particle cannot be observed as an asymptotic, physical state. In this sense also QCD is confining. A definition of confinement based on the Wilson string tension is in no obvious way related to this. In fact, according to the Wilson string tension QCD is not a confining theory. See \cite{Alkofer:2006fu} for a more detailed discussion of this difference.} of matter in QCD is a very long-standing problem \cite{Alkofer:2006fu}. It is especially not yet clear how to read off the confinement of a particle from its elementary correlation functions. This could be both the propagator as well as the vertices \cite{Alkofer:2000wg,Maas:2011se,Fischer:2006ub,Binosi:2009qm,Boucaud:2011ug,Vandersickel:2012tg,Roberts:2015lja}. Of course, the correlation functions describing the elementary particles, matter and gluon alike, are gauge-dependent. Thus, this question requires to fix a gauge, and thus the answer is potentially gauge-dependent. Here, this question will be posed in a particular case, the best-studied one so far, the Landau gauge, in particular the so-called minimal Landau gauge \cite{Maas:2011se}.

A natural quantity to investigate these question is the spectral density. This spectral density is found to be positivity violating for gluons \cite{Alkofer:2000wg,Maas:2011se,Fischer:2006ub,Binosi:2009qm,Boucaud:2011ug,Vandersickel:2012tg,Roberts:2015lja,Strauss:2012as,Cucchieri:2016jwg}. However, the precise form this violation takes, e.\ g.\ by a non-trivial cut structure, complex poles, or otherwise, is not entirely settled. At any rate, any such violation of positivity immediately implies that the particle cannot be part of the physical state space, and thus not observable. Sufficient, but not necessary, conditions for violation of positivity can be either a non-positive definite position-space correlation function or a non-monotonous behavior of the derivatives of the momentum-space correlation functions \cite{Maas:2011se}. If the correlation function is known both in momentum-space and position-space sufficiently well, these results can be used to constrain the type of analytic structure. E.\ g.\ an oscillatory behavior in position space and screened behavior in momentum space points to a complex pole structure \cite{Maas:2011se}.

Results for fermionic matter, especially quarks, are intricate, see \cite{Roberts:1994dr,Roberts:2015lja,Alkofer:2000wg,Alkofer:2003jj,August:2013jia}. However, the results are compatible with a violation of positivity also in fermion propagators. This is true for quarks both in the adjoint and the fundamental representation.

Scalar matter suggests itself as a testbed of this question\footnote{There are multiple subtleties with respect to this question in the dynamical theory due to the possibility of a Brout-Englert-Higgs effect, see \cite{Maas:2017wzi}. In the present quenched case, this is not relevant.}, due to the simpler Lorentz structure. For scalar matter in the fundamental representation there have been various suggestions for its behavior, which have been obtained using continuum methods \cite{Fister:2010yw,Macher:2011ad,Capri:2012ah,Capri:2013oja,Hopfer:2013via,Maas:2011yx,Maas:2013aia,Capri:2017abz}. On the lattice a violation of positivity has been found, though it is not entirely clear whether its remains in the continuum and infinite-volume limit \cite{Maas:2016edk}. Still, the propagator showed the presence of an intrinsic, non-zero mass scale even if massless at tree-level, and consequently exhibits a momentum-space propagator similar to that of a massive particle. Thus, there is no clear indication for confinement in the fundamental scalar sector.

On the other hand, it is naively expected that adjoint scalar matter could show a different behavior in the quenched case \cite{Macher:2011ad,Capri:2017abz}. After all, the Wilson string tension of the adjoint string still vanishes, due to string-breaking by matter-gluon hybrids. In case of the quarks, this does not seem to lead to a differing behavior for the propagator at a qualitative level \cite{Aguilar:2010ad,Maskawa:1974vs,Fukuda:1976zb,August:2013jia}. However, adjoint quarks have also a very different behavior when it comes to chiral symmetry breaking, as their differing finite-temperature behavior shows \cite{Karsch:1998qj,Engels:2005te,Bilgici:2009jy}. This may interfere with a clear picture. Therefore, once more, it becomes interesting to study adjoint scalar matter in the quenched case.

This will be done here. Following \cite{Maas:2016edk}, this will be done for a wide range of lattice parameters, and for two, three, and four dimensions. Considering two dimensions may seem odd at first. However, in this case the violation of positivity for gluons appears similarly as in more dimensions \cite{Maas:2007uv,Cucchieri:2007rg,Maas:2014xma,Cucchieri:2016jwg}. But gluons are not dynamical but only pure gauge in two dimensions. At the same time, a confinement according to the Wilson potential occurs already for purely geometrical reasons \cite{Dosch:1978jt}. Scalar particles are, however, also in two dimensions dynamical. In the fundamental case, this did not lead to any qualitative impact \cite{Maas:2016edk}, and scalar matter behaved in the same way in all dimensions. It is therefore interesting to have a look at two dimensions.

Three dimensions take an intermediate position. While it is a dynamical theory in the quenched case, it has different renormalization properties than in four dimensions. Moreover, while the four-dimensional unquenched case is potentially trivial, this is not true in three dimensions \cite{Callaway:1988ya}. This could have potentially impact as well.

Again, as in \cite{Maas:2016edk}, the quenched calculation will also help to understand lattice artifacts and renormalization properties of the scalar propagator beyond perturbation theory. This is helpful in studies of the dynamical case, which will, e.\ g., be relevant for studies of many kinds of grand-unified theories on the lattice \cite{Maas:2017wzi}, for which a host of predictions await non-perturbative precision tests \cite{Maas:2017xzh} after exploratory investigations in the past \cite{Olynyk:1985tr,Lee:1985yi}.

As technically the study of the adjoint propagator is quite similar to the study of the fundamental propagator, this paper follows closely \cite{Maas:2016edk}. The technical setup is given in section \ref{s:tech}. Renormalization is studied in detail in section \ref{s:ren}. The results in momentum and position space are presented in section \ref{s:ana}. These are the main results of this work. A short summary follows in section \ref{s:sum}. Some preliminary results have been presented in \cite{Maas:2011yx}.

\section{Technical setup}\label{s:tech}

In the following the propagator of a scalar particle in the adjoint representation of SU(2) in the quenched theory will be determined in two, three, and four dimensions. The technical setup is based on \cite{Cucchieri:2006tf,Maas:2007uv,Maas:2010nc,Maas:2016edk}. Thus, the gauge action is the Wilson action for SU(2) Yang-Mills theory. The gauge field configurations are obtained using a cycle of heatbath and overrelaxation updates. The lattice setups are listed in table \ref{tcgf} in appendix \ref{a:ls}. The determination of the lattice spacing has been performed as in \cite{Maas:2014xma}.

Note that the limiting factor in terms of lattice volumes has been the required amount of statistics, especially for the position-space investigation. Though the Schwinger function is found to be positivity-violating in section \ref{ss:ssd}, it still shows (additional) exponential suppression at large times. Since smearing alters the momentum-space properties drastically \cite{Maas:2014tza}, this can only be beaten by an exponential increase in statistics. Hence, the present investigation is primarily statistics-limited. In the same vain, physically small volumes were thus the only possibility to reach the large momenta necessary to investigate the logarithmic behavior of the renormalization constants in section \ref{s:ren}.

Each decorrelated configuration is fixed to minimal Landau gauge \cite{Maas:2011se} using adaptive stochastic overrelaxation \cite{Cucchieri:2006tf}. The quenched adjoint propagator has been obtained in a similar fashion as the quenched fundamental one in \cite{Maas:2010nc,Maas:2016edk}. In the continuum, it is given by the inverse of the covariant adjoint Laplacian including the mass term
\be
-D^2=-\left(\pdm+gf^{abc}A_\mu^c\right)^2\nn+m_0^2,\nn
\ee
\no where the $f^{abc}$ are the structure constants, for SU(2) just the Levi-Civita tensor, the $A_\mu^a$ are the gauge fields, $g=\sqrt{4/\beta}$ the (bare) coupling constant, and $m_0$ the bare mass of the scalar. As the lattice version of this operator its naive discretization \cite{Greensite:2006ns}
\bea
-D^2_L&=&-\sum_\mu\left(U^a_\mu(x)\delta_{y(x+e_\mu)}+U_\mu^{a\dagger}(x-\mu)\delta_{y(x-e_\mu)}-2\delta_{xy}\right)+m_0^2\delta_{xy}\label{cov}\nn\\
U_{\mu bc}^a&=&\frac{1}{2}\tr\left(\tau^b U_\mu^\dagger\tau^c U_\mu\right)\nn,
\eea
\no has been used, where $U^a_\mu$ are the link variables in the adjoint representation and $U_\mu$ the usual links in the fundamental representation. The links are transformed between both representations using the generators $\tau^a$, in the present case the Pauli matrices. The $e_\mu$ are lattice unit vectors in the corresponding directions. Since this operator is positive semi-definite, it can be inverted. This has been done using the same method as for the Faddeev-Popov operator in \cite{Cucchieri:2006tf}. It should be noted that even a zero mass is not a problem for this method\footnote{In contrast to the Faddeev-Popov operator, this operator has no trivial zero modes, and thus an inversion even at zero momentum is possible. However, since constant modes affect the result on a finite lattice, this is not done here, as in \cite{Maas:2016edk}.}. The final result is averaged over color. The momenta are evaluated along the\footnote{Note that, where possible, the momentum directions are not averaged, as this would require additional expensive inversions but introducing additional correlations.} $x$-axis as edge momenta and along the $xy$, $xyz$, and $xyzt$ diagonal directions, when available in a given number of dimensions. This provides access to both the lowest and highest possible momenta for all dimensions with the least corresponding lattice artifacts \cite{Maas:2014xma} without employing additional improvements \cite{Sternbeck:2012qs,Boucaud:2008gn}. The latter would require again higher statistics at all momenta, e.\ g.\ to obtain sufficiently precise renormalization constants for all volumes as well to make effective use of.

Fixing the bare mass $m_0$ in \pref{cov} is done as in \cite{Maas:2016edk}: Using the known lattice spacings, it is set to the desired tree-level value $m=am_0$ at the ultraviolet cutoff $1/a$. Four different values will be used, zero, 100 MeV, 1 GeV, and 10 GeV. The bare values $m_0$ for 1 GeV physical tree-level mass are listed in table \ref{tcgf}. In \cite{Maas:2016edk} it was found that the lattice artifacts were for all masses comparatively small, even for zero and 10 GeV. As will be seen, this is not the case here, and substantial discretization artifacts are encountered independent of the bare mass. In this respect, adjoint matter is different than fundamental matter.

\section{Renormalization}\label{s:ren}

\subsection{Definition of the renormalization scheme}\label{ss:scheme}

For the adjoint case, the same scheme will be used as for the fundamental case \cite{Maas:2016edk}. For completeness, it will be briefly repeated here. It assumes that the renormalization can be performed as in the perturbative case \cite{Bohm:2001yx}, i.\ e.\ a wave-function renormalization constant and a mass renormalization is sufficient. While discretization effects are large it seems that this is indeed possible for sufficiently fine lattices.

There are then two necessary renormalization constants, a multiplicative wave-function renormalization $Z$, and an additive mass renormalization $\delta m^2$. The renormalized propagator is
\be
D^{ij}(p^2)=\frac{\delta^{ij}}{Z(p^2+m_r^2)+\Pi(p^2)+\delta m^2}\label{dr},
\ee
\no where $m_r^2$ is the renormalized mass, $p^2$ is the momentum and $\Pi(p^2)$ is the self-energy obtained from the unrenormalized color-averaged propagator $D_u=D_u^{ii}/N_c$,
\be
\Pi(p^2)=\frac{1-p^2D_u(p^2)}{D_u(p^2)}\label{pi}
\ee
\no and therefore encodes the deviation from the tree-level propagator
\be
D_u=\frac{1}{p^2+\Pi(p^2)}\nn.
\ee
\no The inclusion of the tree-level mass $m^2$ in the self-energy is technically convenient, as it avoids to use explicitly the scale $a$.

The renormalization scheme is
\bea
D^{ij}(\mu^2)&=&\frac{\delta^{ij}}{\mu^2+m_r^2}\label{prc}\\
\frac{\pd D^{ij}}{\pd p}(\mu^2)&=&-\frac{2\mu\delta^{ij}}{(\mu^2+m_r^2)^2}\label{dprc}\\
Z&=&\frac{2\mu-\frac{d\Pi(p^2)}{dp}(\mu^2)}{2\mu}\nn\\
\delta m^2&=&\frac{(\mu^2+m_r^2)\frac{d\Pi(p^2)}{dp}(\mu^2)-2\mu\Pi(\mu^2)}{2\mu}\nn,
\eea
\no with the renormalization scale $\mu$. In most of the paper the choice $\mu=1.5$ GeV and $m_r=m$ will be made. The effect of different choices will be investigated in sections \ref{ss:ssd} and \ref{ss:ssda}. Numerically, the constants are determined by linear interpolation between the two momenta values along the $x$-axis between which the actual value of $\mu$ is. The derivative of $\Pi$ is obtained by deriving the linear interpolation of $\Pi$ between both points analytically. Errors are determined by error propagation from the statistical bootstrap errors of the propagator \cite{Cucchieri:2006tf}.

Note that the statistical errors of $Z$ and $\delta m^2$ had been propagated back into the renormalized propagator in \cite{Maas:2016edk}. This is not done here. The reason is that the mass renormalization $\delta m^2$ is found to be very large in comparison to the fundamental case. If the error is back-propagated, this yielded very large errors, but those were highly correlated, i.\ e.\ substantially overestimated the statistical error. This was seen as having almost identical values for different derived quantities, but with error bars much larger than the curves suggest. To avoid this correlation, only the error on $\Pi$, determined by error propagation from the only direct lattice measurement of the bare propagator $D_u$ in \pref{pi}, was propagated into the renormalized propagator \pref{dr}. That can be equally well seen as defining the renormalization constants, rather than to determine them from the data.

Note that if at the pole location the propagator depends only on $|p|^2$ this is the analytically continued pole scheme. However, as will be seen in section \ref{s:ana}, this is not the case in general in the quenched theory.

\subsection{Numerical results and discretization dependence}\label{ss:numr}

\begin{figure}
\includegraphics[width=0.95\linewidth]{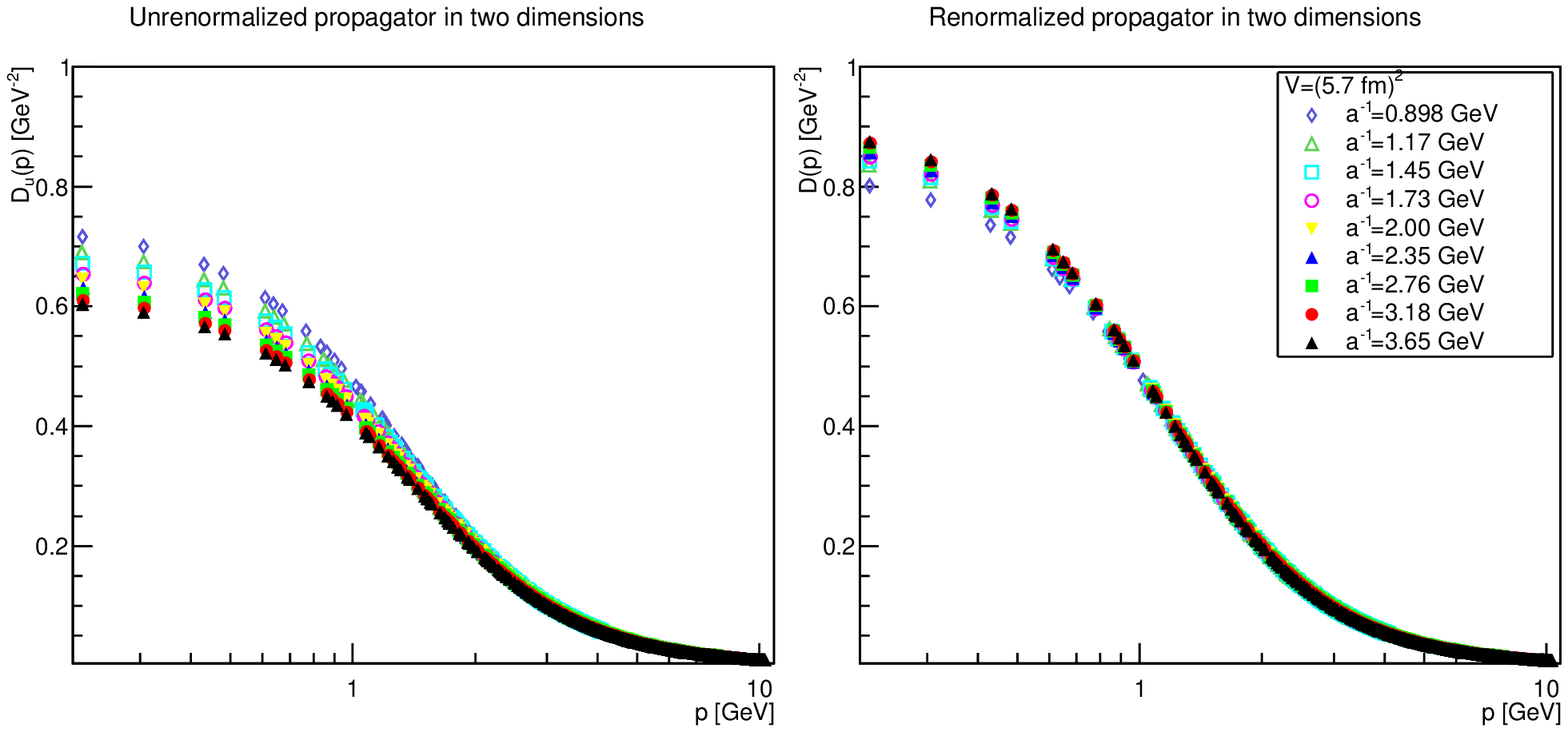}\\
\includegraphics[width=0.95\linewidth]{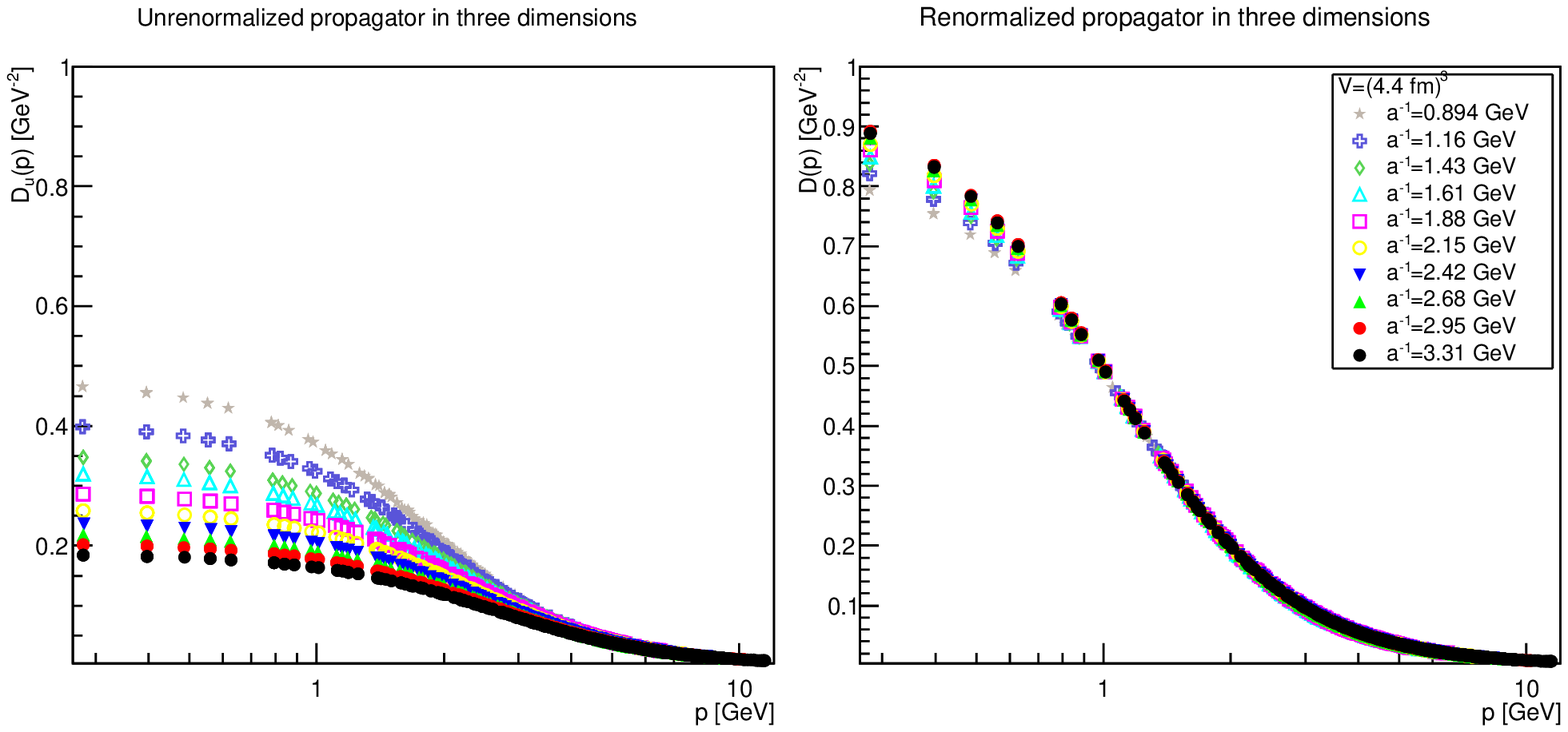}\\
\includegraphics[width=0.95\linewidth]{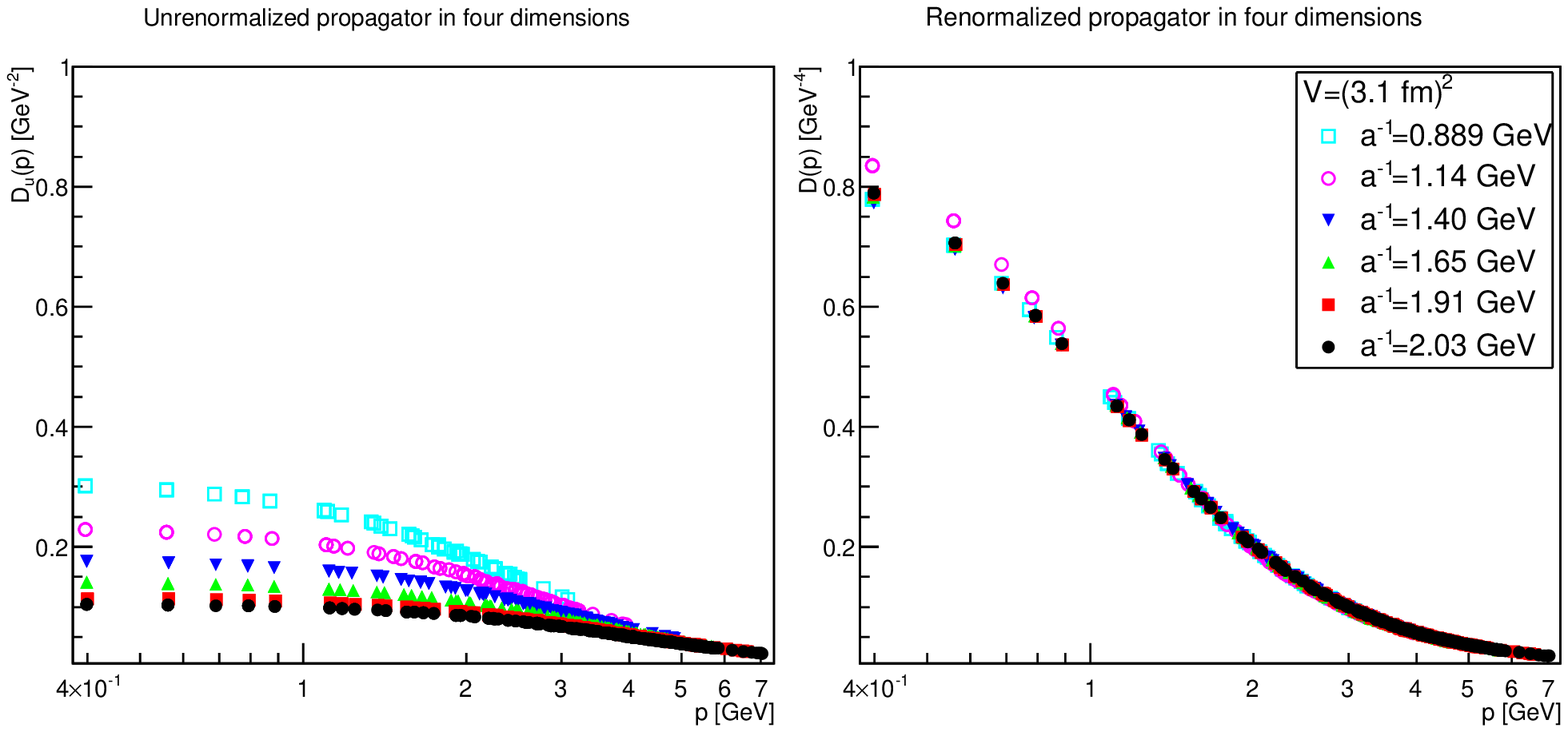}
\caption{\label{fig:ur}Unrenormalized propagator (left panels) and renormalized propagator (right panels). The top panels are for two dimensions, the middle panels for three dimensions, and the bottom panels for four dimensions. The values are $m_r=m=1$ GeV and $\mu=1.5$ GeV. If the (statistical 1$\sigma$) error bars here and hereafter are not visible then they are smaller than the symbol size.}
\end{figure}

\afterpage{\clearpage}

The effect of renormalization is shown in figure \ref{fig:ur}. It is clearly seen that the unrenormalized propagator depends substantially on $a$, the more the higher the dimension. This dependency is almost removed by the renormalization prescription of section \ref{ss:scheme} if $a^{-1}\gtrapprox 2$ GeV. Only a slight dependency is left afterwards, decreasing the finer the lattice gets. The residual dependency on $a$ is actually slightly larger the smaller the masses, but the general trend is the same. This is likely also affected by a mixing with finite-volume effects.

Nonetheless, this remainder systematic discretization error is substantially larger than for the fundamental case \cite{Maas:2016edk}, even though the dependency on $a$ without renormalization is in both cases similar. This will have consequences throughout the rest of the investigation. Therefore, it will be necessary to often assess the volume-dependence and $a$-dependence independently, rather than just looking at the finest lattices as it was possible in the fundamental case \cite{Maas:2016edk}. To the author, it is not clear where this difference originates from, and will be taken here as an observation which has to be taken into account when judging the results.

Note that, like for the fundamental case \cite{Maas:2016edk}, the differences mainly arise in the infrared, indicating that the mass renormalization is stronger $a$-dependent than the wave-function renormalization, as expected from perturbation theory. This will be confirmed below in the systematic analysis in section \ref{ss:renvc}

As in the fundamental case \cite{Maas:2016edk} any attempt to improve systematic uncertainties by using more points in the interpolation for the determination of the renormalization constants were more than offset by the increase of the statistical errors. In fact, the statistical fluctuation are found to be stronger for the adjoint case than for the fundamental case. Thus, the linear interpolation described in section \ref{ss:scheme} will be used throughout. Still, in cases where the renormalization point is mainly dominated by a single of the two points in the interpolation this can induce an additional systematic error, as is seen, e.\ g., at $a^{-1}=1.14$ GeV in four dimensions in figure \ref{fig:ur}. This case will happen less and less the closer the lattice parameters are to the thermodynamic limit.

\subsection{Scale and scheme dependence}\label{ss:ssd}

Of course, the choice of scheme in section \ref{ss:scheme} is completely arbitrary. To test the impact of this choice on the propagator in momentum space both the renormalized mass and the renormalization scale will be varied. However, this can give only then a reasonable estimate of the effects if the range is not probing the extremes of the lattice, requiring a sufficiently fine resolution. Also, the bare mass should be sufficiently far away from the extremes of the lattice. Thus, in the following the case $m=1$ GeV will be considered, requiring volumes for which lattice spacings $a^{-1}\gtrsim(2$ GeV)$^{-1}$ are available. Furthermore, for the sake of comparability the same physical volumes will be used as in \cite{Maas:2016edk}.

\begin{figure}
\includegraphics[width=0.475\linewidth]{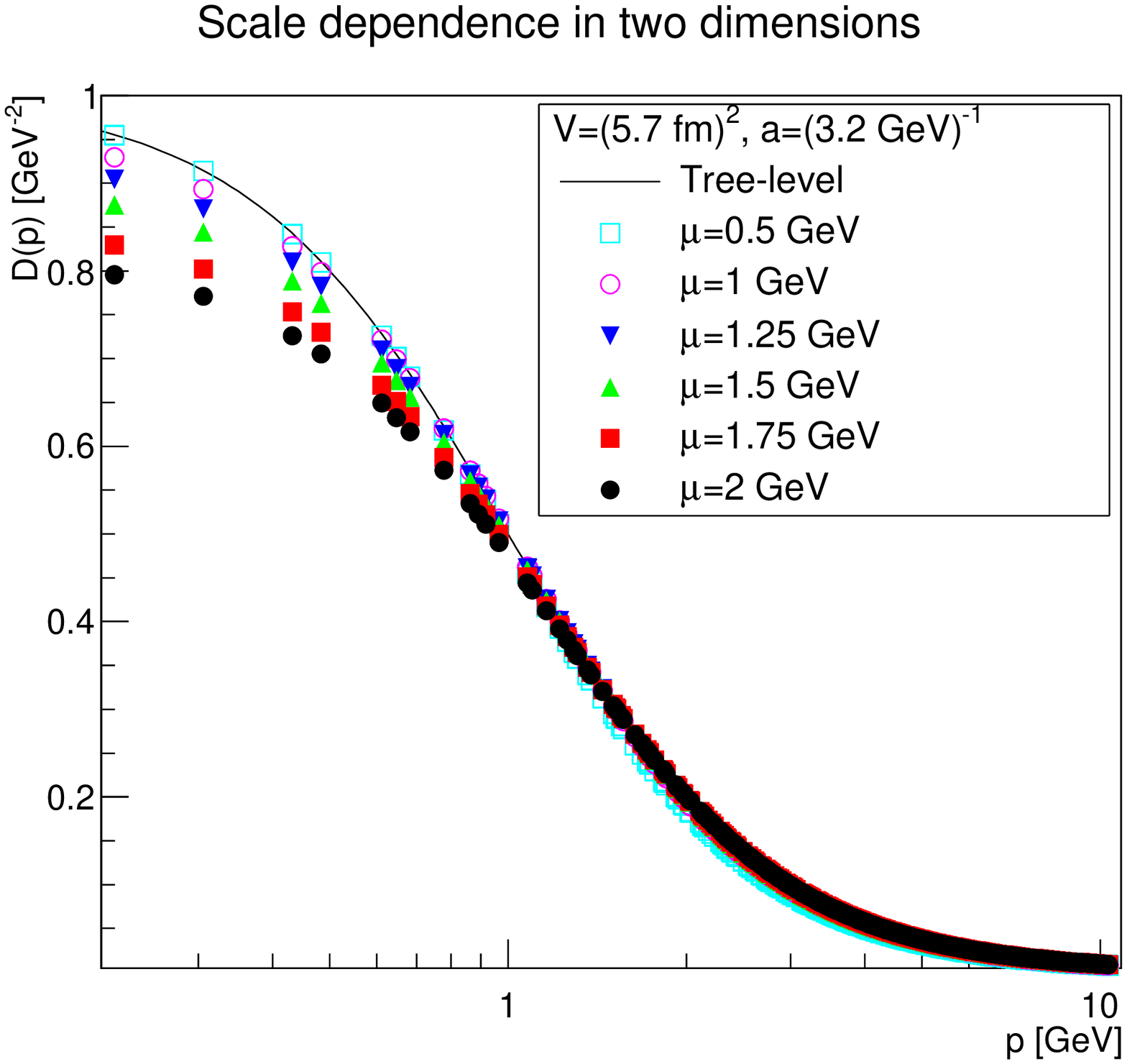}\includegraphics[width=0.475\linewidth]{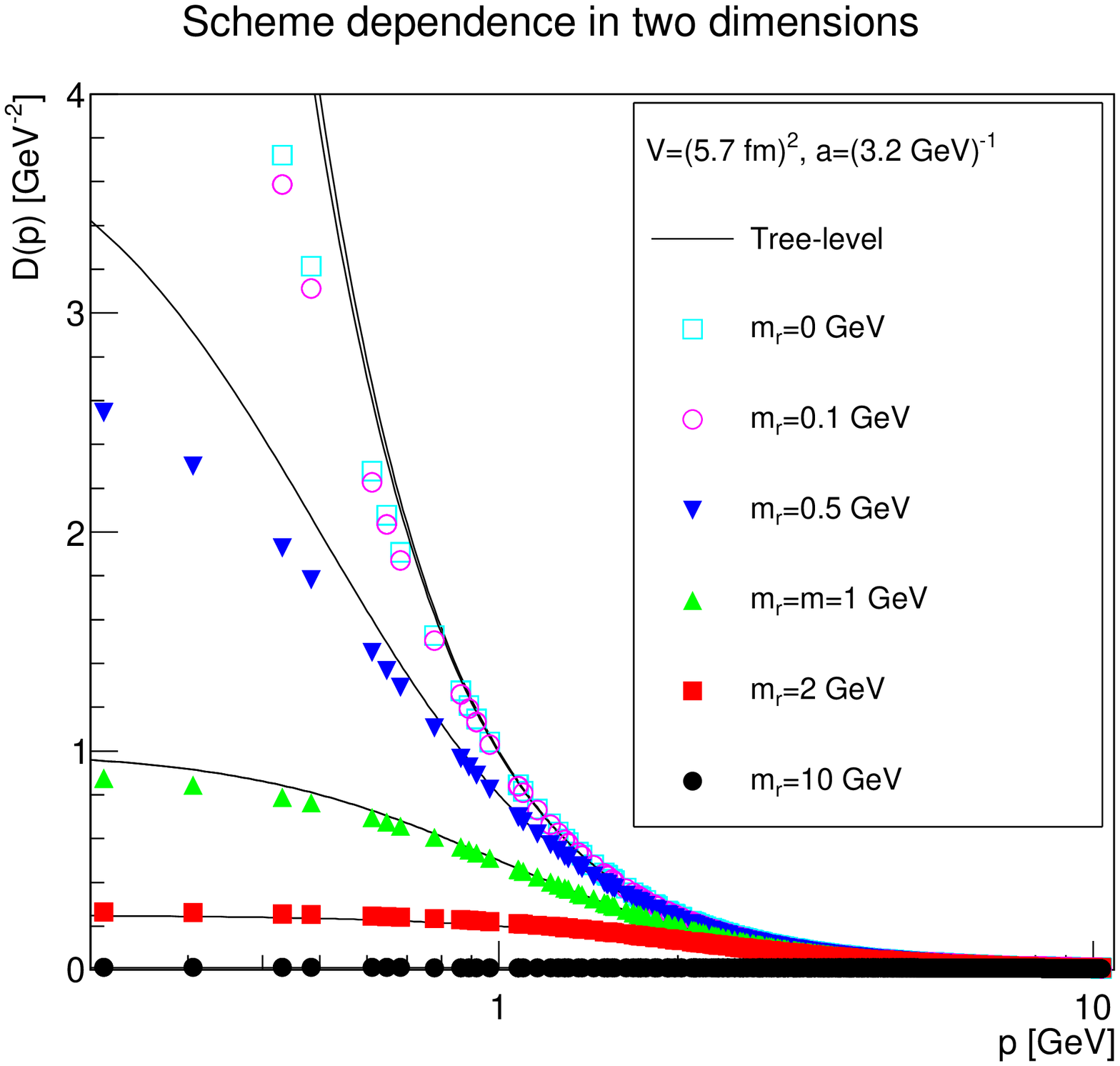}\\
\includegraphics[width=0.475\linewidth]{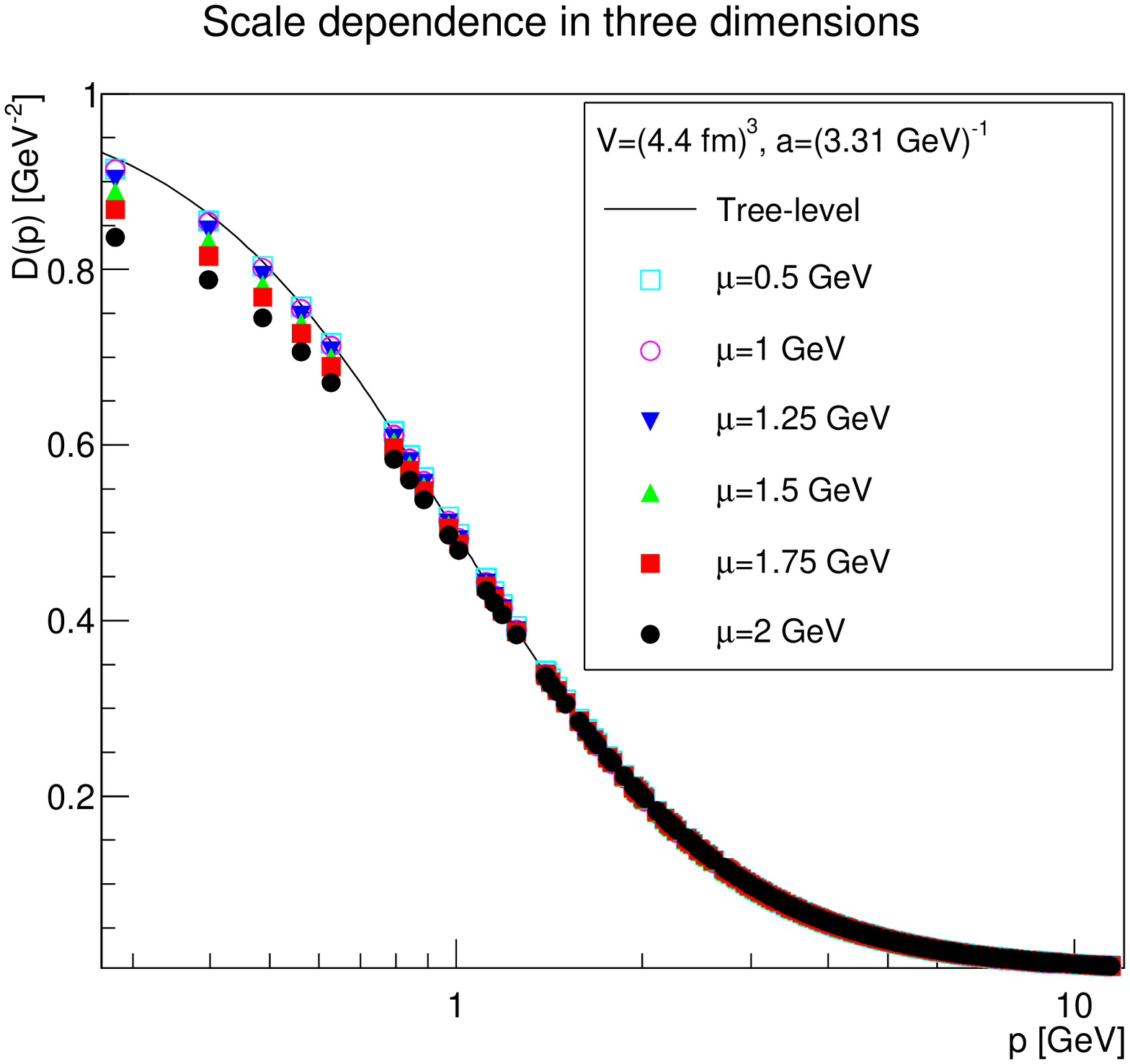}\includegraphics[width=0.475\linewidth]{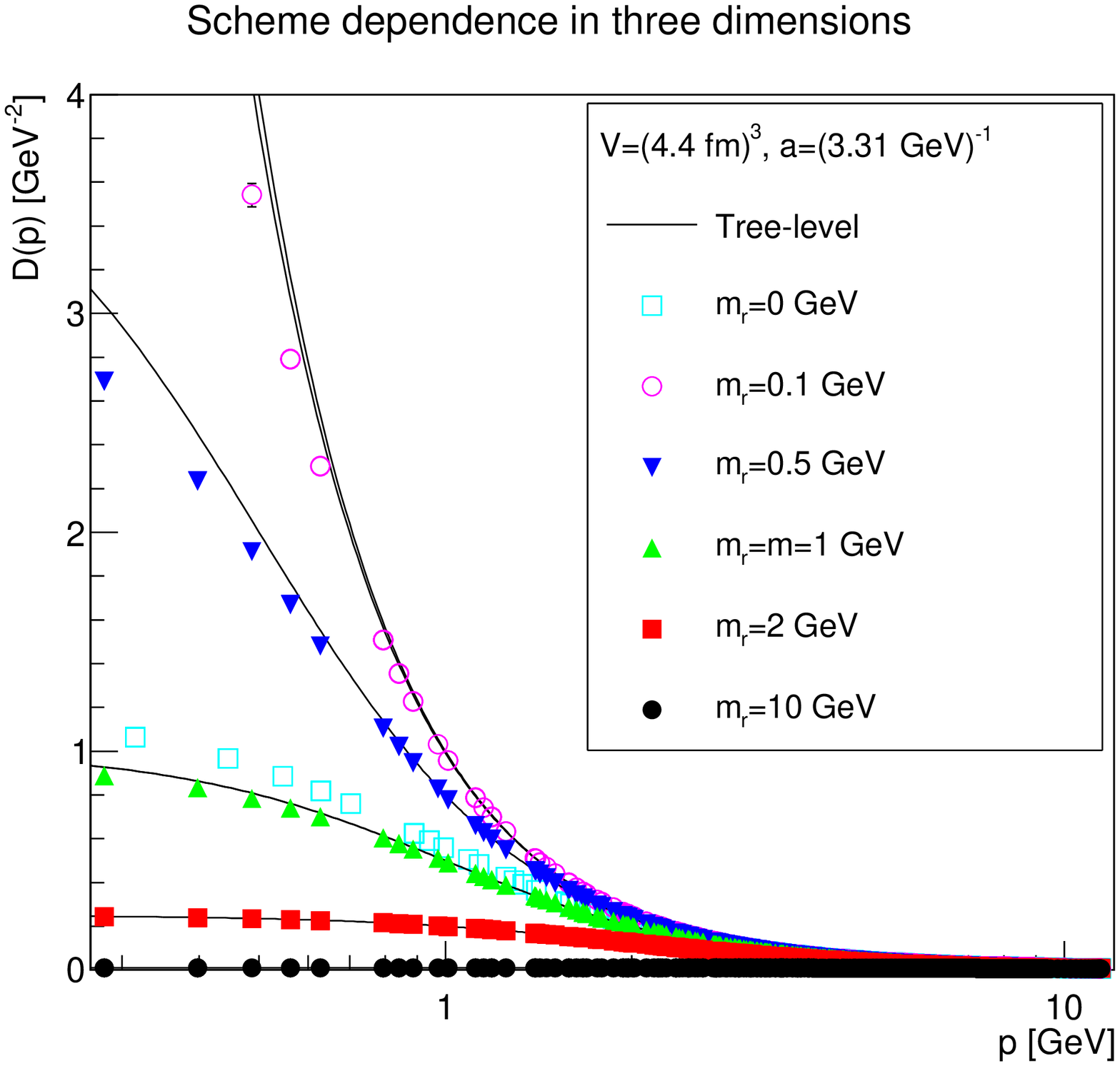}\\
\includegraphics[width=0.475\linewidth]{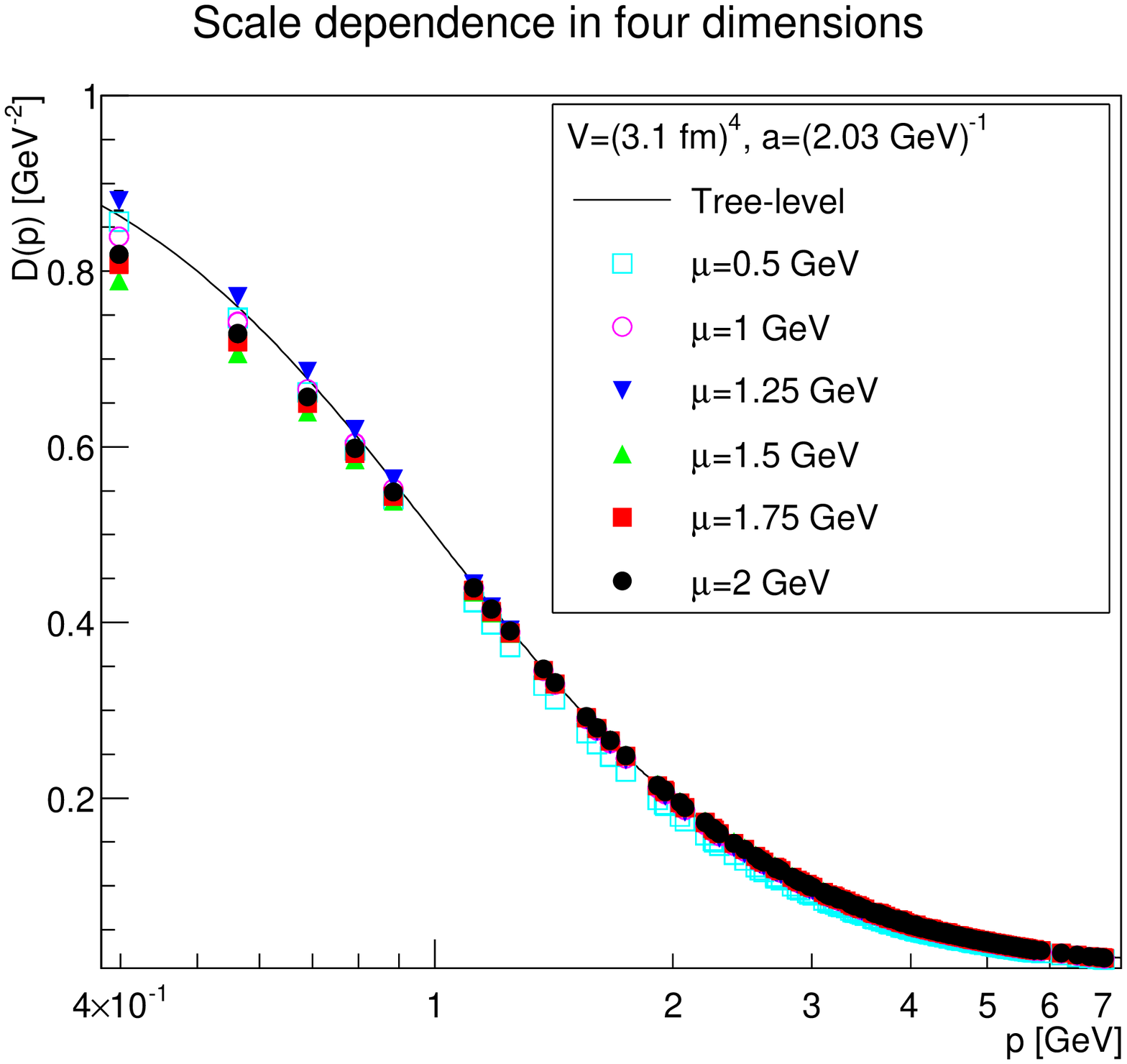}\includegraphics[width=0.475\linewidth]{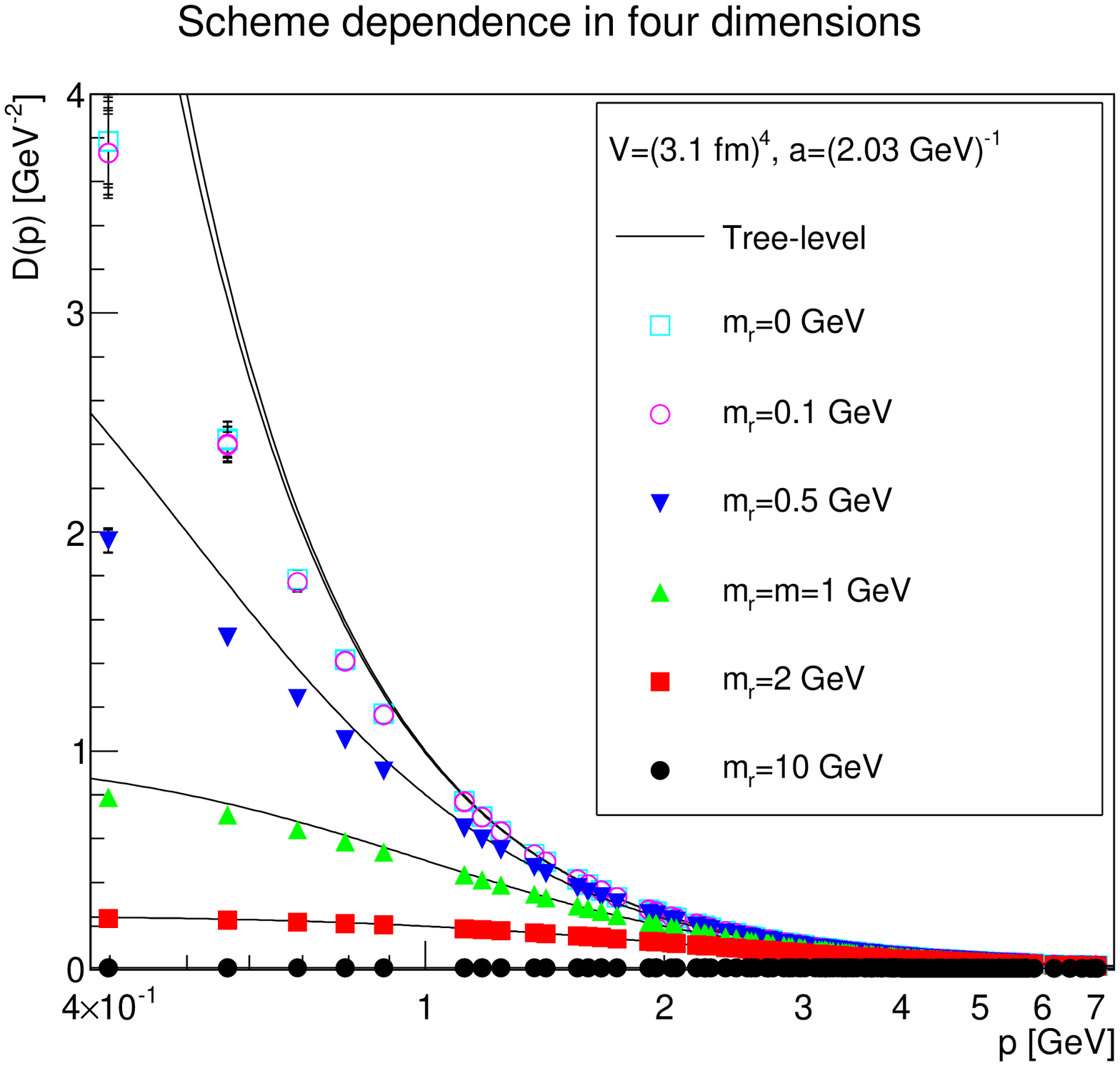}
\caption{\label{fig:ren}Scale-dependence at $m_r=m=1$ GeV (left panels) and scheme dependence at $\mu=1.5$ GeV (right panels) of the renormalized propagator. The top panels are two dimensions, the middle panels three dimensions, and the bottom panels four dimensions. The tree-level propagator is shown for comparison as a full line.}
\end{figure}

\afterpage{\clearpage}

The results are shown for both scale and scheme dependence in figure \ref{fig:ren}. The dependence on the scale is relatively mild. Because of the derivative condition \pref{dprc}, the change of scale leads to a tilting of the propagator, due to its monotonous behavior. That the strongest effect is seen in the infrared indicates already that the largest deviation from tree-level will be encountered there. This is also emphasized by the scheme dependence. When introducing a large mass scale by the renormalization a closer resemblance to tree-level is obtained. However, if introducing a smaller mass scale larger deviations are seen. It appears that there is an intrinsic mass-scale, similar to the fundamental case \cite{Maas:2016edk}, which adds to the mass scale introduced by the scheme. This will be investigated, and confirmed, further in section \ref{s:ana}. At any rate, there is little difference between the different dimensionality.

\subsection{Dependence of the renormalization constants on the volume and the cutoff}\label{ss:renvc}

For unquenched simulations \cite{Maas:2013aia} it is much harder to find lines-of-constant physics. In addition, investigating multiple volumes is expensive due to the amount of configurations necessary for spectroscopy. Hence, it is quite important to know how renormalization needs to be performed as a function of lattice parameters, and how it is influenced by discretization and finite-volume artifacts. For the fundamental case \cite{Maas:2016edk}, it was found that the functional dependence on $a^{-1}$ was of the qualitative form expected from perturbation theory \cite{Bohm:2001yx}, and a dependence on volume was quasi non-existent, even for rather small volumes. This permits to obtain high-precision renormalization constants on small volumes to be used on larger volumes. It will be seen that the same is true for the adjoint case. Note that in the following only the standard scheme $m_r=m$ with a renormalization scale $\mu=1.5$ GeV will be investigated. In other schemes this finding could not be true.

\begin{figure}
\includegraphics[width=0.475\linewidth]{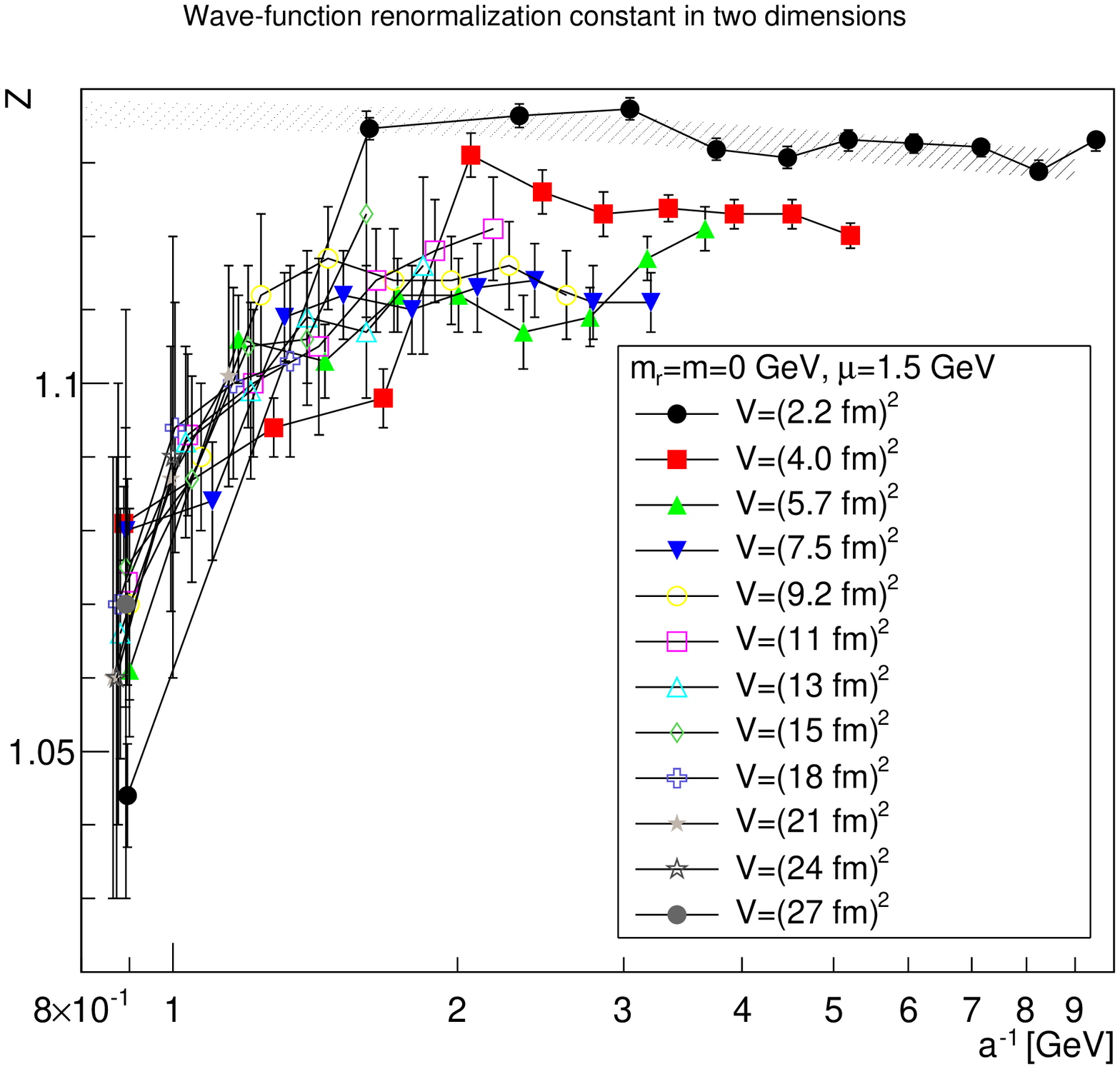}\includegraphics[width=0.475\linewidth]{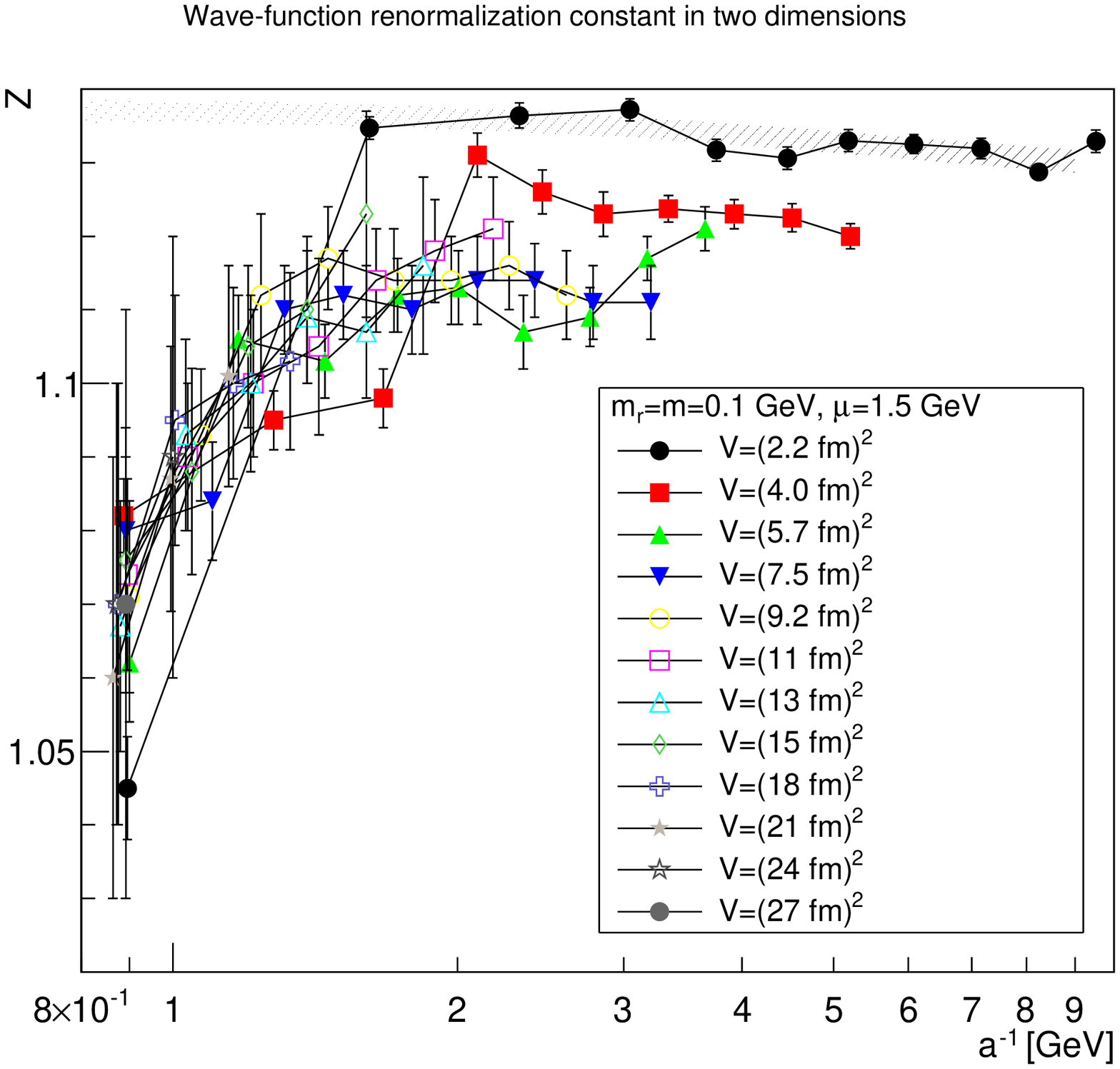}\\
\includegraphics[width=0.475\linewidth]{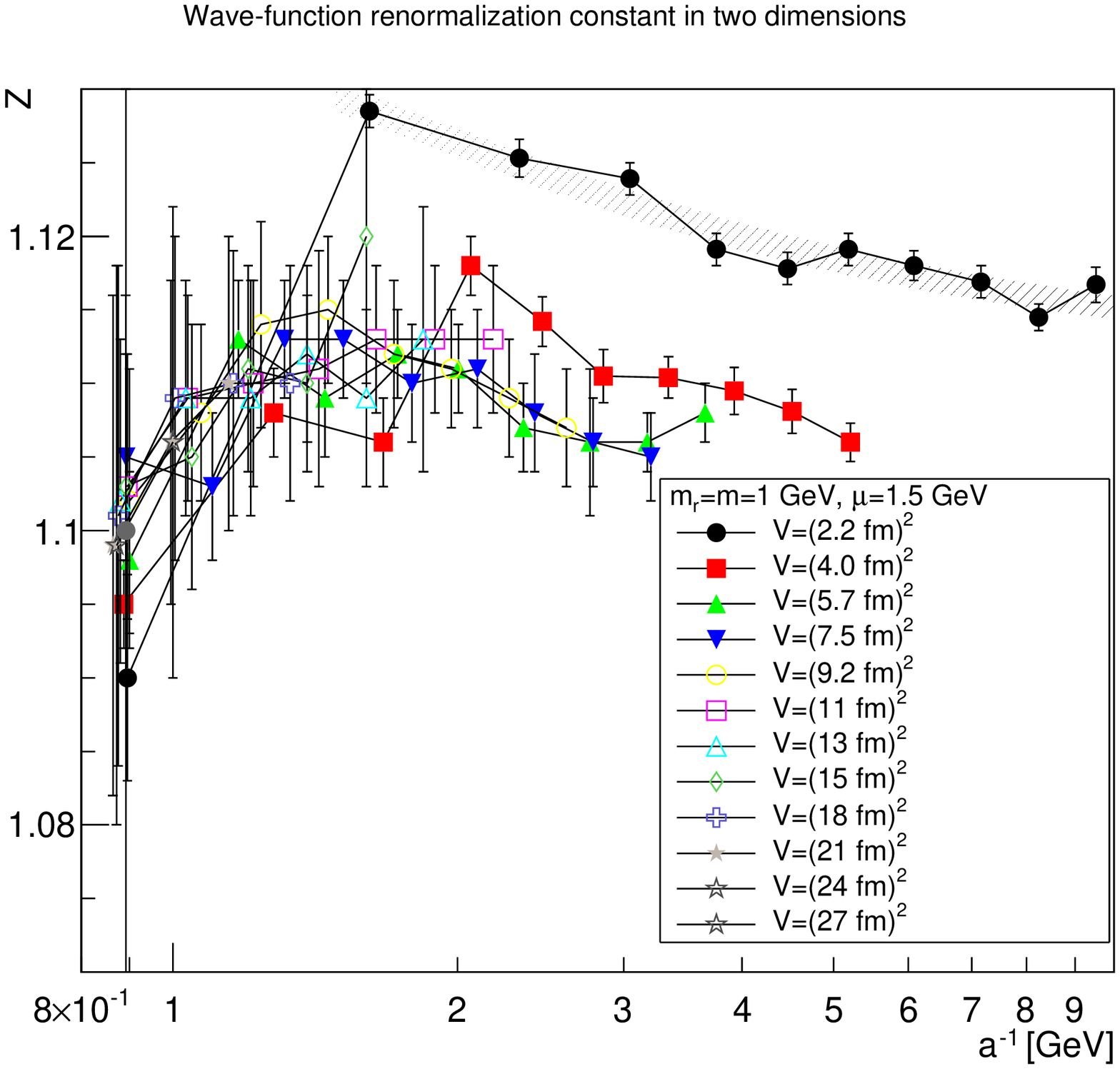}\includegraphics[width=0.475\linewidth]{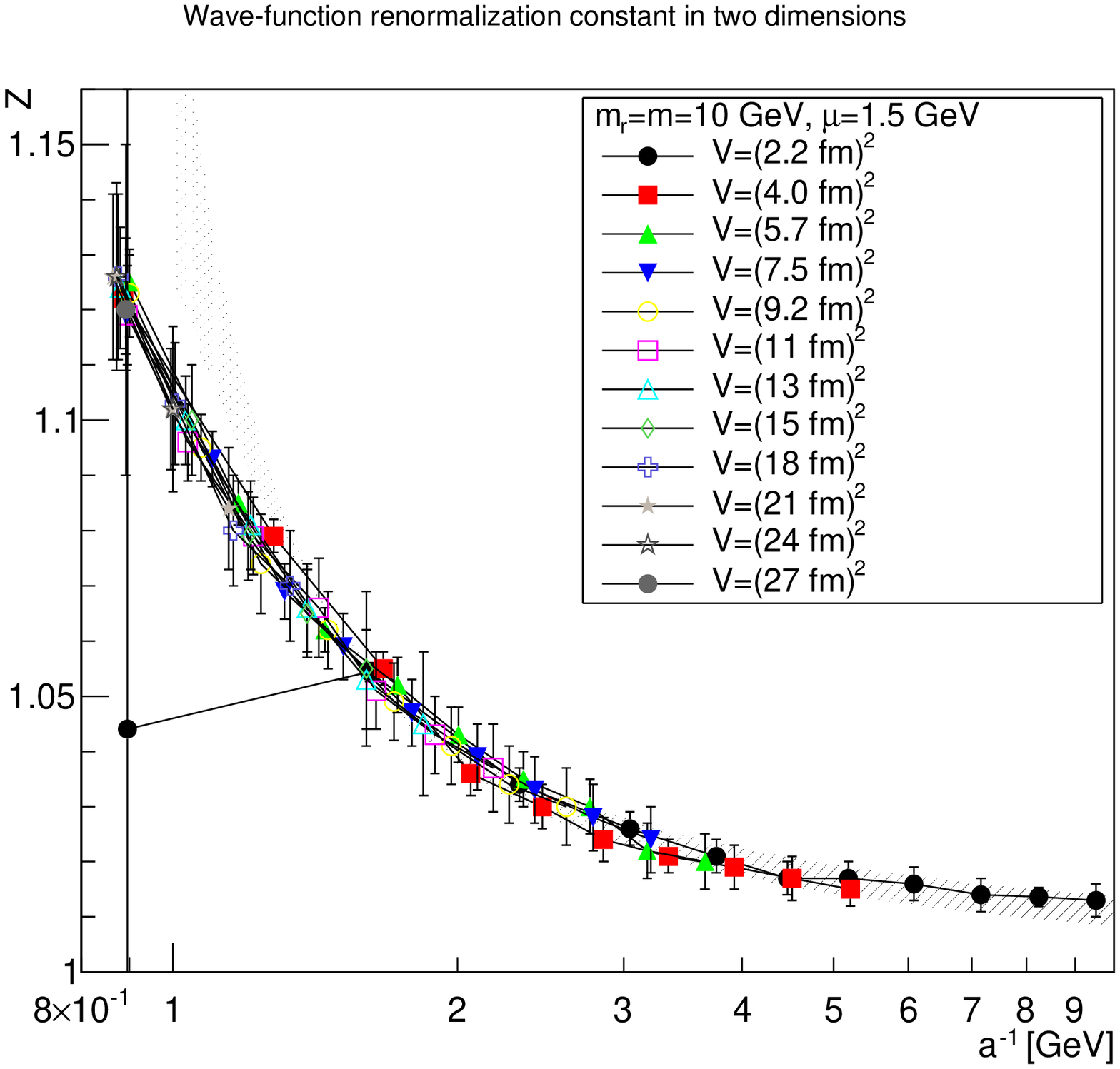}
\caption{\label{fig:z2}The wave-function renormalization constant as a function of the lattice cutoff and the lattice volume in two dimensions for $\mu=1.5$ GeV. The top-left panel shows the case of $m=m_r=0$ GeV, the top-right panel of $m=m_r=0.1$ GeV, the bottom-left panel of $m=m_r=1$ GeV, and the bottom-right panel of $m=m_r=10$ GeV. The hatched band is the fit \pref{zfit} with the parameters given in table \ref{fitsz}.}
\end{figure}

\begin{figure}
\includegraphics[width=0.475\linewidth]{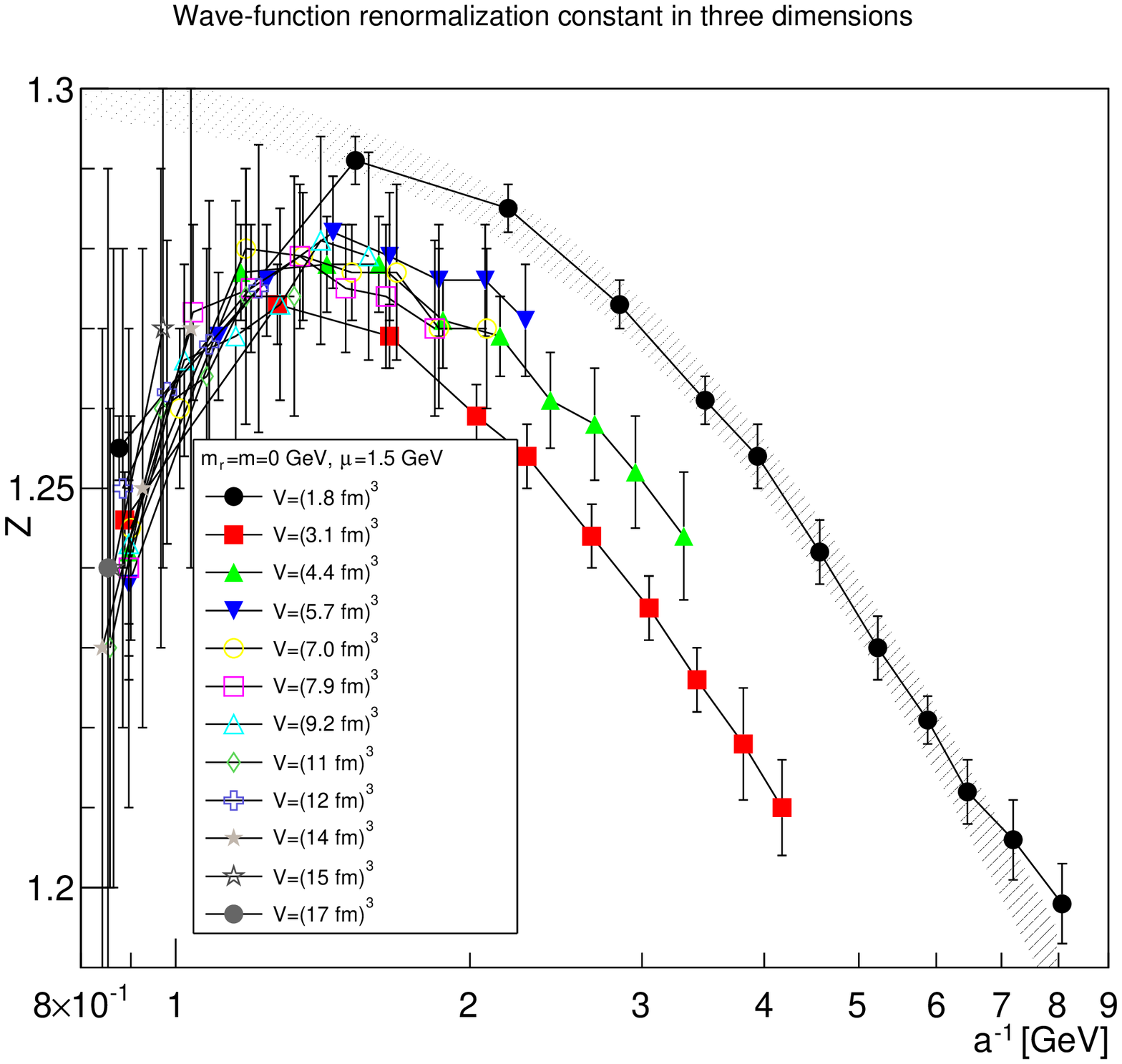}\includegraphics[width=0.475\linewidth]{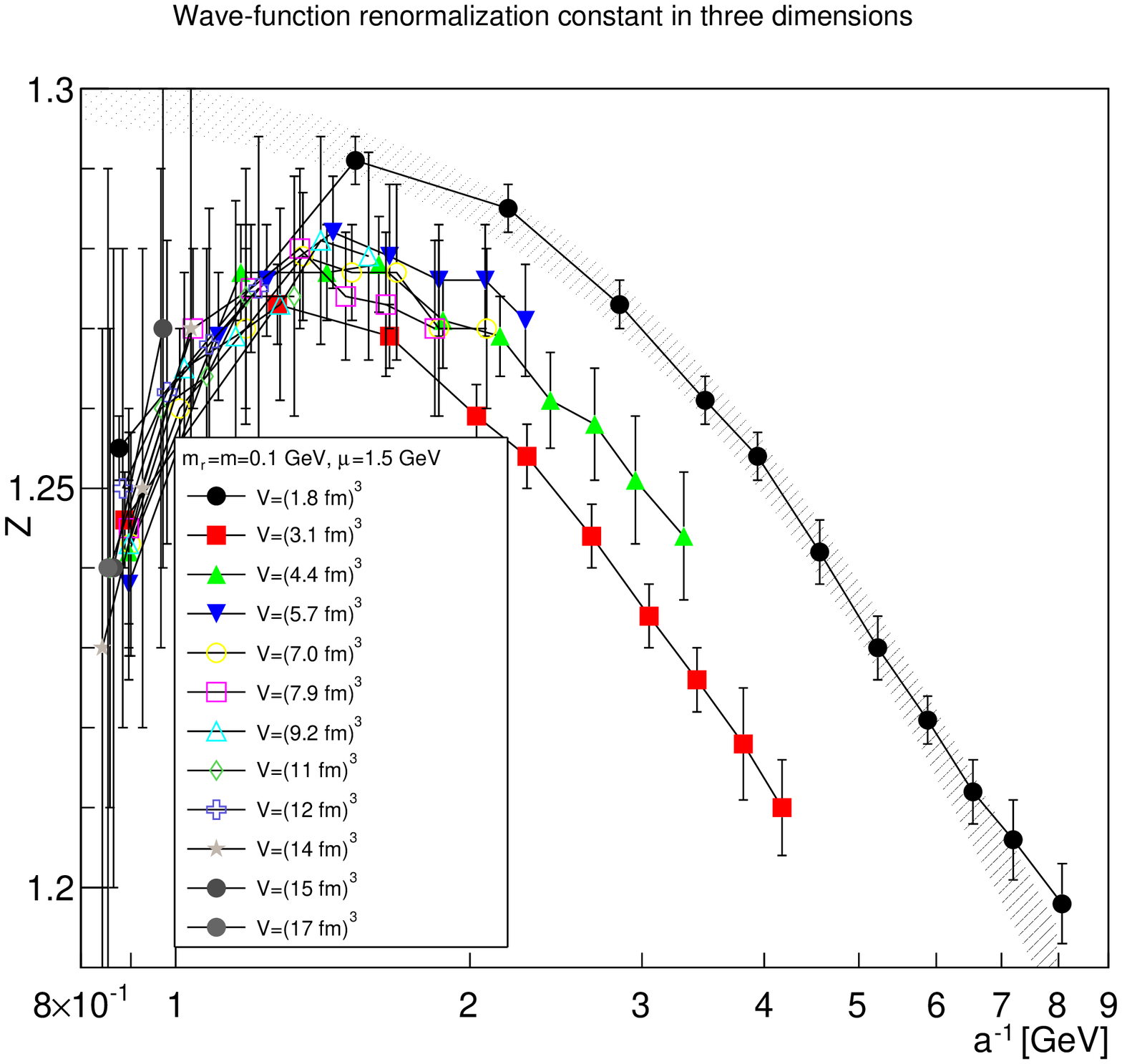}\\
\includegraphics[width=0.475\linewidth]{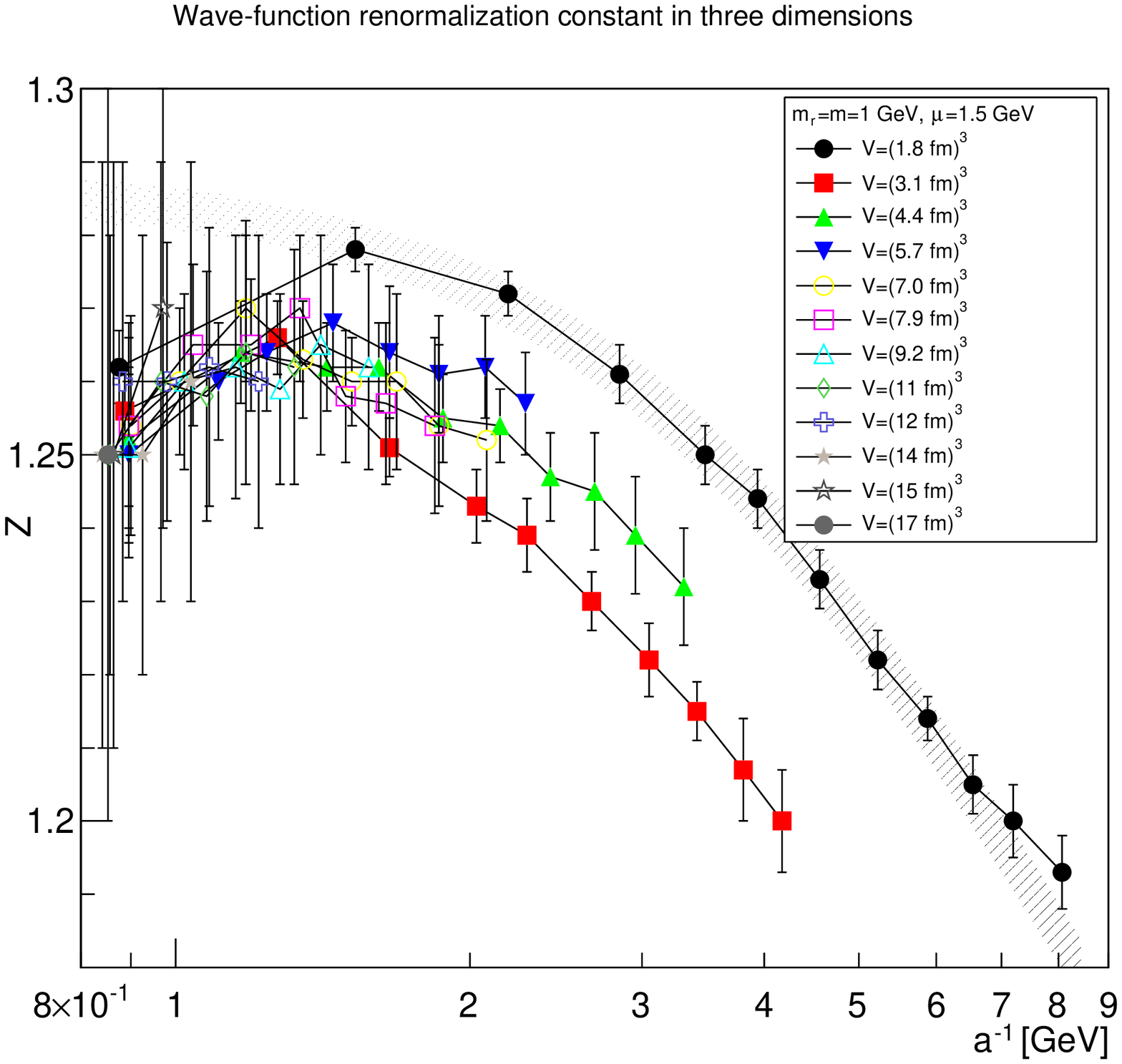}\includegraphics[width=0.475\linewidth]{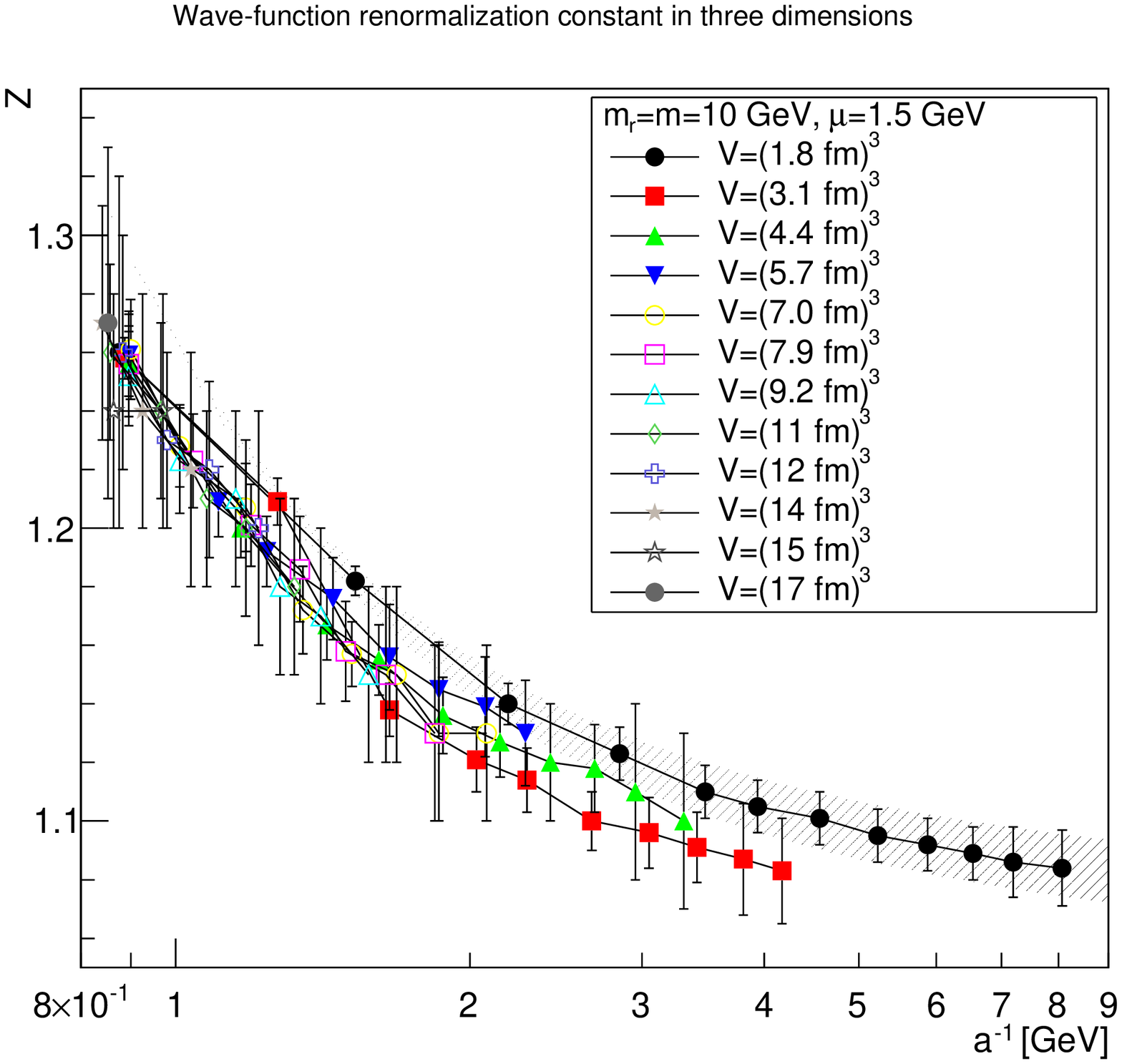}
\caption{\label{fig:z3}The wave-function renormalization constant as a function of the lattice cutoff and the lattice volume in three dimensions for $\mu=1.5$ GeV. The top-left panel shows the case of $m=m_r=0$ GeV, the top-right panel of $m=m_r=0.1$ GeV, the bottom-left panel of $m=m_r=1$ GeV, and the bottom-right panel of $m=m_r=10$ GeV. The hatched band is the fit \pref{zfit} with the parameters given in table \ref{fitsz}.}
\end{figure}

\begin{figure}
\includegraphics[width=0.475\linewidth]{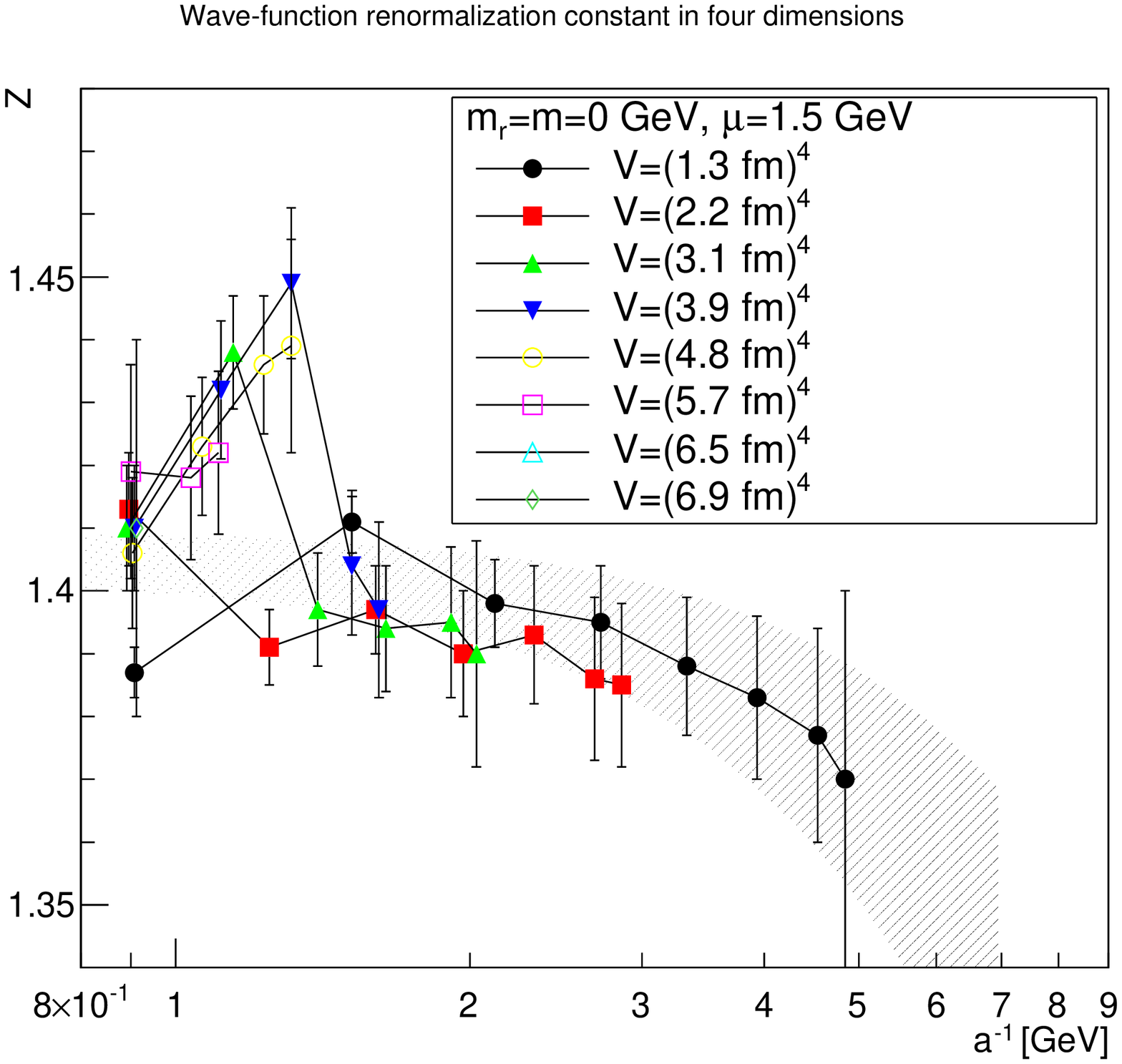}\includegraphics[width=0.475\linewidth]{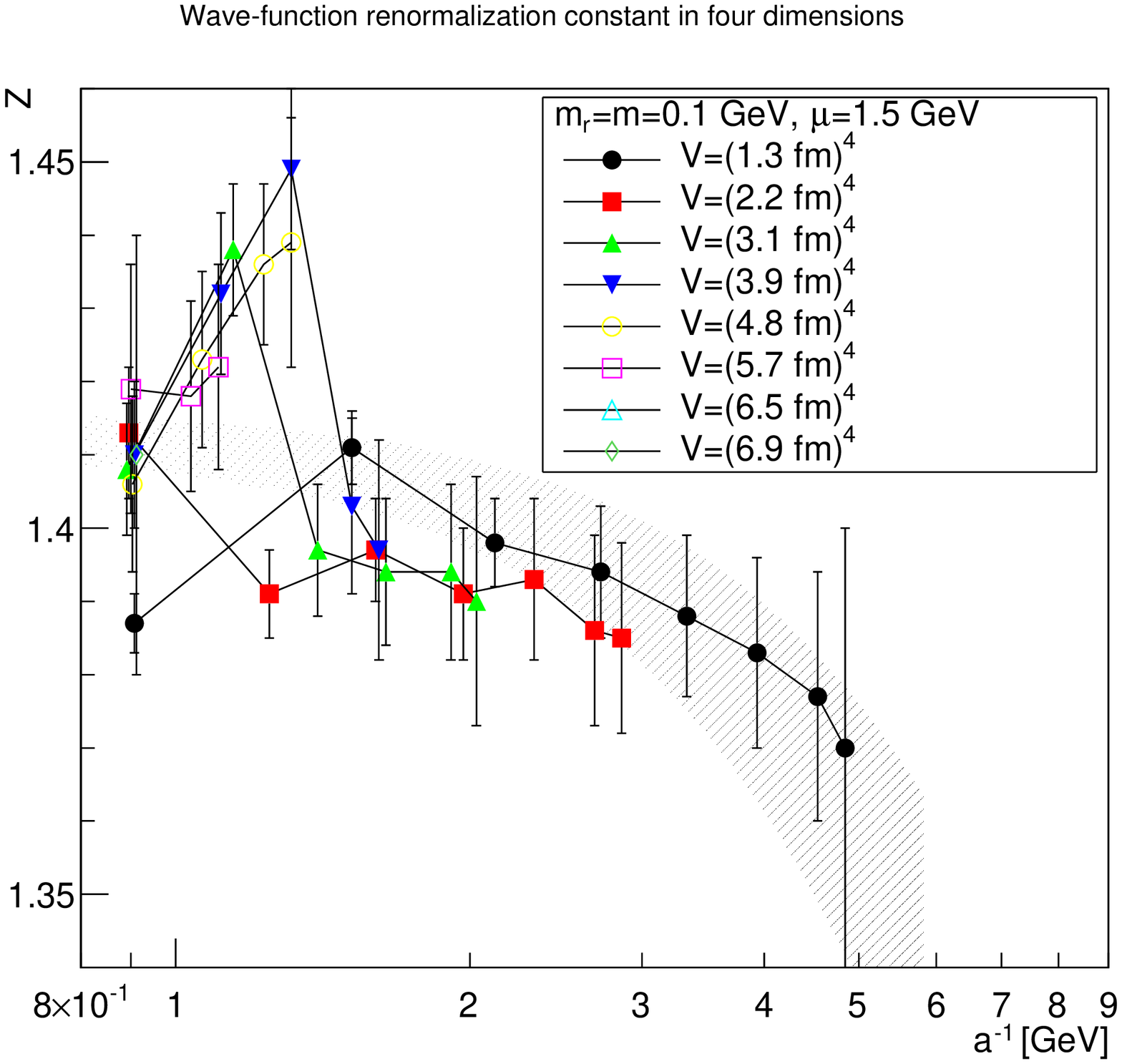}\\
\includegraphics[width=0.475\linewidth]{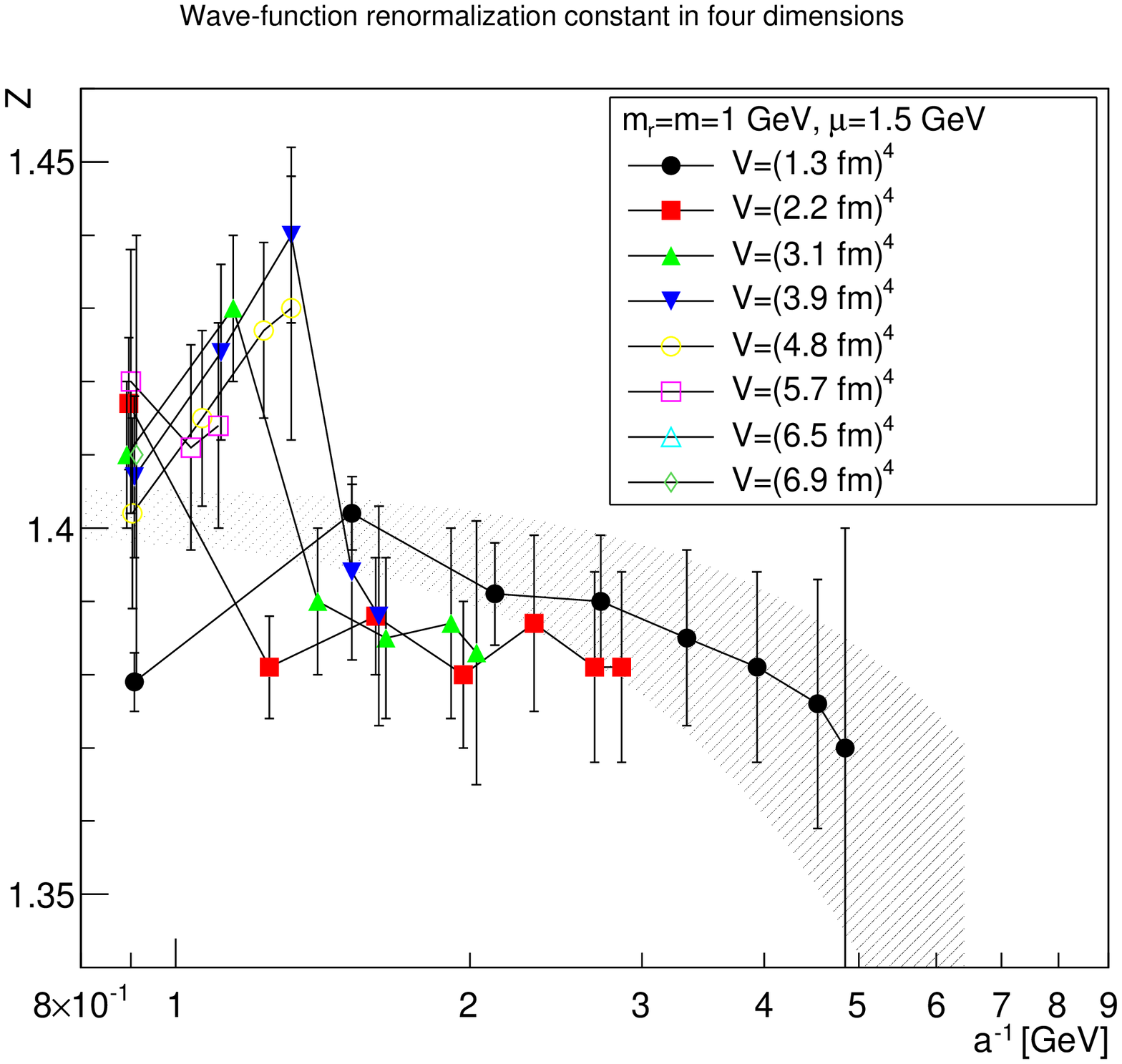}\includegraphics[width=0.475\linewidth]{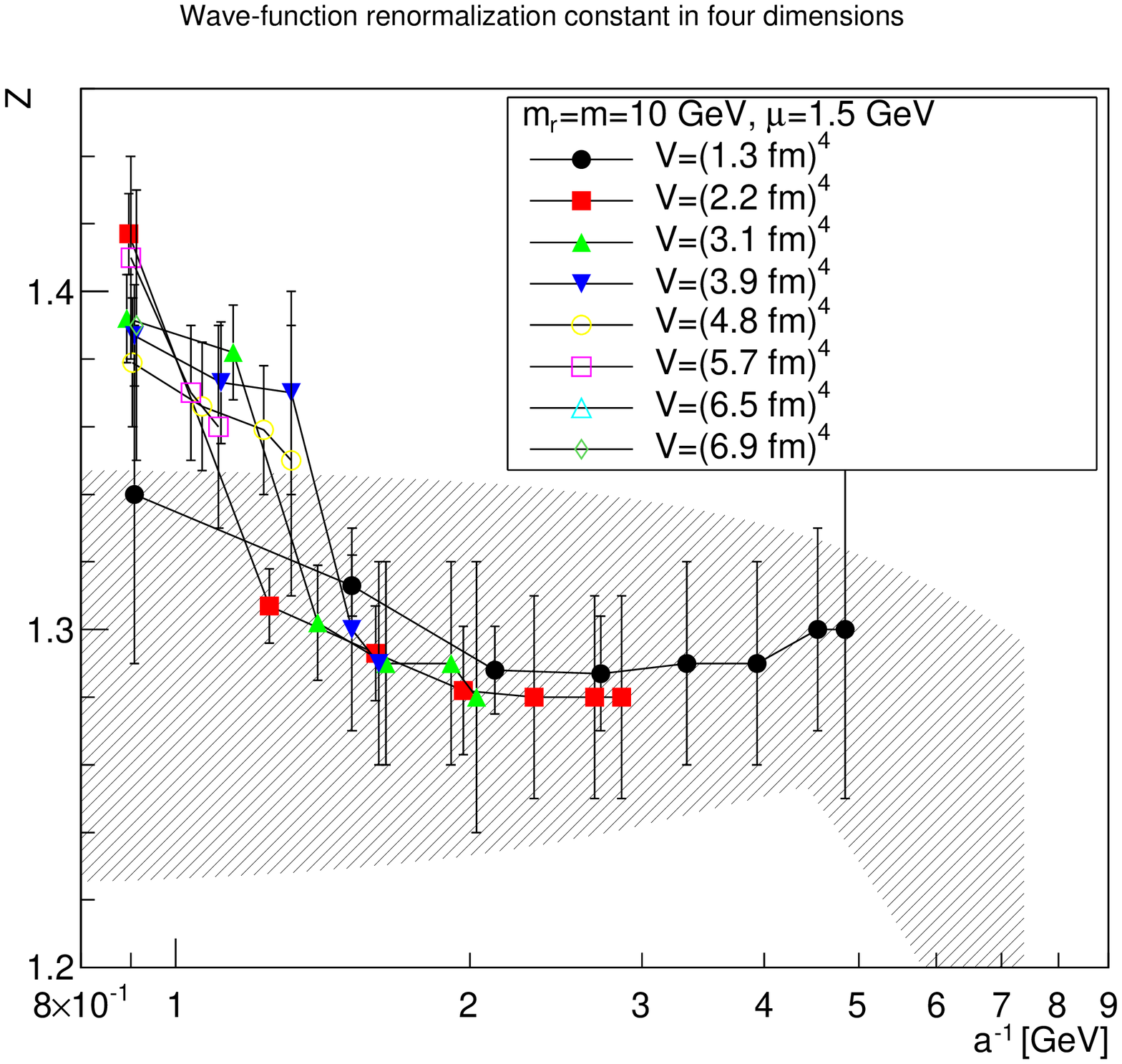}
\caption{\label{fig:z4}The wave-function renormalization constant as a function of the lattice cutoff and the lattice volume in four dimensions for $\mu=1.5$ GeV. The top-left panel shows the case of $m=m_r=0$ GeV, the top-right panel of $m=m_r=0.1$ GeV, the bottom-left panel of $m=m_r=1$ GeV, and the bottom-right panel of $m=m_r=10$ GeV. The hatched band is the fit \pref{zfit} with the parameters given in table \ref{fitsz}.}
\end{figure}

The wave-function renormalization is shown in figures \ref{fig:z2}-\ref{fig:z4} for varying dimensionality. In all cases it decreases eventually with increasing $1/a$. There is some finite-volume dependence visible, especially in three dimensions. However, this seems to be essentially only a rescaling, as for the fundamental case \cite{Maas:2016edk}. The qualitative behavior seems to be volume-independent. There is also a jumping behavior in four dimensions when $a^{-1}$ crosses a critical, volume-dependent value. This hints to a mass-scale, which increases with increasing volume, which has to be crossed before the asymptotic behavior can be reached. Indeed, there is a mass scale different from the renormalized mass present, as will be discussed in section \ref{s:ana}.

\begin{longtable}[!H]{|c|c|c|c|c|}
\caption{\label{fitsz}Fit parameters of \pref{zfit} for the wave-function renormalization constants in the standard scheme. A value of 0 for $Z_\infty$ indicates that no stable fit with a non-zero value for $Z_\infty$ could be found.}\cr
\hline
$d$	& m [GeV]	& $Z_\infty$	& $c$ & $\Lambda$ [GeV]\cr
\hline	\endfirsthead
\hline
\multicolumn{5}{|l|}{Table \ref{fitsz} continued}\cr
\hline
$d$	& m [GeV]	& $Z_\infty$	& $c$ & $\Lambda$ [GeV]\cr
\hline\endhead
\hline
\multicolumn{5}{|r|}{Continued on next page}\cr
\hline\endfoot
\endlastfoot
\hline
2 & 0 &	1.116(6) & 0.058(1) & 4.4(4) \cr
\hline
2 & 0.1 & 1.118(3) & 0.053(6) & 4.1(3) \cr
\hline
2 & 1 & 1.1067(8) & 0.0400(14) & 1.88(5) \cr
\hline
2 & 10 & 0.997(3) & 0.06401(13) & 0.7(1) \cr
\hline
\hline
3 & 0 & 0.847(16) & 1.56(5) & 5.56(4) \cr
\hline
3 & 0.1 & 0.85(3) & 1.56(12) & 5.56(16) \cr
\hline
3 & 1 & 0.869(5) & 1.44(5) & 5.58(12) \cr
\hline
3 & 10 & 1.043(13) & 0.176(11) & 1.101(6) \cr
\hline
\hline
4 & 0 & 0 & 7.2(7) & 13(4) \cr
\hline
4 & 0.1 & 0 & 6.7(5) & 11(2) \cr
\hline
4 & 1 & 0 & 7.1(8) & 12(4) \cr
\hline
4 & 10 & 0 & 6.2$^{+1.1}_{-0.9}$ & 11(4) \cr
\hline
\end{longtable}

It is therefore useful to investigate the asymptotic behavior at large cutoffs, thus utilizing the smallest volume. The slow evolution and the fundamental case \cite{Maas:2016edk} suggest a logarithmic behavior. Indeed, an ansatz
\be
Z(a)=Z_\infty+\frac{b}{\ln\left(\frac{a^{-2}+\Lambda^2}{(1\text{ GeV})^2}\right)}\label{zfit}
\ee
\no provides a reasonable good approximation, as is visible in figures \ref{fig:z2}-\ref{fig:z4}. The fit parameters can be found in table \ref{fitsz}.

There are two interesting observations. The first is that in four dimensions $Z_\infty$ is zero, while it is non-zero in lower dimensions, in agreement with perturbative expectations \cite{Bohm:2001yx}. Incidentally, this already hints that the adjoint scalar is not a physical particle, due to the Oehme-Zimmermann superconvergence relation \cite{Oehme:1979ai}. It is also the same pattern as in the fundamental case \cite{Maas:2016edk}.

The second is that the fit parameters become increasingly mass-in\-de\-pen\-dent the higher the dimension. Especially, within errors, they are completely mass-independent in four dimensions. This is the more remarkable as $a^{-1}$ is in four dimensions at most of the order of the largest mass. Interestingly, also the scale $\Lambda$ is quite large, much larger than in the fundamental case \cite{Maas:2016edk}, where it is essentially 1 GeV. This is in-line with results from the fermion sector \cite{Aguilar:2010ad,August:2013jia} as well as from glueball masses \cite{Mathieu:2008me} which suggest an intrinsically higher value for adjoint scales than for fundamental scales.

\begin{figure}
\includegraphics[width=0.475\linewidth]{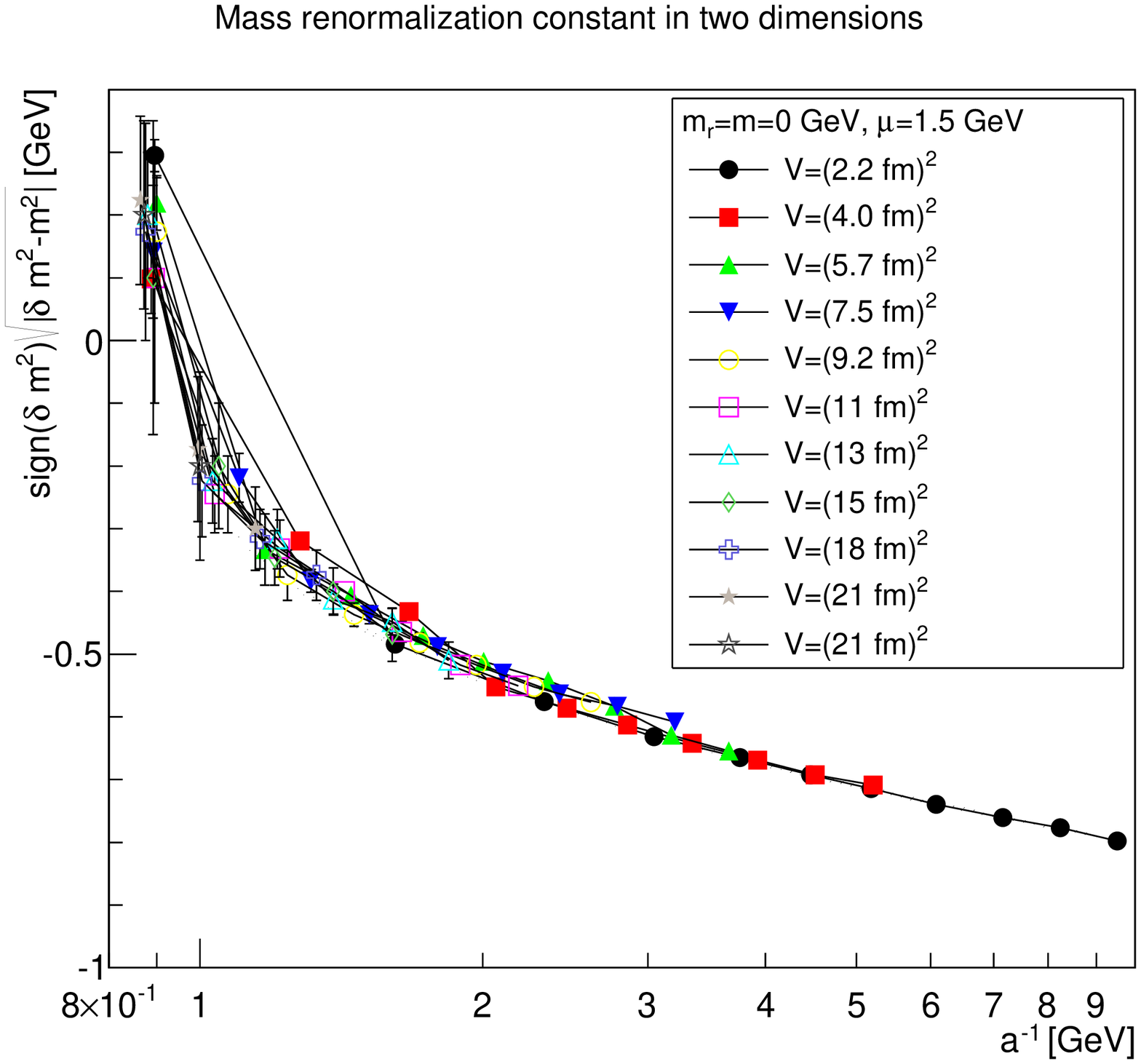}\includegraphics[width=0.475\linewidth]{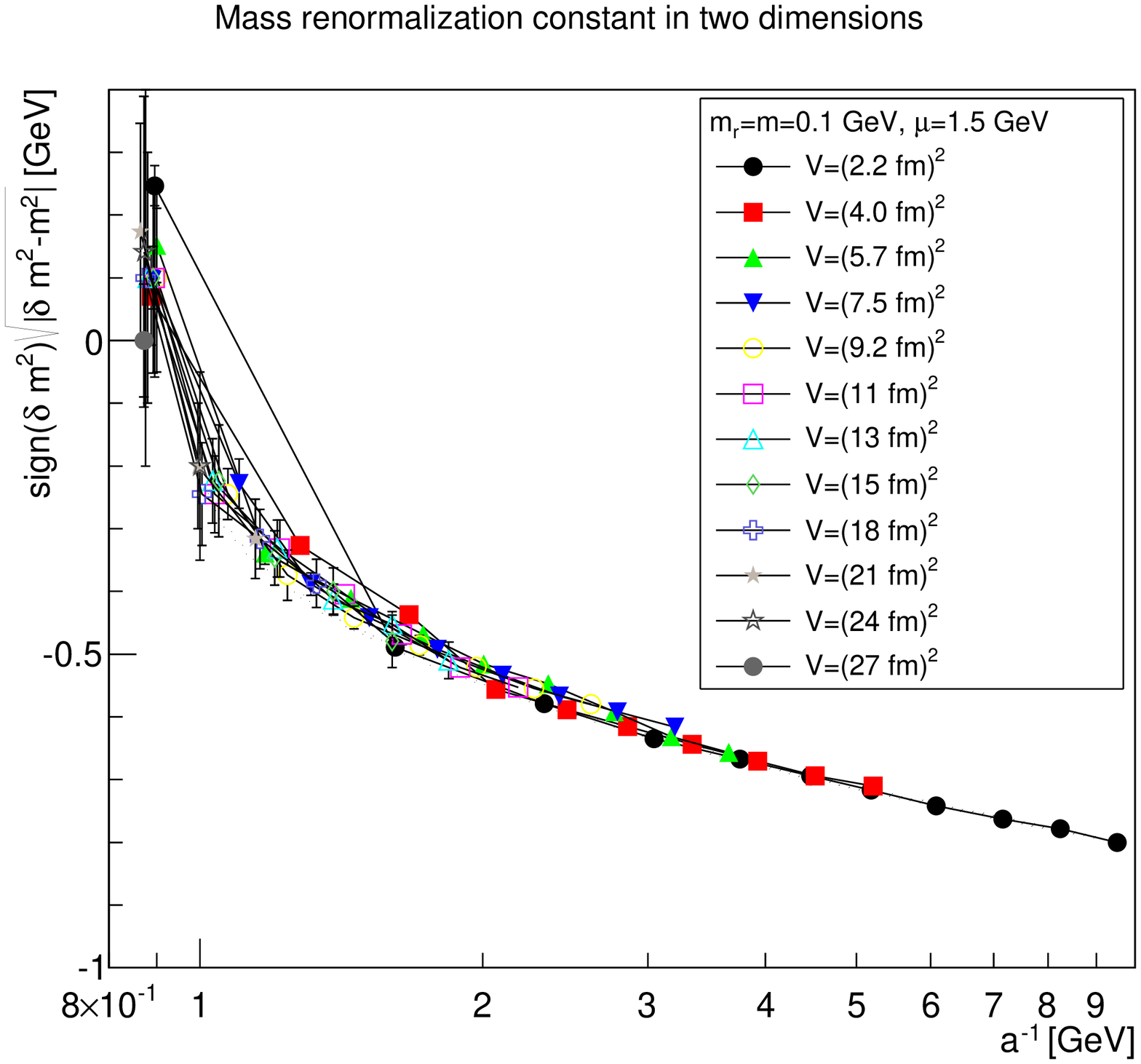}\\
\includegraphics[width=0.475\linewidth]{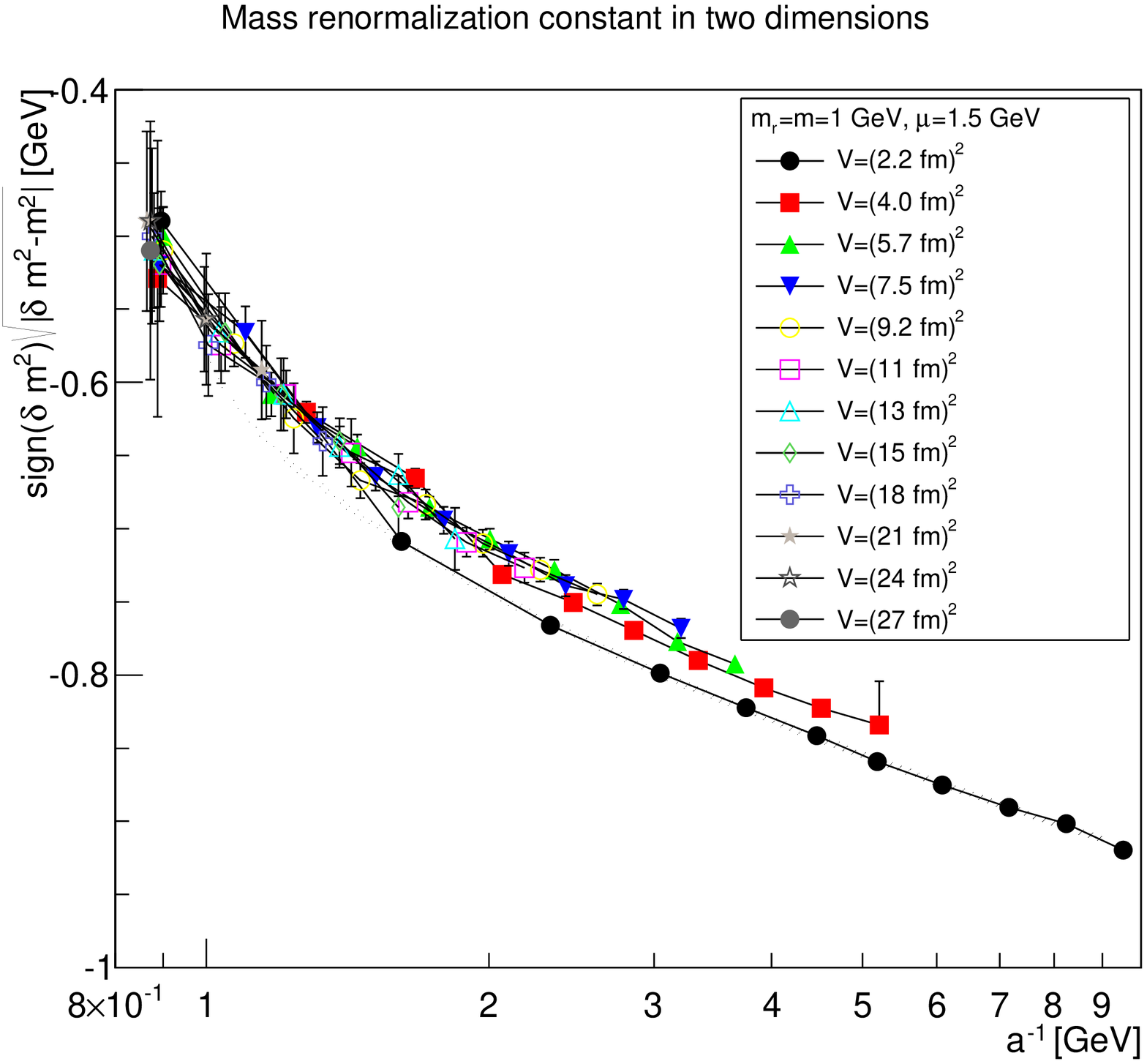}\includegraphics[width=0.475\linewidth]{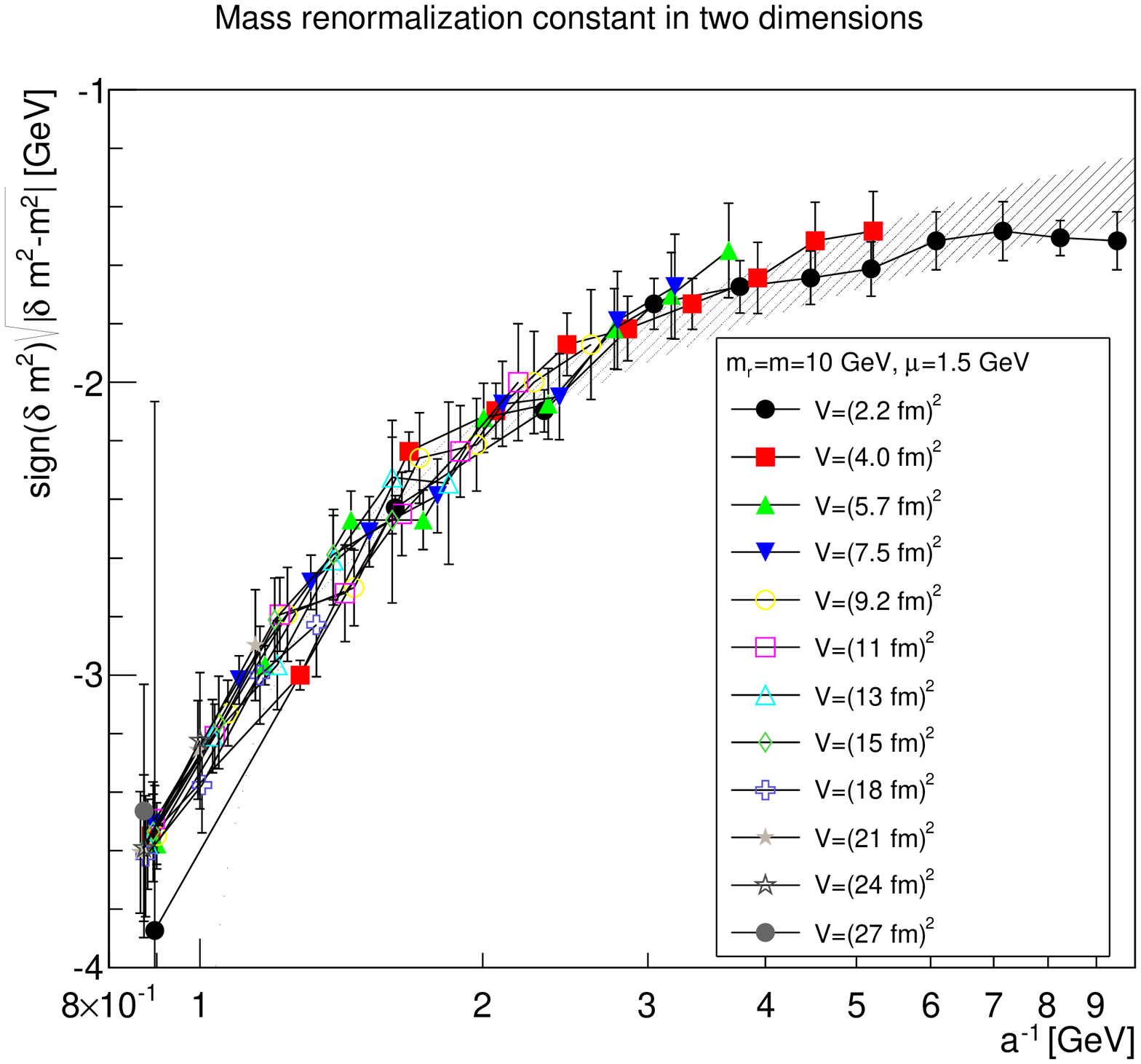}
\caption{\label{fig:m2}The mass renormalization constant as a function of the lattice cutoff and the lattice volume in two dimensions for $\mu=1.5$ GeV. The top-left panel shows the case of $m=m_r=0$ GeV, the top-right panel of $m=m_r=0.1$ GeV, the bottom-left panel of $m=m_r=1$ GeV, and the bottom-right panel of $m=m_r=10$ GeV. The hatched band is the fit \pref{mfit} with the parameters given in table \ref{fitsm}.}
\end{figure}

\begin{figure}
\includegraphics[width=0.475\linewidth]{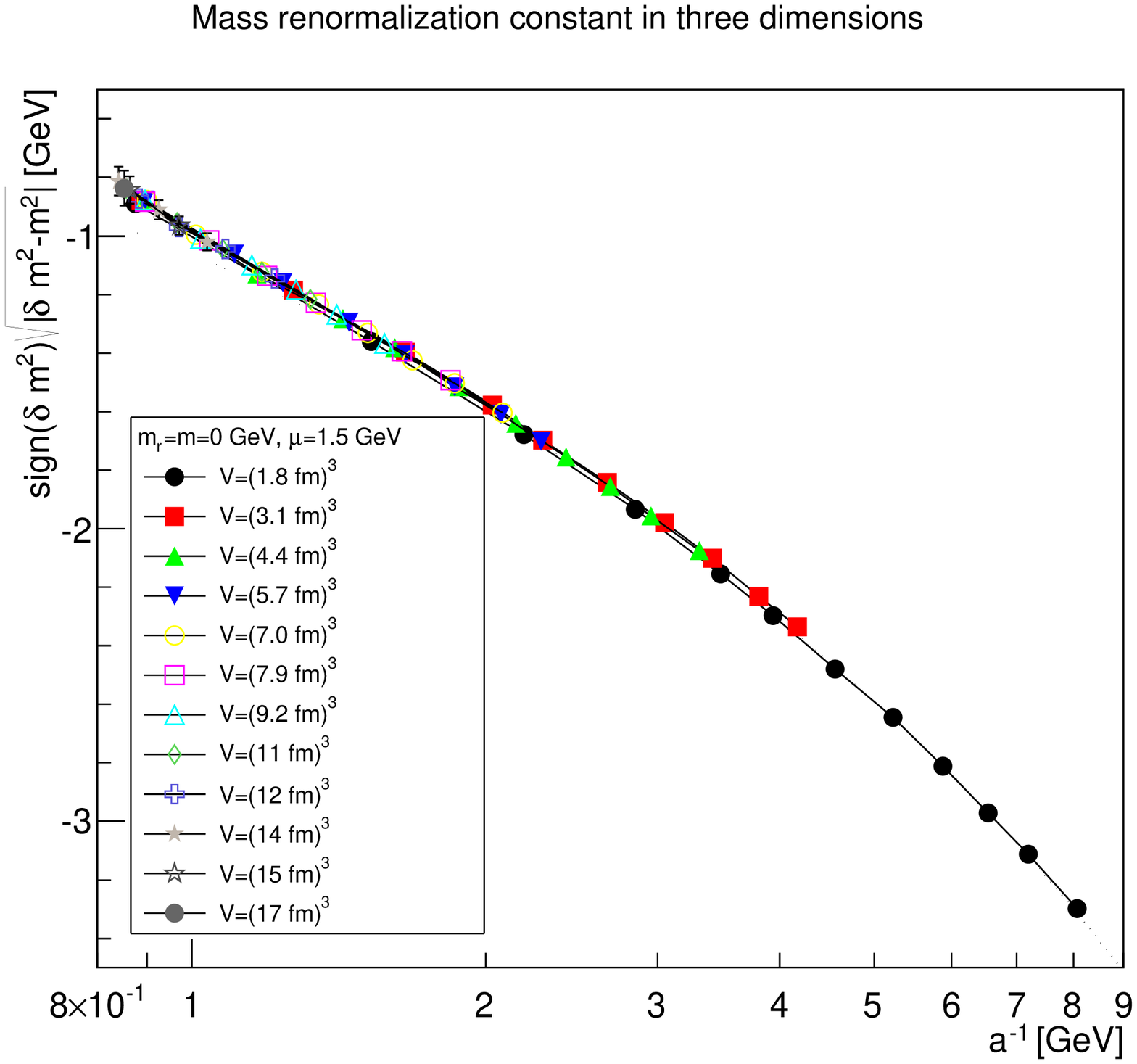}\includegraphics[width=0.475\linewidth]{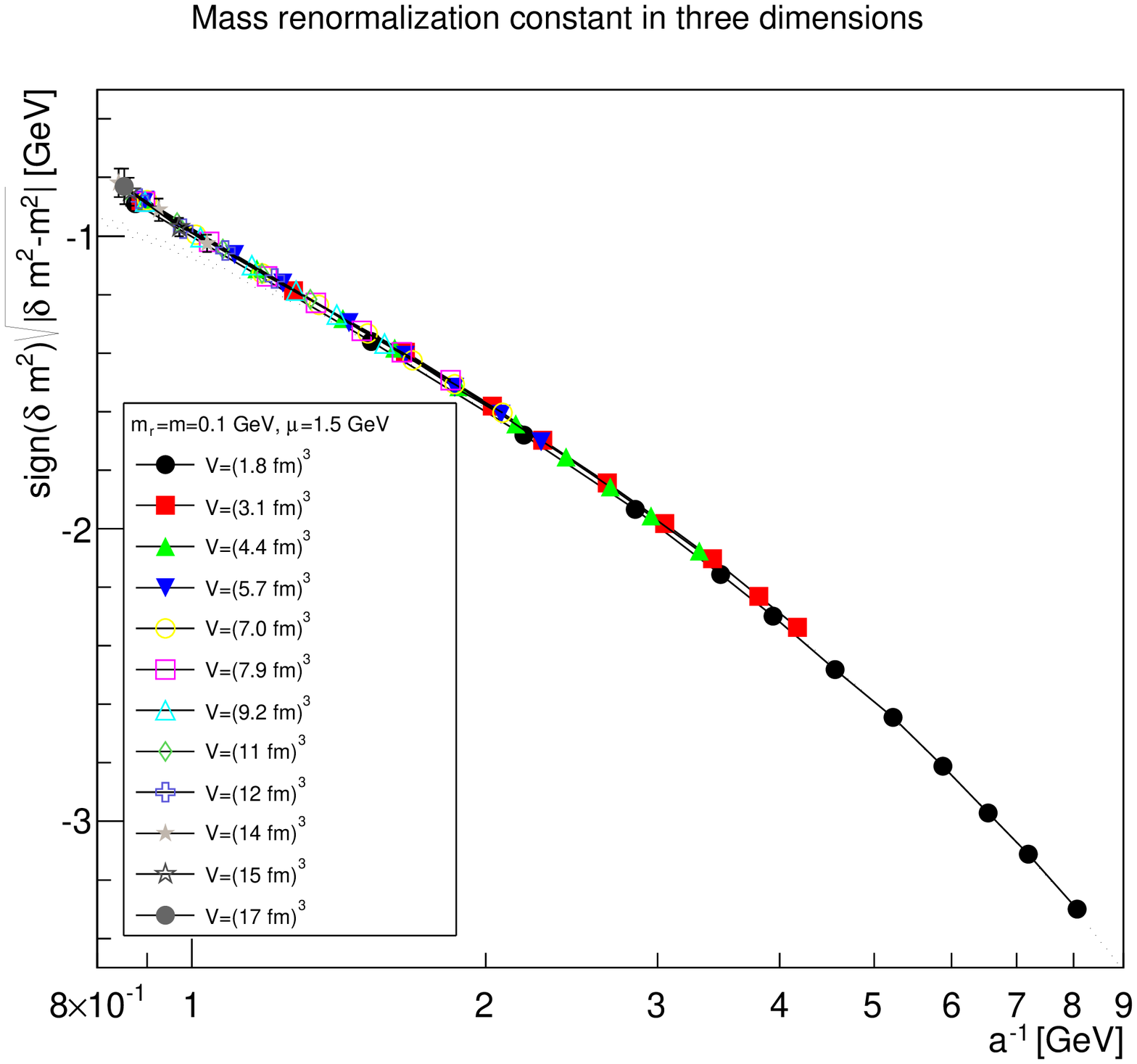}\\
\includegraphics[width=0.475\linewidth]{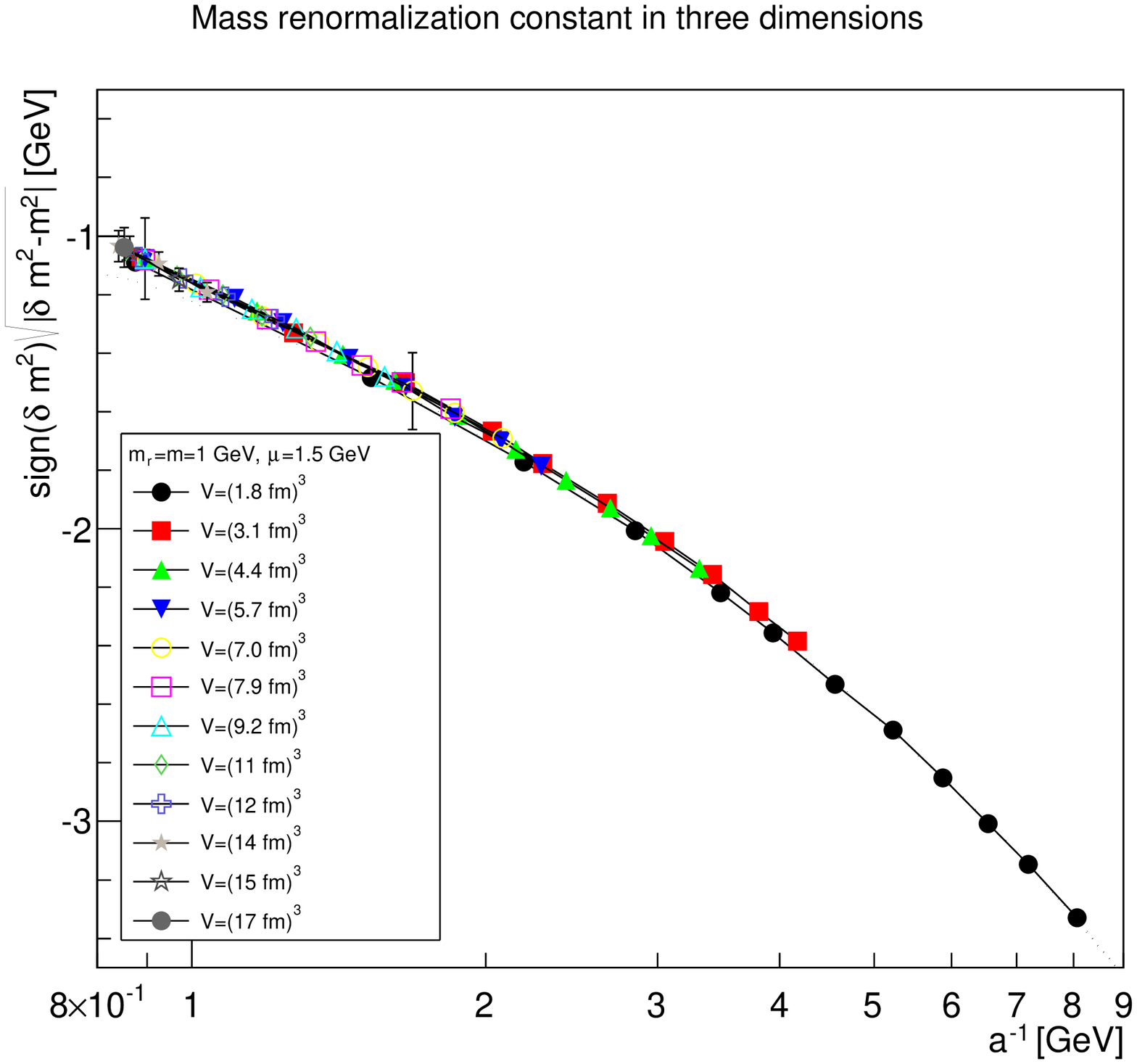}\includegraphics[width=0.475\linewidth]{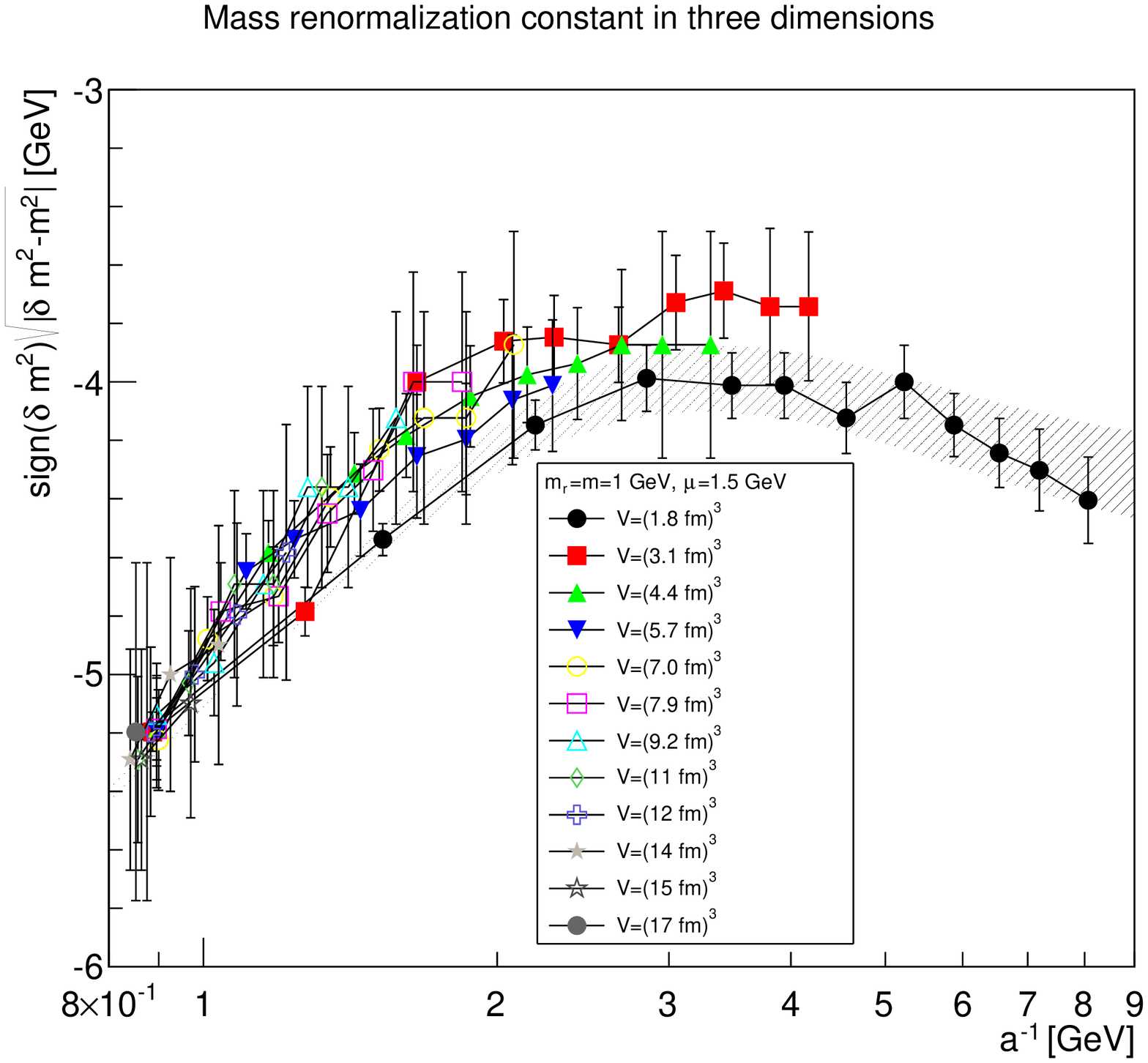}
\caption{\label{fig:m3}The mass renormalization constant as a function of the lattice cutoff and the lattice volume in three dimensions for $\mu=1.5$ GeV. The top-left panel shows the case of $m=m_r=0$ GeV, the top-right panel of $m=m_r=0.1$ GeV, the bottom-left panel of $m=m_r=1$ GeV, and the bottom-right panel of $m=m_r=10$ GeV. The hatched band is the fit \pref{mfit} with the parameters given in table \ref{fitsm}. Note that the hatched band can be as narrow as the lines, and therefore not be visible.}
\end{figure}

\begin{figure}
\includegraphics[width=0.475\linewidth]{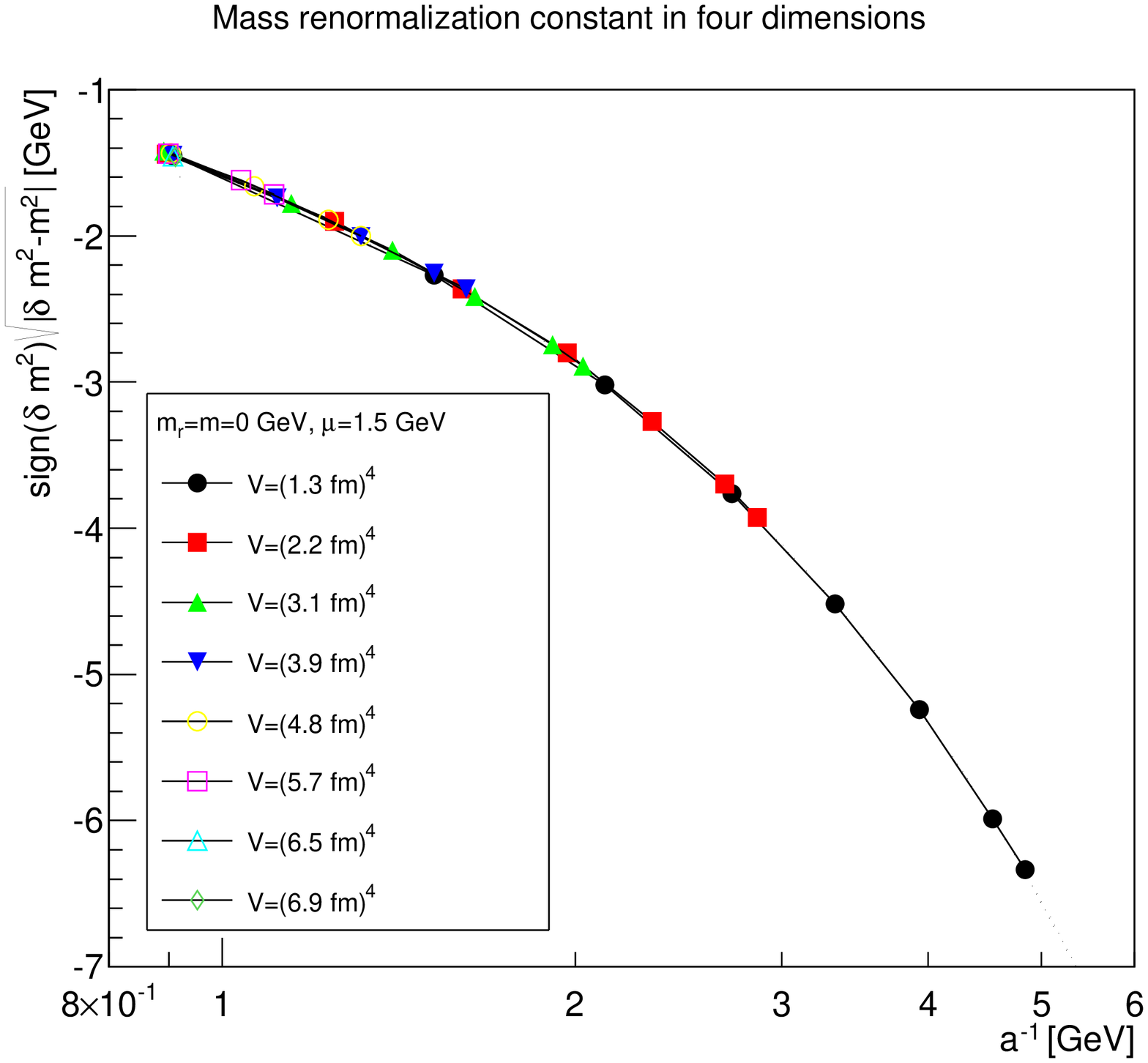}\includegraphics[width=0.475\linewidth]{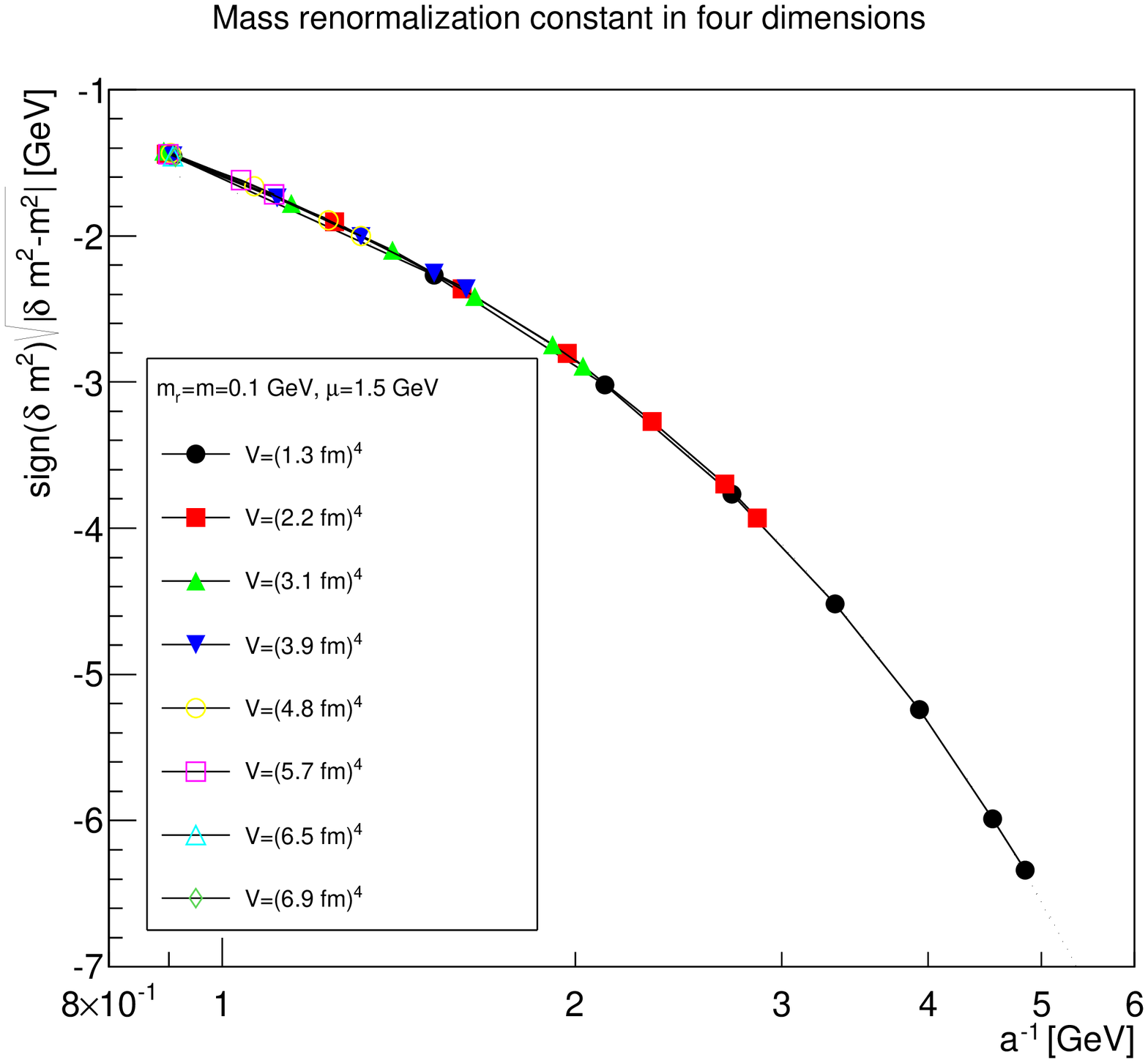}\\
\includegraphics[width=0.475\linewidth]{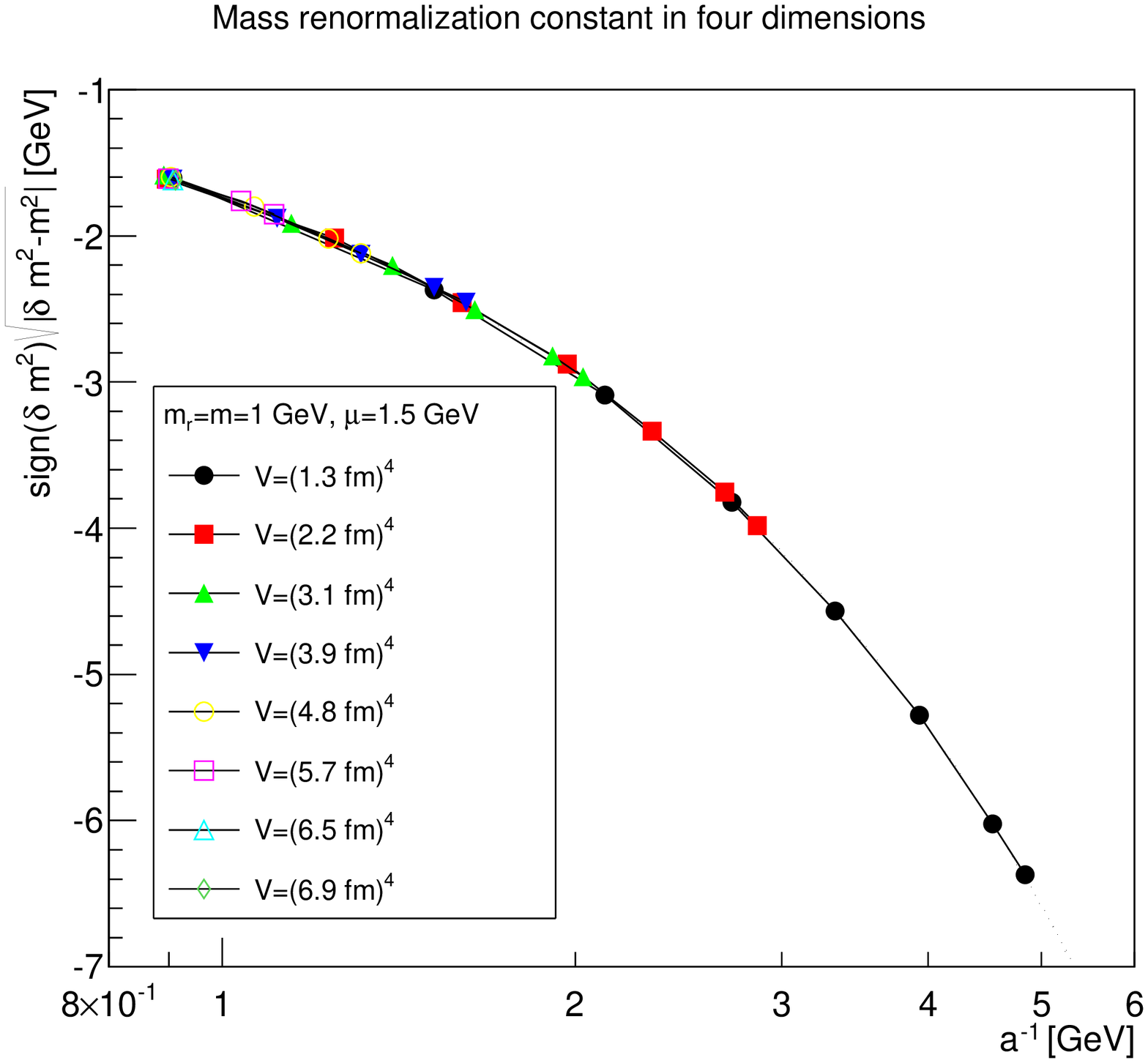}\includegraphics[width=0.475\linewidth]{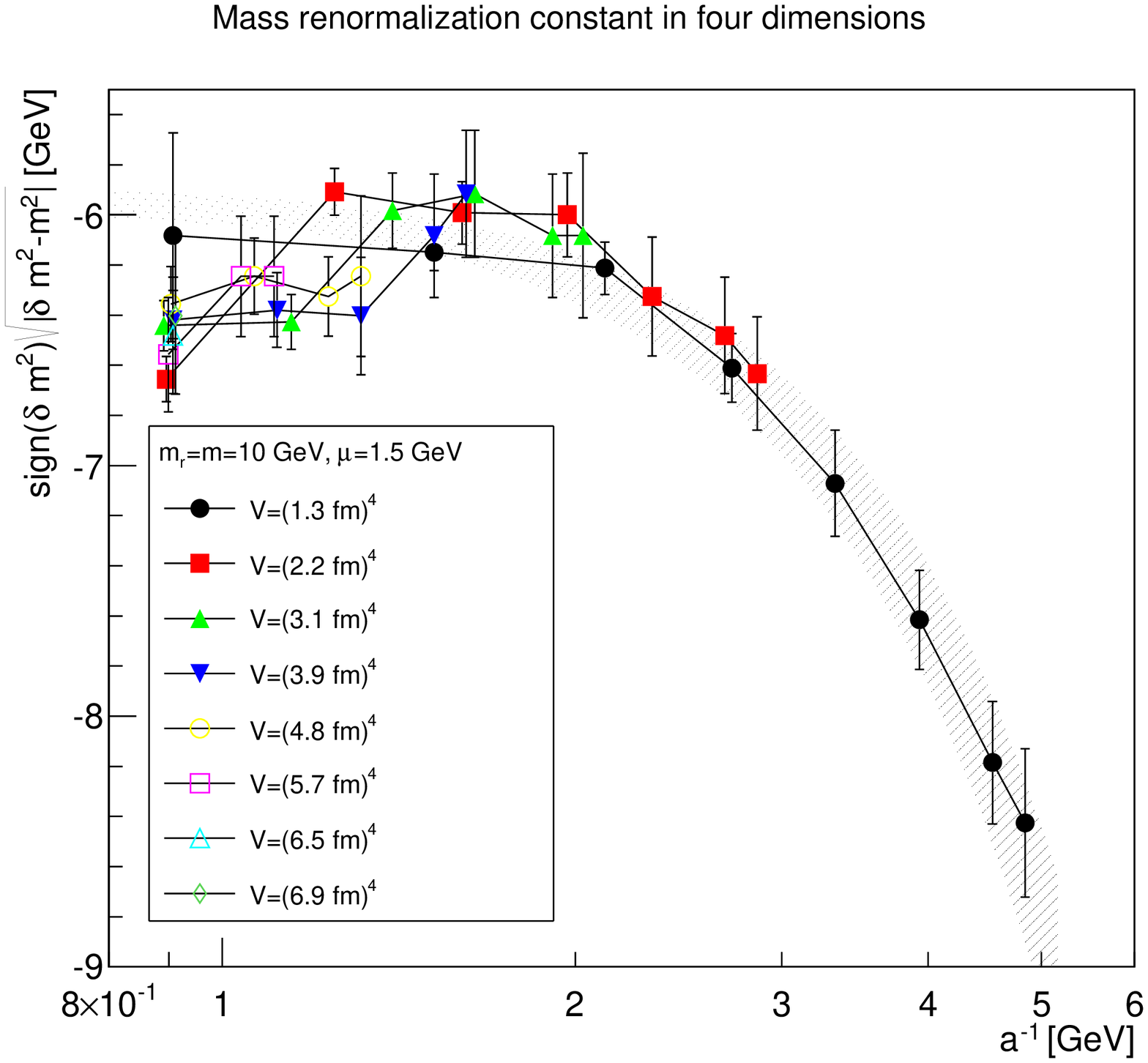}
\caption{\label{fig:m4}The mass renormalization constant as a function of the lattice cutoff and the lattice volume in four dimensions for $\mu=1.5$ GeV. The top-left panel shows the case of $m=m_r=0$ GeV, the top-right panel of $m=m_r=0.1$ GeV, the bottom-left panel of $m=m_r=1$ GeV, and the bottom-right panel of $m=m_r=10$ GeV. The hatched band is the fit \pref{mfit} with the parameters given in table \ref{fitsm}. Note that the hatched band can be as narrow as the lines, and therefore not be visible.}
\end{figure}

The situation for the mass renormalization constant, more precisely for $(|\delta m^2-m^2|)^{1/2}$, is shown in figures \ref{fig:m2}-\ref{fig:m4}. 

\begin{longtable}{|c|c|c|c|c|c|}
\caption{\label{fitsm}Fit parameters of \pref{mfit} for the mass renormalization constants at $\mu=1.5$ GeV.}\\
\hline
$d$	& m [GeV] & M [GeV]	& $c$ [GeV$^{3-d}$]	& $\Lambda$ [GeV] & $\epsilon$ \cr
\hline\endfirsthead
\hline
\multicolumn{6}{|l|}{Table \ref{fitsm} continued}\\
\hline
$d$	& m [GeV] & M [GeV]	& $c$ [GeV$^{3-d}$]	& $\Lambda$ [GeV] & $\epsilon$ \cr
\hline\endhead
\multicolumn{6}{|r|}{Continued on next page}\\
\hline\endfoot
\endlastfoot
\hline
2	& 0	& -0.138(4) & -0.353(8) & {\cal O}(10$^{-9}$) & -0.415(9) \cr 
\hline
2	& 0.1	& -0.160(4) & -0.3348(19) & {\cal O}(10$^{-7})$ & -0.428(5) \cr
\hline
2	& 1	& -0.570(2) & -0.1427(11) & {\cal O}(10$^{-8}$) & -0.591(5) \cr
\hline
2	& 10	& -16(2) & 1.18(17) & {\cal O}(10$^{-4}$) & 0.008(3) \cr
\hline
\hline
3	& 0	& -0.4117(4) & -1.489(5) & 3.055(19) & 0.973(3) \cr
\hline
3	& 0.1	& -0.42(5) & -1.50(3) & 3.09(12) & 0.978(5) \cr
\hline
3	& 1	& -0.670(3) & -1.259(6) & 3.266(14) & 0.912(3) \cr
\hline
3	& 10	& -6.9(3) & 18.8(13) & 3.1(1) & 2.79(15) \cr
\hline
\hline
4	& 0	& -1.092(11) & -2.25(7) & 2.59(5) & 1.88(3) \cr
\hline
4	& 0.1	& -1.094(11) & -2.26(7) & 2.59(5) & 1.88(3) \cr
\hline
4	& 1	& -1.283(6) & -2.20(7) & 2.71(5) & 1.88(3) \cr
\hline
4	& 10	& -5.93(5) & -0.065(12) & 0.95(5) & -0.48(8) \cr
\hline
\end{longtable}

The results can be fitted rather well by the form
\be
\delta m^2(a)-m^2=-\left(M+ca^{2-d}\left(\ln\left(\frac{\Lambda^2+a^{-2}}{(1\textrm{ GeV})^2}\right)\right)^\epsilon\right)^2\label{mfit}.
\ee
\no The fit parameters are listed in table \ref{fitsm}. As expected, the dependence on $a^{-1}$ is purely logarithmic in two dimensions, linear in three dimensions, and quadratic in four dimensions. In the latter cases also logarithmic corrections appear, as expected \cite{Bohm:2001yx,Jackiw:1980kv}. The logarithms also exhibit anomalous dimensions. In general, except in two dimensions and aside from $M$, the fit parameters are almost independent of the bare mass. Only for the largest bare mass this is not true, but this is conceivably a discretization artifact. In fact, the $a$ dependence in the latter case is different at small $1/a$, but at large $1/a$ the behavior starts to change and to become similar to the ones at smaller bare mass. This behavior is once more similar to the fundamental case \cite{Maas:2016edk}\footnote{Note that for the fundamental case it was possible to fit the leading $a$ dependence in \pref{mfit}, rather than to set it to $2-d$. This did not yield stable fits in the present case, and therefore this behavior was fixed.}. Hence, the asymptotic behavior seems to emerge only for $a^{-1}\gtrsim m$.

Interestingly, the only parameter showing a pronounced dependence on the mass is $-M$. It behaves roughly like constant$+m/d$, with the characteristic constant being roughly 0.15, 0.4, and 1 GeV in two, three, and four dimensions, respectively. This once more indicates an intrinsic scale.\\

\section{Analytic structure}\label{s:ana}

\subsection{Momentum space properties}\label{ss:mom}

The results of the previous section, especially figure \ref{fig:ur}, suggests that discretization artifacts are sizable, particularly in the infrared. Thus, in the following not only the results for the finest lattices but also at fixed $a^{-1}$ will be considered.

In addition to the propagator also the dressing function, defined as
\be
H(p^2)=(p^2+m_r^2)D(p^2)\label{df},
\ee
\no will be presented. The dressing function therefore describes the deviation from the (renormalized) tree-level form. By construction at $\mu$ all dressing functions equal 1.

\afterpage{\clearpage}

\begin{figure}
\includegraphics[width=\linewidth]{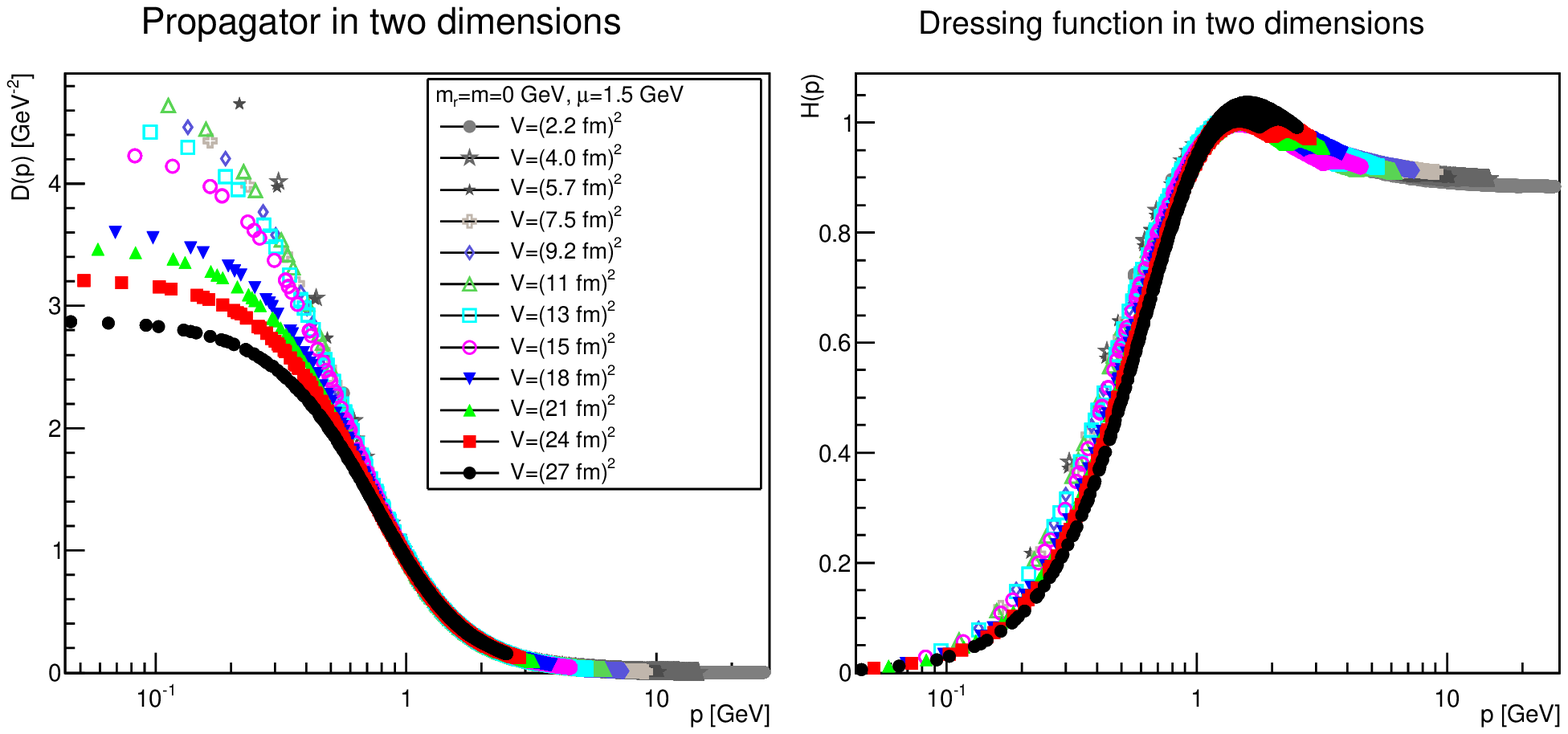}
\caption{\label{fig:d20}The propagator (left panel) and the dressing function \pref{df} (right panel) in two dimensions for $m=m_r=0$ GeV and $\mu=1.5$ GeV.}
\end{figure}

\begin{figure}
\includegraphics[width=\linewidth]{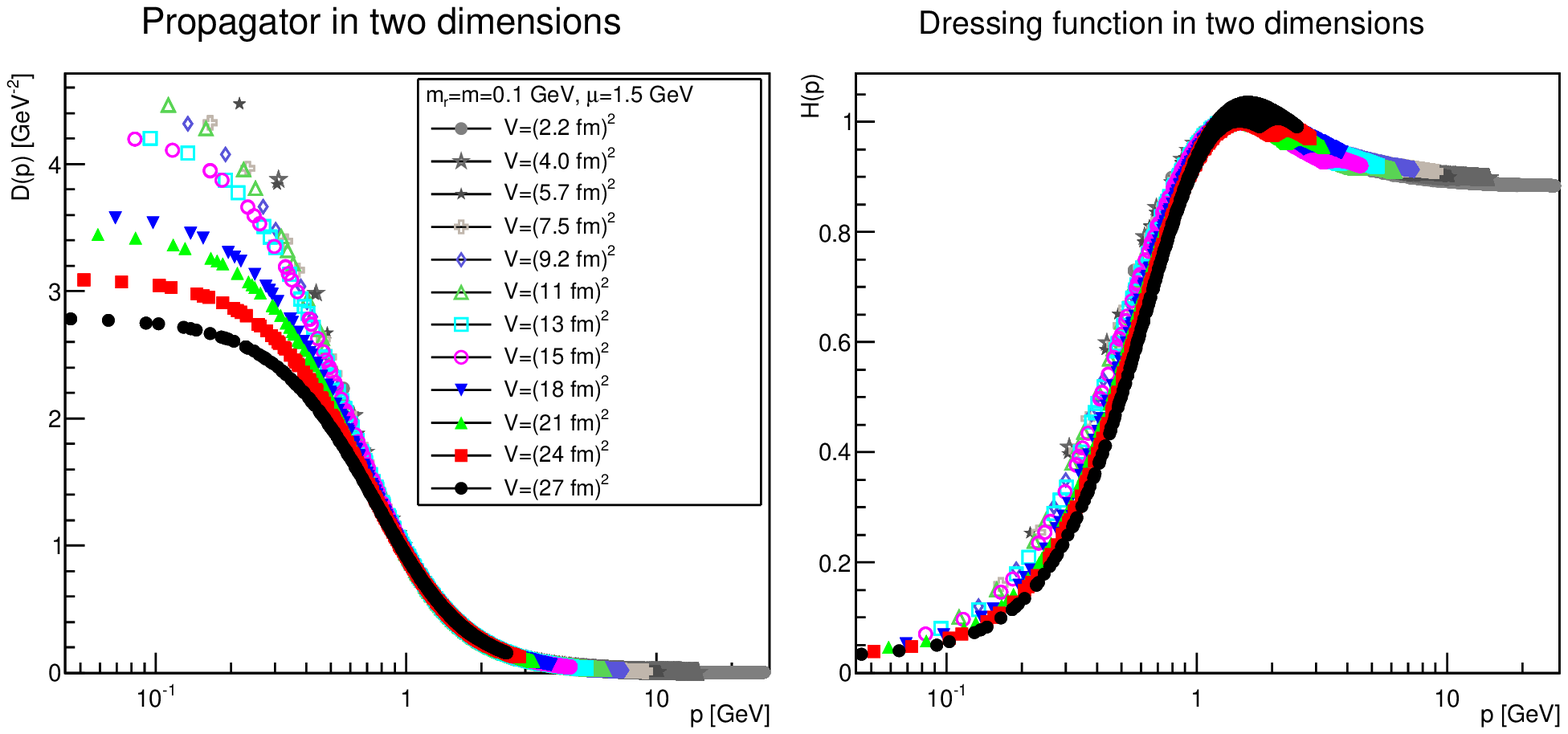}
\caption{\label{fig:d21}The propagator (left panel) and the dressing function \pref{df} (right panel) in two dimensions for $m=m_r=0.1$ GeV and $\mu=1.5$ GeV.}
\end{figure}

\begin{figure}
\includegraphics[width=\linewidth]{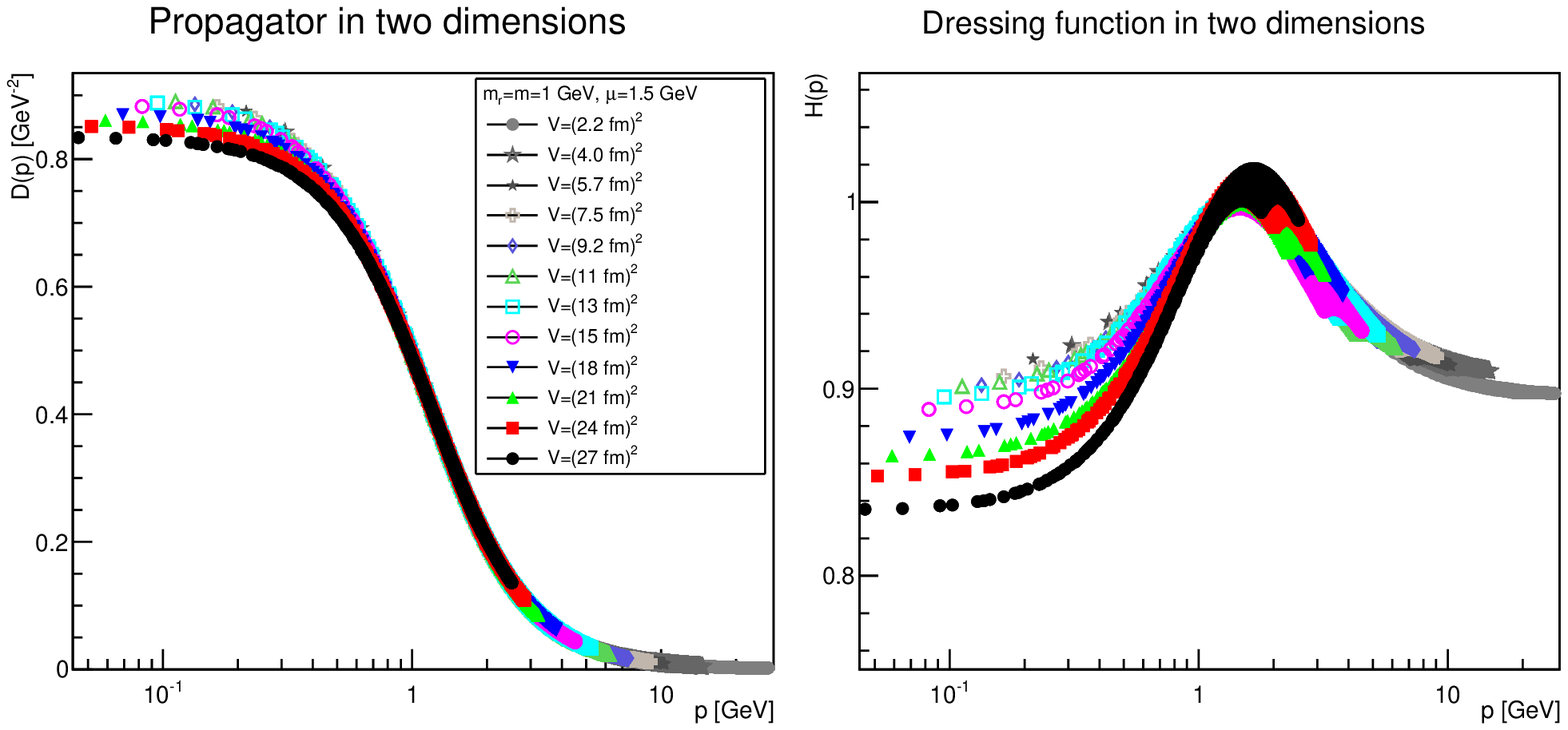}
\caption{\label{fig:d22}The propagator (left panel) and the dressing function \pref{df} (right panel) in two dimensions for $m=m_r=1$ GeV and $\mu=1.5$ GeV. Note the different scale in the right-hand panel compared to figures \ref{fig:d20} and \ref{fig:d21}.}
\end{figure}

\begin{figure}
\includegraphics[width=\linewidth]{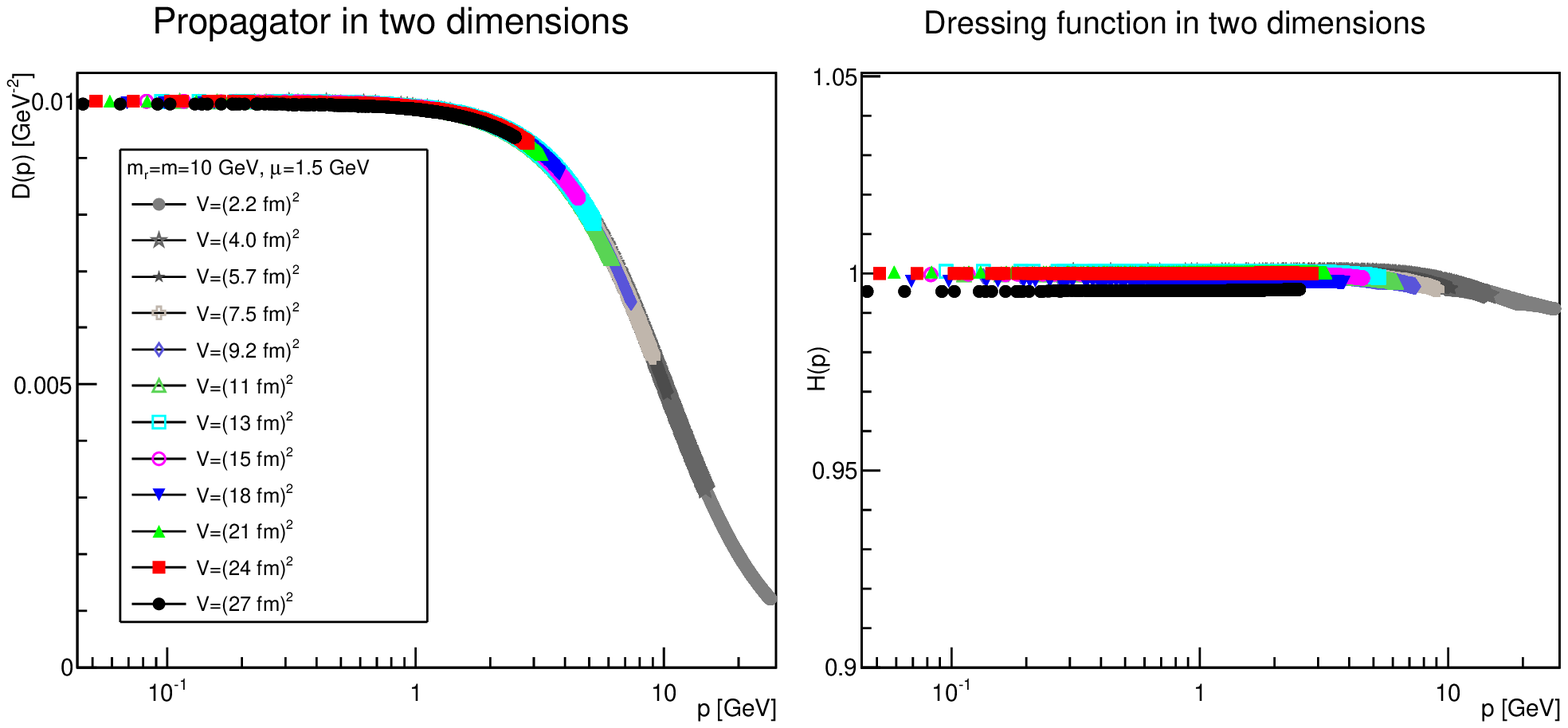}
\caption{\label{fig:d23}The propagator (left panel) and the dressing function \pref{df} (right panel) in two dimensions for $m=m_r=10$ GeV and $\mu=1.5$ GeV. Note the different scale in the right-hand panel compared to figures \ref{fig:d20} and \ref{fig:d21}.}
\end{figure}

\afterpage{\clearpage}

\begin{figure}
\includegraphics[width=\linewidth]{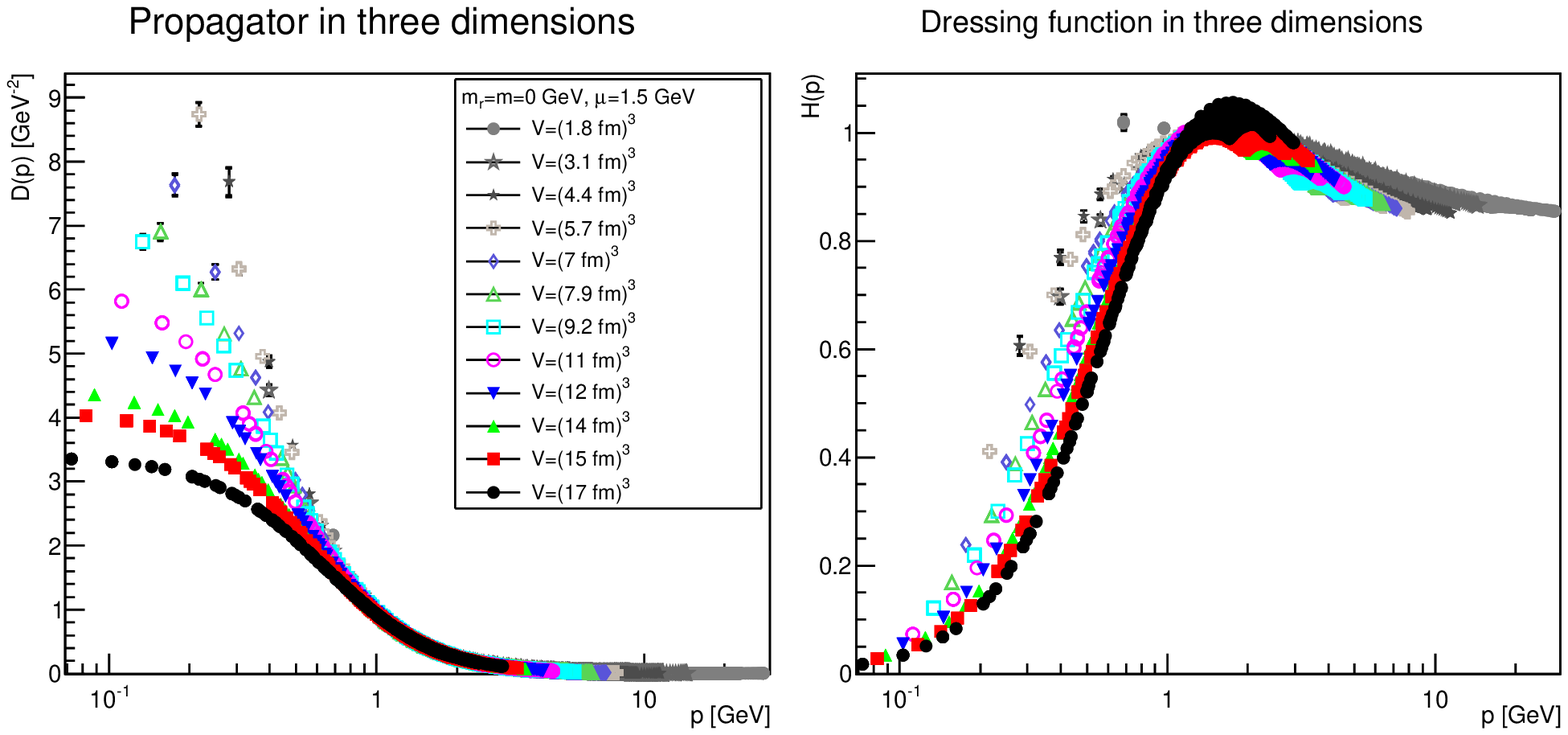}
\caption{\label{fig:d30}The propagator (left panel) and the dressing function \pref{df} (right panel) in three dimensions for $m=m_r=0$ GeV and $\mu=1.5$ GeV.}
\end{figure}

\begin{figure}
\includegraphics[width=\linewidth]{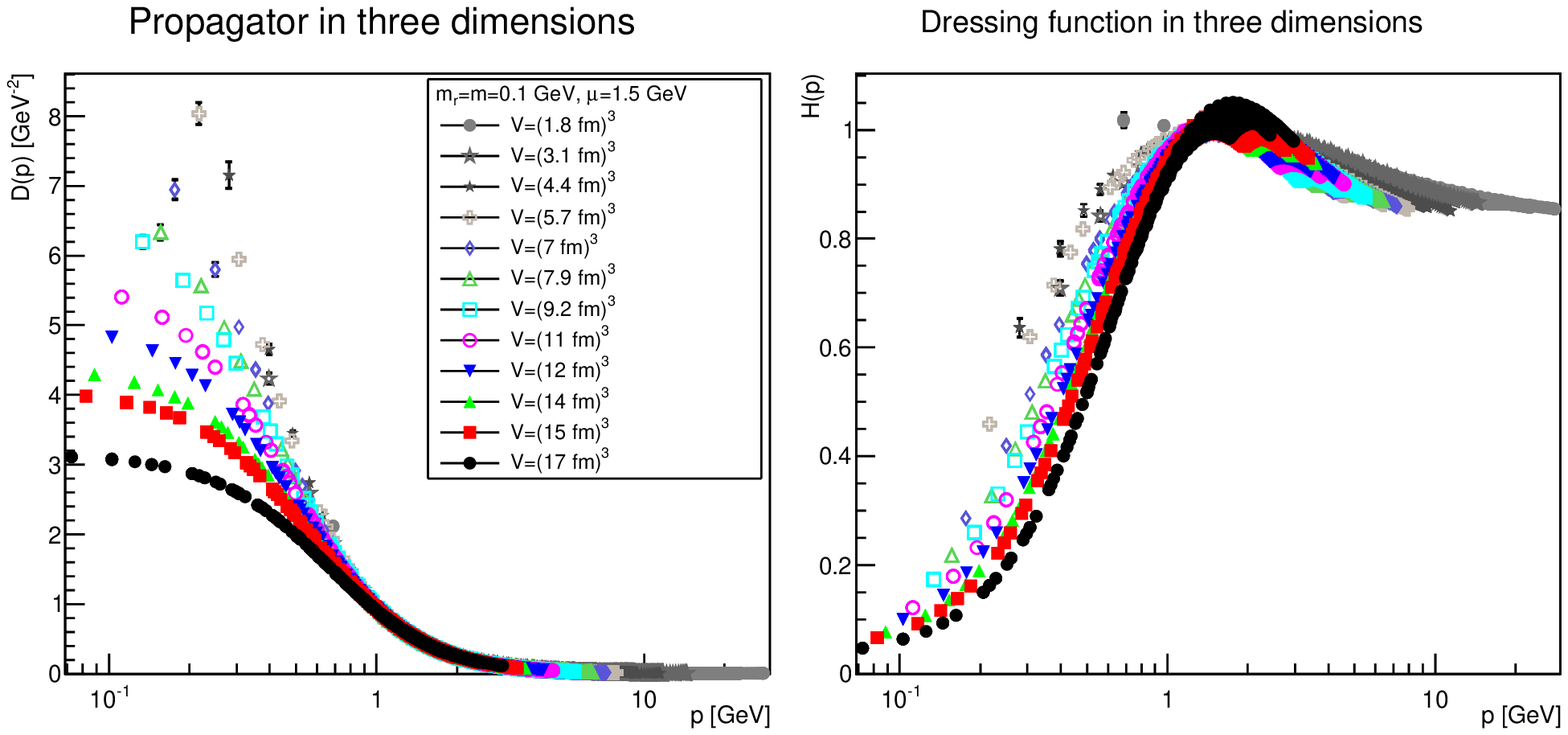}
\caption{\label{fig:d31}The propagator (left panel) and the dressing function \pref{df} (right panel) in three dimensions for $m=m_r=0.1$ GeV and $\mu=1.5$ GeV.}
\end{figure}

\begin{figure}
\includegraphics[width=\linewidth]{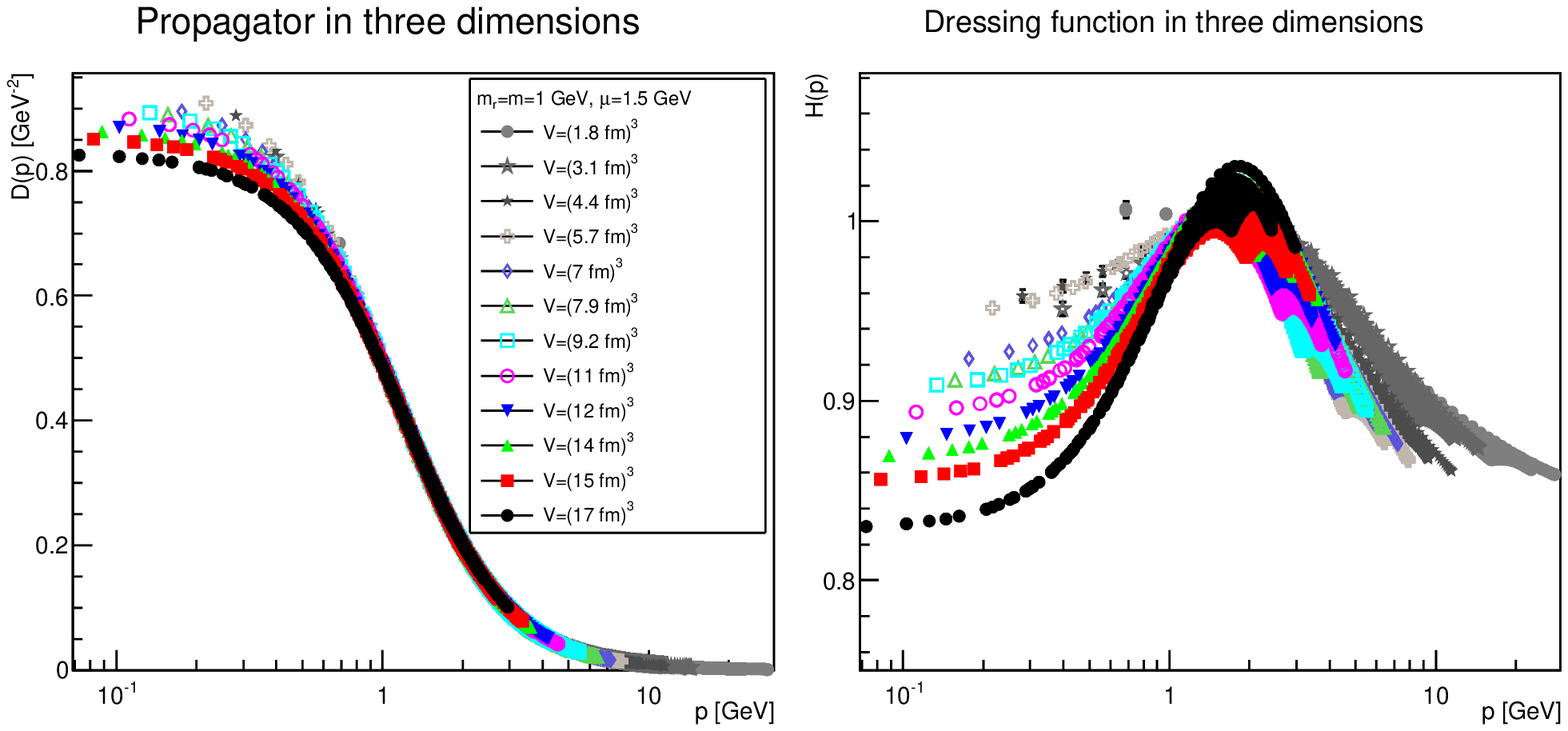}
\caption{\label{fig:d32}The propagator (left panel) and the dressing function \pref{df} (right panel) in three dimensions for $m=m_r=1$ GeV and $\mu=1.5$ GeV. Note the different scale in the right-hand panel compared to figures \ref{fig:d30} and \ref{fig:d31}.}
\end{figure}

\begin{figure}
\includegraphics[width=\linewidth]{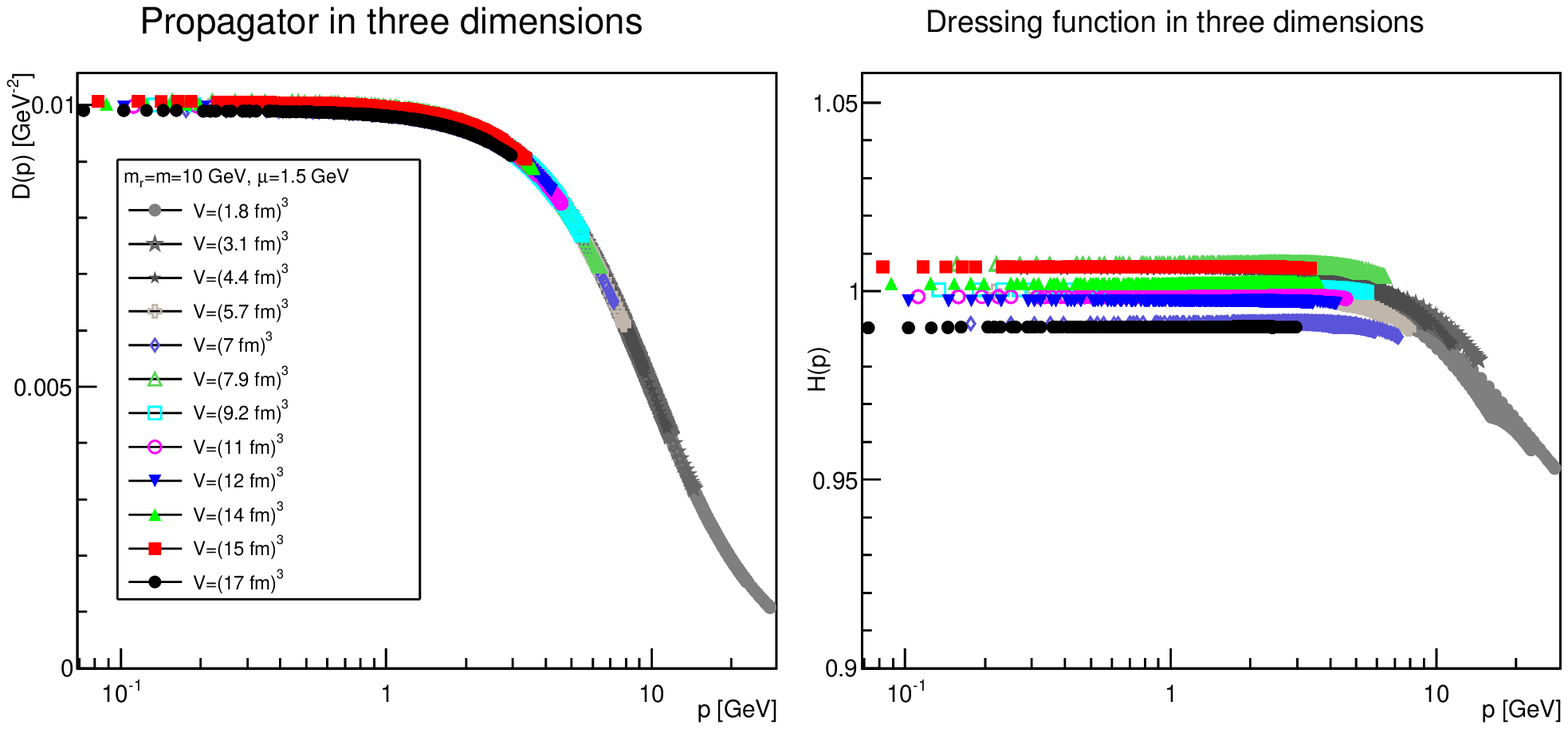}
\caption{\label{fig:d33}The propagator (left panel) and the dressing function \pref{df} (right panel) in three dimensions for $m=m_r=10$ GeV and $\mu=1.5$ GeV. Note the different scale in the right-hand panel compared to figures \ref{fig:d30} and \ref{fig:d31}.}
\end{figure}

\afterpage{\clearpage}

\begin{figure}
\includegraphics[width=\linewidth]{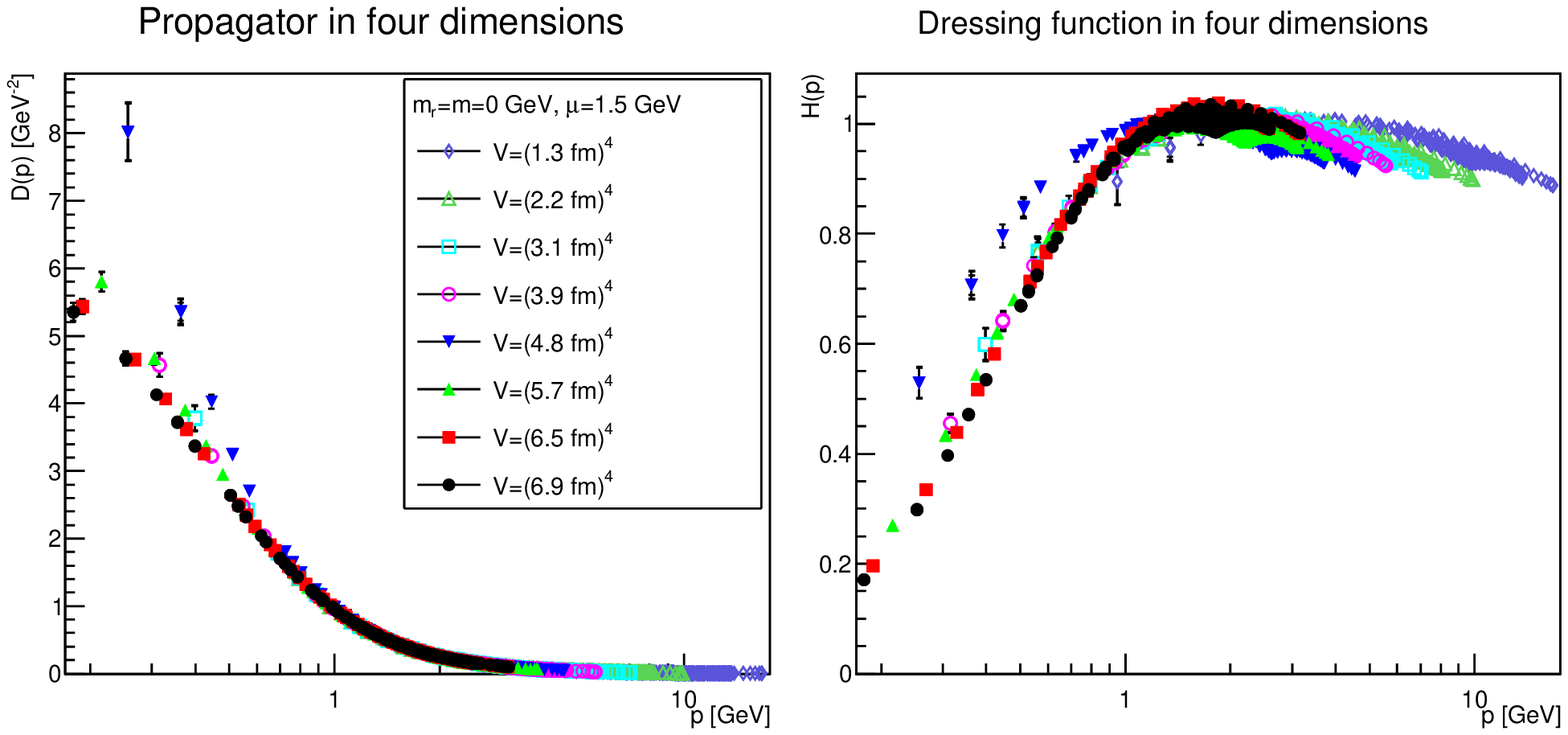}
\caption{\label{fig:d40}The propagator (left panel) and the dressing function \pref{df} (right panel) in four dimensions for $m=m_r=0$ GeV and $\mu=1.5$ GeV.}
\end{figure}

\begin{figure}
\includegraphics[width=\linewidth]{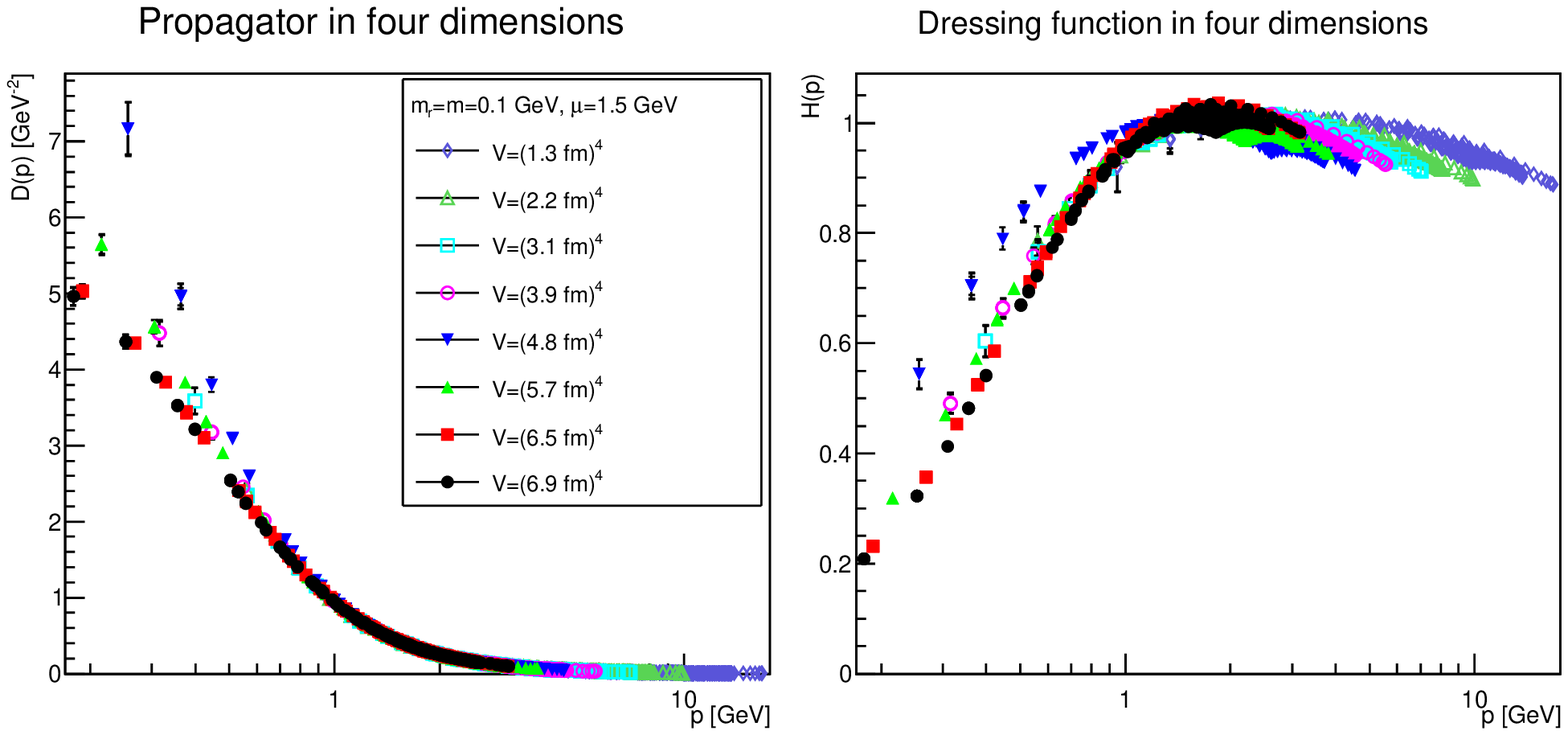}
\caption{\label{fig:d41}The propagator (left panel) and the dressing function \pref{df} (right panel) in four dimensions for $m=m_r=0.1$ GeV and $\mu=1.5$ GeV.}
\end{figure}

\begin{figure}
\includegraphics[width=\linewidth]{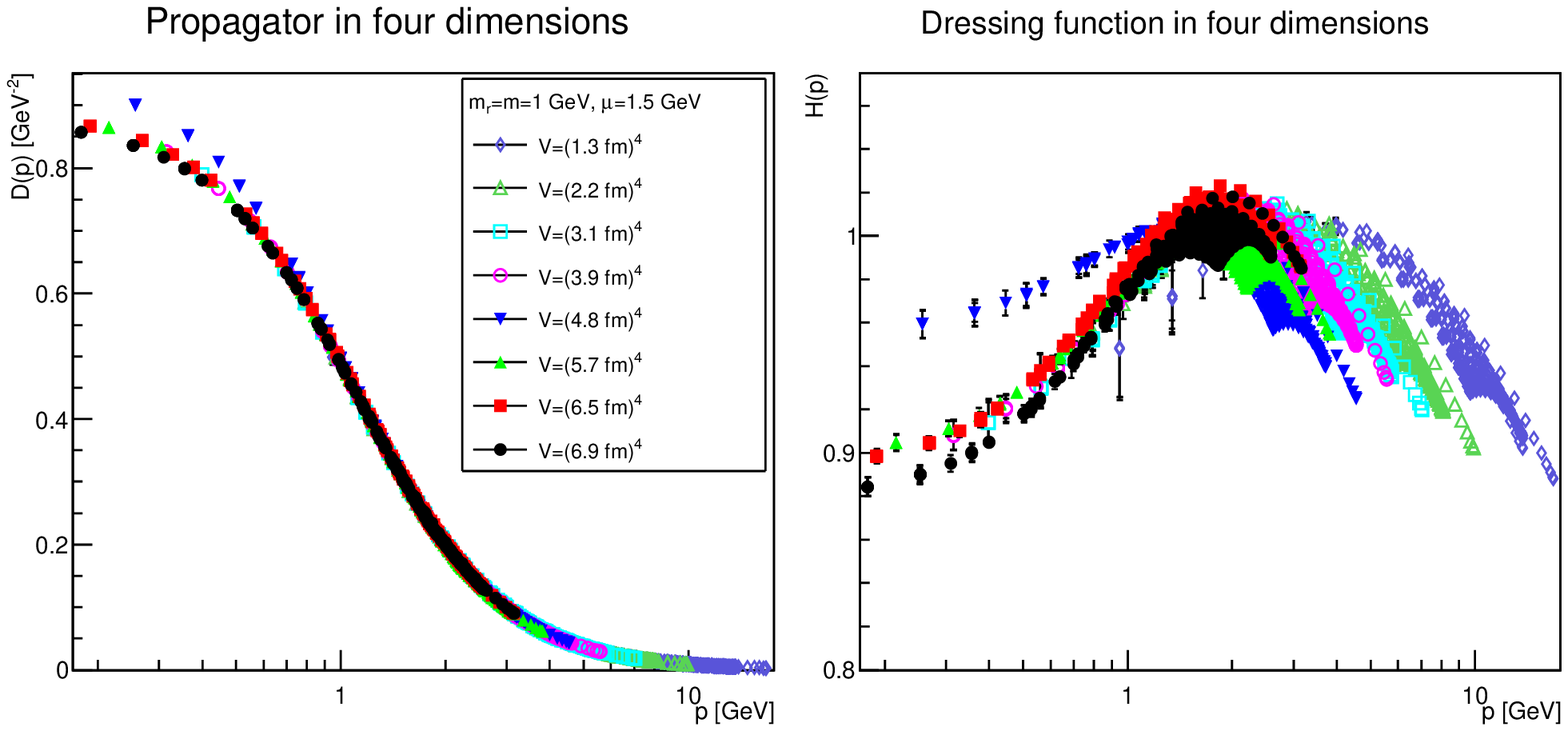}
\caption{\label{fig:d42}The propagator (left panel) and the dressing function \pref{df} (right panel) in four dimensions for $m=m_r=1$ GeV and $\mu=1.5$ GeV. Note the different scale in the right-hand panel compared to figures \ref{fig:d40} and \ref{fig:d41}.}
\end{figure}

\begin{figure}
\includegraphics[width=\linewidth]{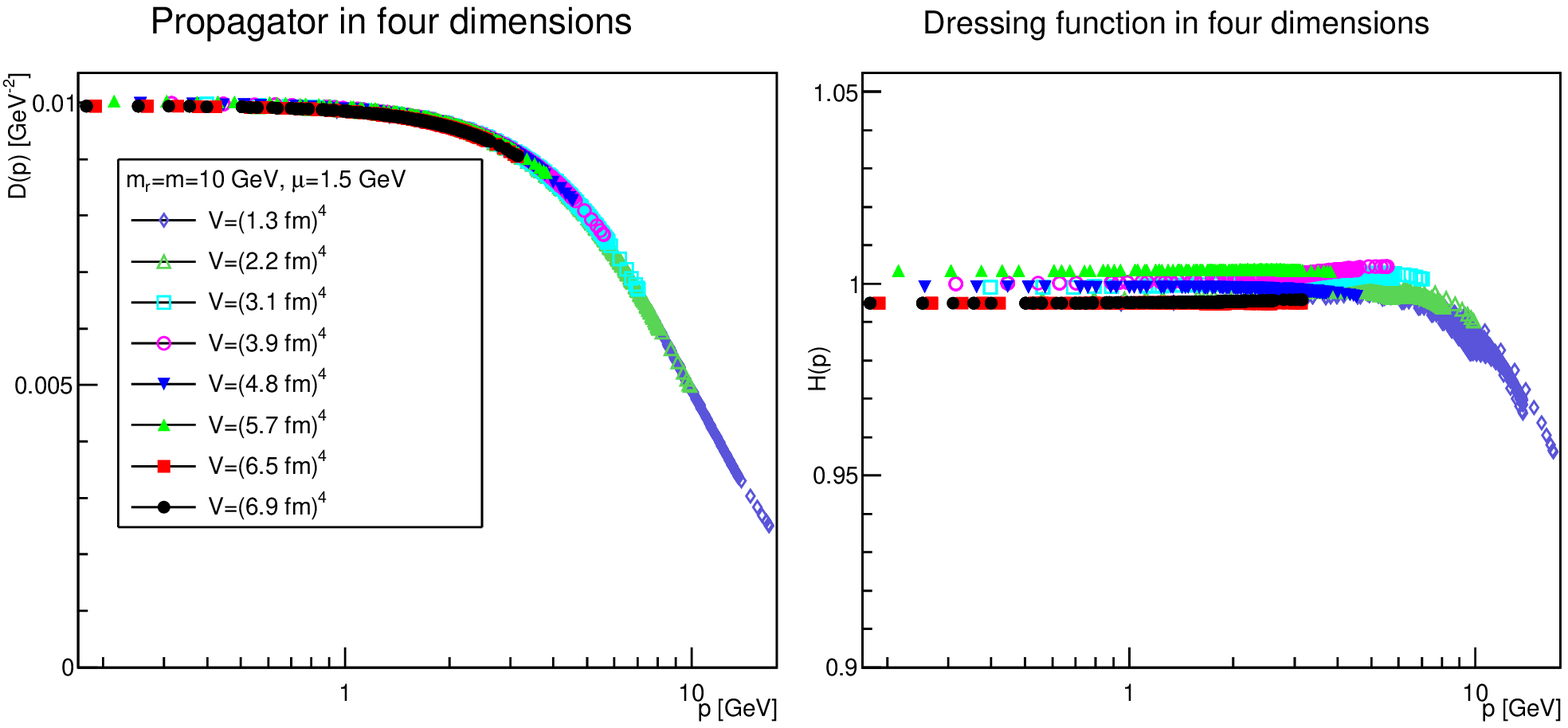}
\caption{\label{fig:d43}The propagator (left panel) and the dressing function \pref{df} (right panel) in four dimensions for $m=m_r=10$ GeV and $\mu=1.5$ GeV. Note the different scale in the right-hand panel compared to figures \ref{fig:d40} and \ref{fig:d41}.}
\end{figure}

\afterpage{\clearpage}

The results are shown for two dimensions in figures \ref{fig:d20}-\ref{fig:d23}, for three dimensions in figures \ref{fig:d30}-\ref{fig:d33}, and for four dimensions in figures \ref{fig:d40}-\ref{fig:d43}. From the dressing functions at large momenta substantial discretization artifacts are visible. They lead to a deviation away from the continuum limit, especially in four dimensions. This is quite similar in kind to the other propagators \cite{Maas:2014xma} and could be improved using the techniques described, e.\ g., in \cite{Sternbeck:2012qs,Boucaud:2008gn}.

At low momenta, a marked infrared suppression compared to the renormalized tree-level behavior is visible from the propagators. This is the stronger the smaller $m$. Only for $m=m_r=10$ GeV (almost) no such effect is seen. A similar effect was also observed for the fundamental case \cite{Maas:2016edk}, but it is much stronger here, suggesting a much larger screening effect. A stronger screening in the adjoint case than in the fundamental case is also observed for fermions \cite{Roberts:1994dr,Roberts:2015lja,Alkofer:2000wg,Alkofer:2003jj,August:2013jia}. In addition, there are stronger finite-volume effects than in the fundamental case, especially at low $m$. They therefore extend to larger momenta, up to a few hundred MeV. However, this is intertwined with the discretization artifacts.

\begin{figure}
\includegraphics[width=0.5\linewidth]{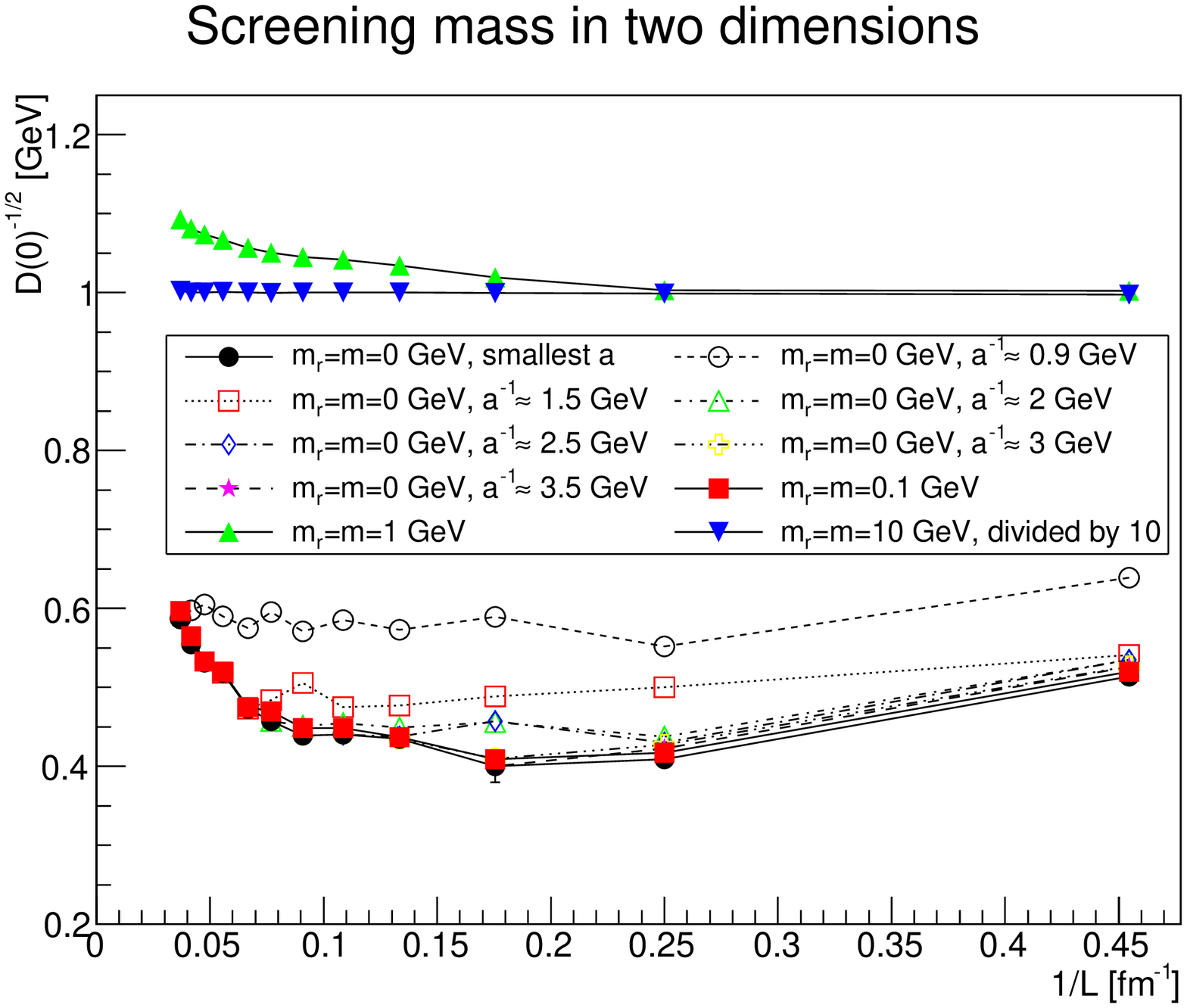}\includegraphics[width=0.5\linewidth]{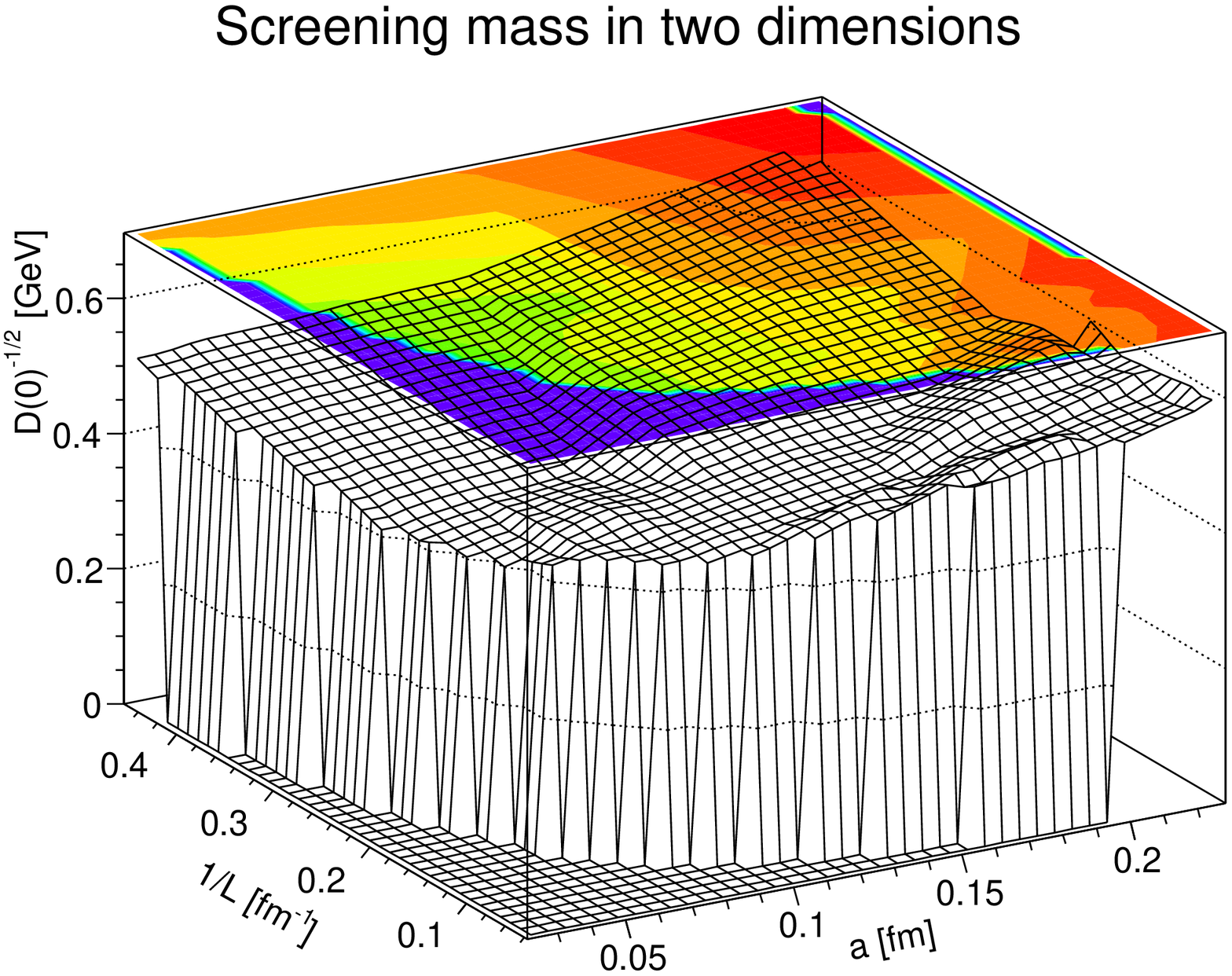}\\
\includegraphics[width=0.5\linewidth]{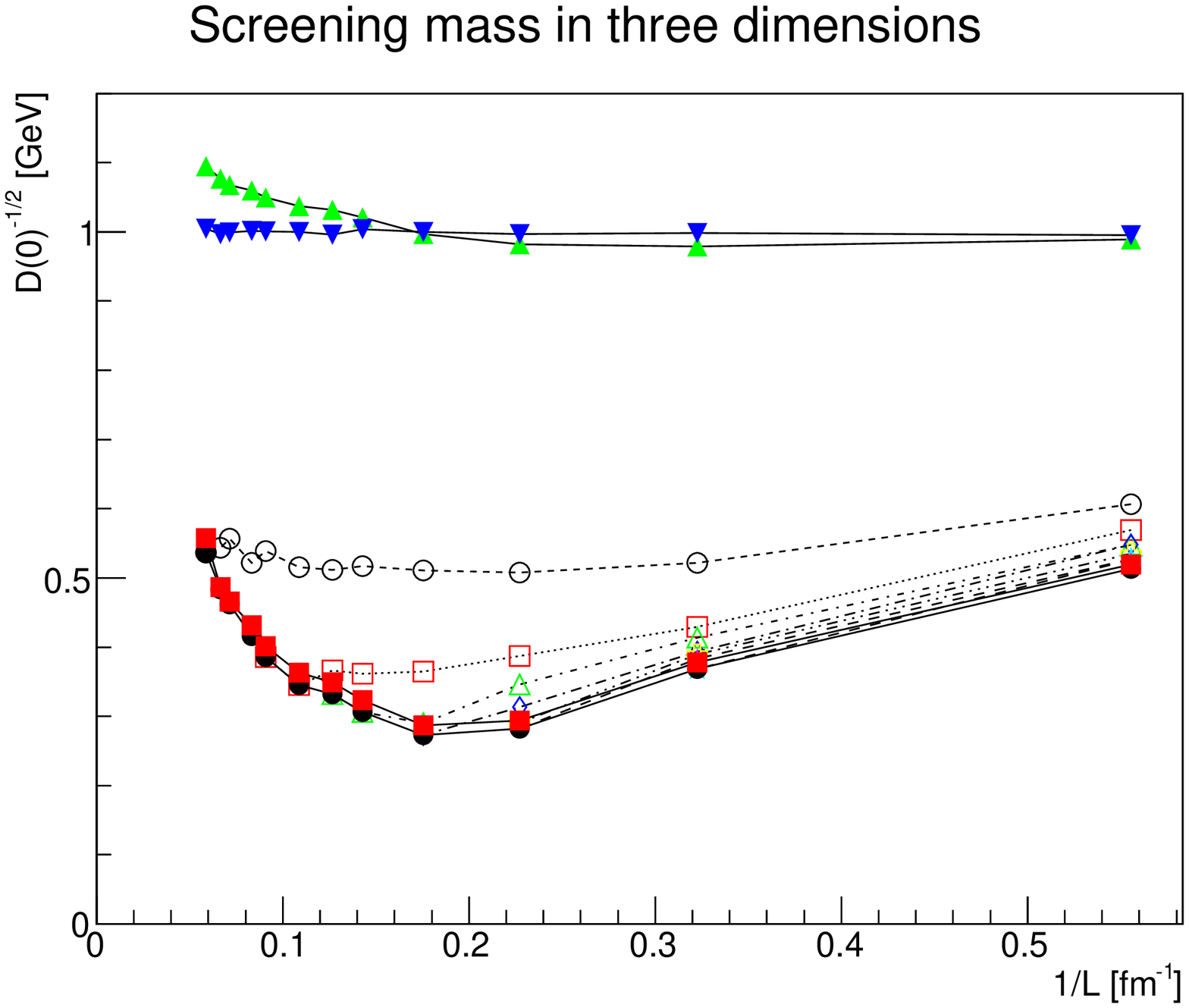}\includegraphics[width=0.5\linewidth]{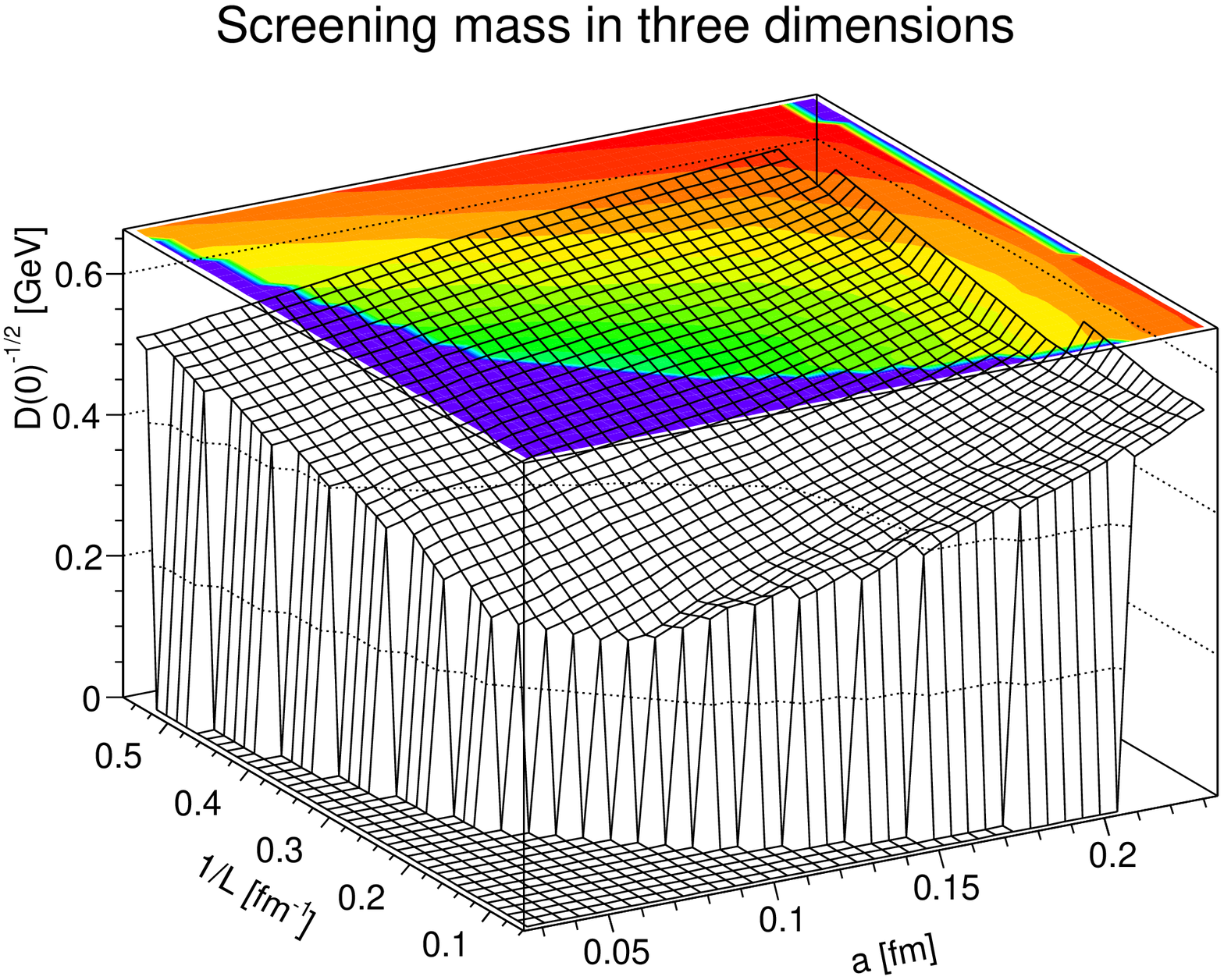}\\
\includegraphics[width=0.5\linewidth]{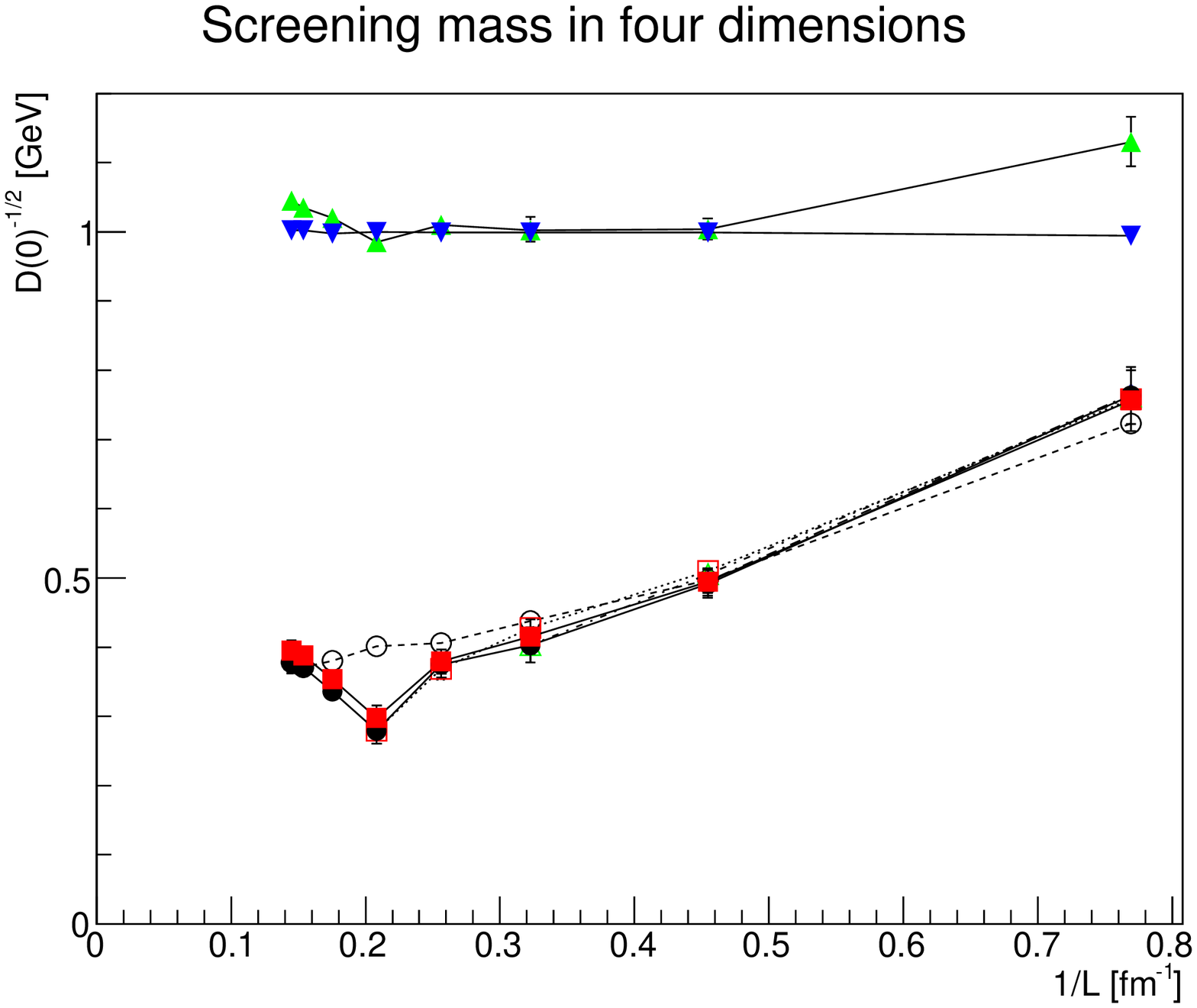}\includegraphics[width=0.5\linewidth]{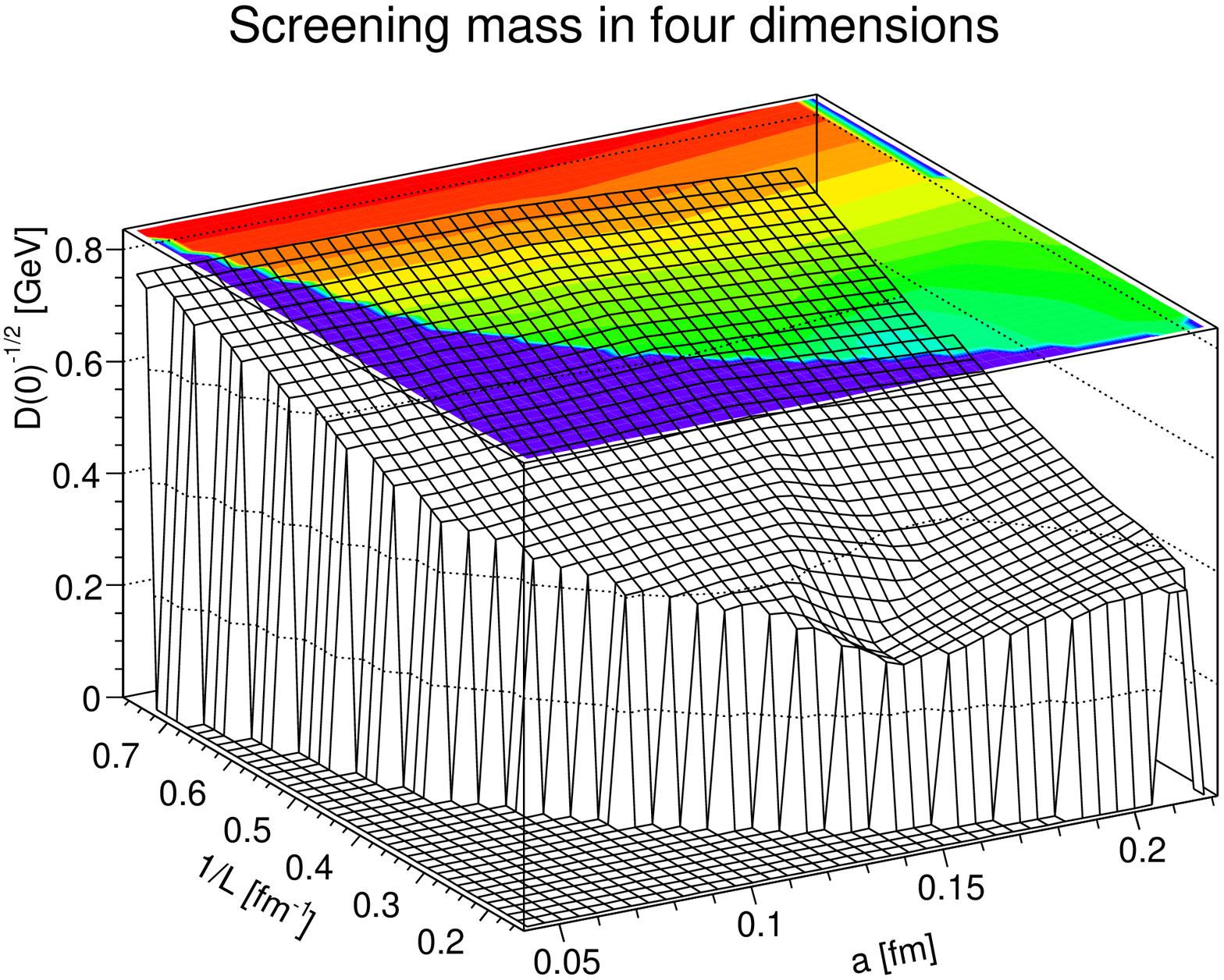}
\caption{\label{fig:d0}The value of the screening mass $D(0)^{-1/2}$ as a function of lattice extent in two (top panel), three (middle panel), and four dimensions (bottom panel). The various dashed lines show the dependence at roughly fixed $a$. The right-hand-side is at $m_r=m=0$.}
\end{figure}

To study the combination of both effects in more detail, in figure \ref{fig:d0} the screening mass $D(0)^{-1/2}$ is shown as a function of physical lattice extension and discretization. Note that since the propagator has not been evaluated at zero momentum, this value is obtained by a linear extrapolation of the propagator at the two lowest non-zero momenta. There is a quite different trend than in the fundamental case \cite{Maas:2016edk}. There, after some initial effects, the screening mass became essentially volume-independent. 

The situation here in the adjoint case is quite different. It is also quite different for the different number of dimensions. In two dimensions, there is, except for the coarsest lattices, little dependency on volume at fixed lattice spacing. But there is a pronounced dependency on the lattice spacing at fixed volume. Still, The results tend visibly towards a finite value of about 400-500 MeV in the continuum limit. The situation is far less obvious in three dimensions. There a pronounced dependency on both the lattice spacing and the physical lattice extension is seen. It is not yet sure that the screening mass tends towards a finite value in the thermodynamic limit. However, for the finest and largest lattices the value is still at about 300 MeV, and thus of comparable size to the one in two dimensions. If it would vanish, it would need to do so substantially faster than linear. In four dimensions the situation changes once more. Now the results are much less dependent on the discretization, but show a strong dependency on the physical volume, reaching down again to values of about 300 MeV. Once more, it is not clear if it surely extrapolates to a non-zero value in the thermodynamic limit, but again if it vanishes it needs to do so much faster than linear.

Note that the trends are also surprisingly true for the large tree-level mass of 1 GeV, though here the effect is much weaker than for the lighter tree-level masses. Still, this indicates a screening which intensifies with larger volumes, while it was, more or less, constant in the fundamental case. In particular, the screening mass is substantially larger than zero, even for the case of zero tree-level and renormalized mass. The size of this screening mass in the massless case of 300-500 MeV for the largest volume is somewhat larger than in the fundamental case, where it was of order 200-250 MeV. The same effect is also seen for adjoint quarks in comparison to fundamental quarks: The effective screening mass for the adjoint quarks is much larger than for the fundamental ones, almost a factor of three \cite{Aguilar:2010ad,Maskawa:1974vs,Fukuda:1976zb,August:2013jia}. In this sense, the situation for the adjoint scalars is less drastic.

At larger momenta the behavior is far less drastic. As seen in figures \ref{fig:d20}-\ref{fig:d43} the propagators follow at higher momenta more or less the expected pattern. At momenta much larger than the renormalization scale the propagators start again to deviate from the tree-level one. In four dimensions, this follows from the usual logarithmic running. In lower dimensions, this is somewhat unexpected, and in contrast to the gauge propagators \cite{Maas:2011se}. This is, however, likely due to the additional wave-function renormalization, which compensates partly for a self-energy contribution, and this discrepancy yields the observed effect: Due to asymptotic freedom, at large momenta all propagators in two and three dimensions tend to $D=1/(Zp^2)$, yielding $H(p)=1/Z$, rather than unity. In addition, also in lower dimensions additional logarithmic corrections can arise on top of the usual power-law \cite{Jackiw:1980kv}, which could potentially also contribute.

\subsection{Schwinger function and effective mass}\label{ss:ssda}

The Schwinger function
\be
\Delta(t)=\frac{1}{\pi}\int_0^\infty dp_0\cos(tp_0)D(p_0^2)=\frac{1}{a\pi}\frac{1}{N_t}\sum_{P_0=0}^{N_t-1}\cos\left(\frac{2\pi tP_0}{N_t}\right)D(P_0^2)\nn,
\ee
\no essentially the temporal correlator, is obtained from the renormalized propagator. The calculation is straightforward in principle, though requires obtaining the removed value at zero momentum. As above, this is obtained by a linear extrapolation of the propagator at the two lowest momenta. Because of the relatively large statistical noise, this induces a corresponding larger error. Systematically, any uncertainty in this constant will vanish as a function of the physical volume, as the extrapolation is done over a smaller and smaller distance in momentum. It can therefore be considered an additional finite-volume effect.

From the Schwinger function the effective (time-dependent) mass
\be
m_\text{eff}(t)=-\ln\frac{\Delta(t+a)}{\Delta(t)}\label{effmass},
\ee
\no can be derived. If the Schwinger function decays strictly like an exponential this mass will be time-independent, and coincides with the pole mass \cite{Seiler:1982pw}. On a finite lattice, for any physical particle with a positive spectral function this effective mass is a monotonously decreasing function for $t\le L/2$. Eventually, at sufficiently long time, it is just the energy of the lightest state with which the operator has a non-zero overlap. If the effective mass is non-monotonously decreasing, the spectral function has necessarily negative contributions. Therefore it then does not describe a physical particle.

Because of the larger systematic uncertainties, especially with respect to discretization, the interpretation of these quantities is more involved than in the fundamental case. In the latter case \cite{Maas:2016edk}, the effective masses approached at long times a, more or less, physical behavior, indicating a would-be pole mass of about 200-250 MeV, with indications of positivity violations remaining at short times. The only necessity was to be sufficiently close to the thermodynamic limit to observe a universal behavior.

\begin{figure}
\includegraphics[width=\linewidth]{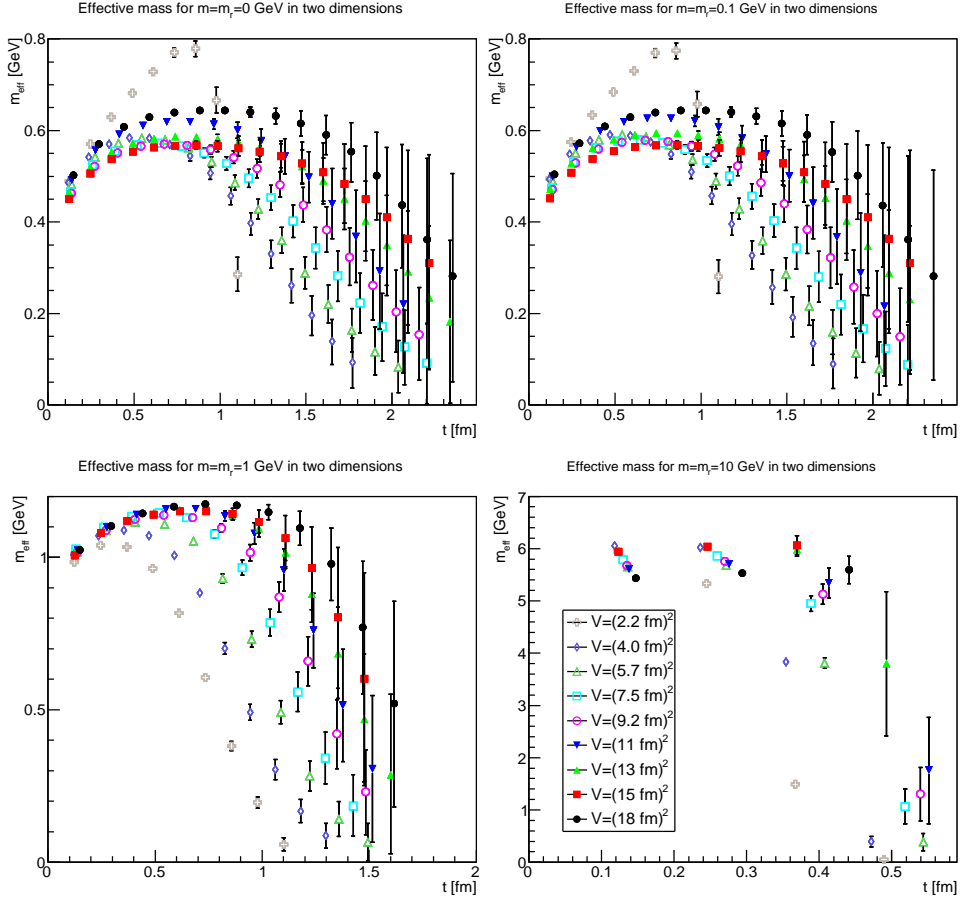}
\caption{\label{fig:m2d}The effective mass \pref{effmass} in two dimensions at lattice spacing $a^{-1}\approx 1.5$ GeV. The top-left panel shows the case of $m=m_r=0$ GeV, the top-right panel of $m=m_r=0.1$ GeV, the bottom-left panel of $m=m_r=1$ GeV, and the bottom-right panel of $m=m_r=10$ GeV. Points with more than 100\% relative error are suppressed.}
\end{figure}

\begin{figure}
\includegraphics[width=\linewidth]{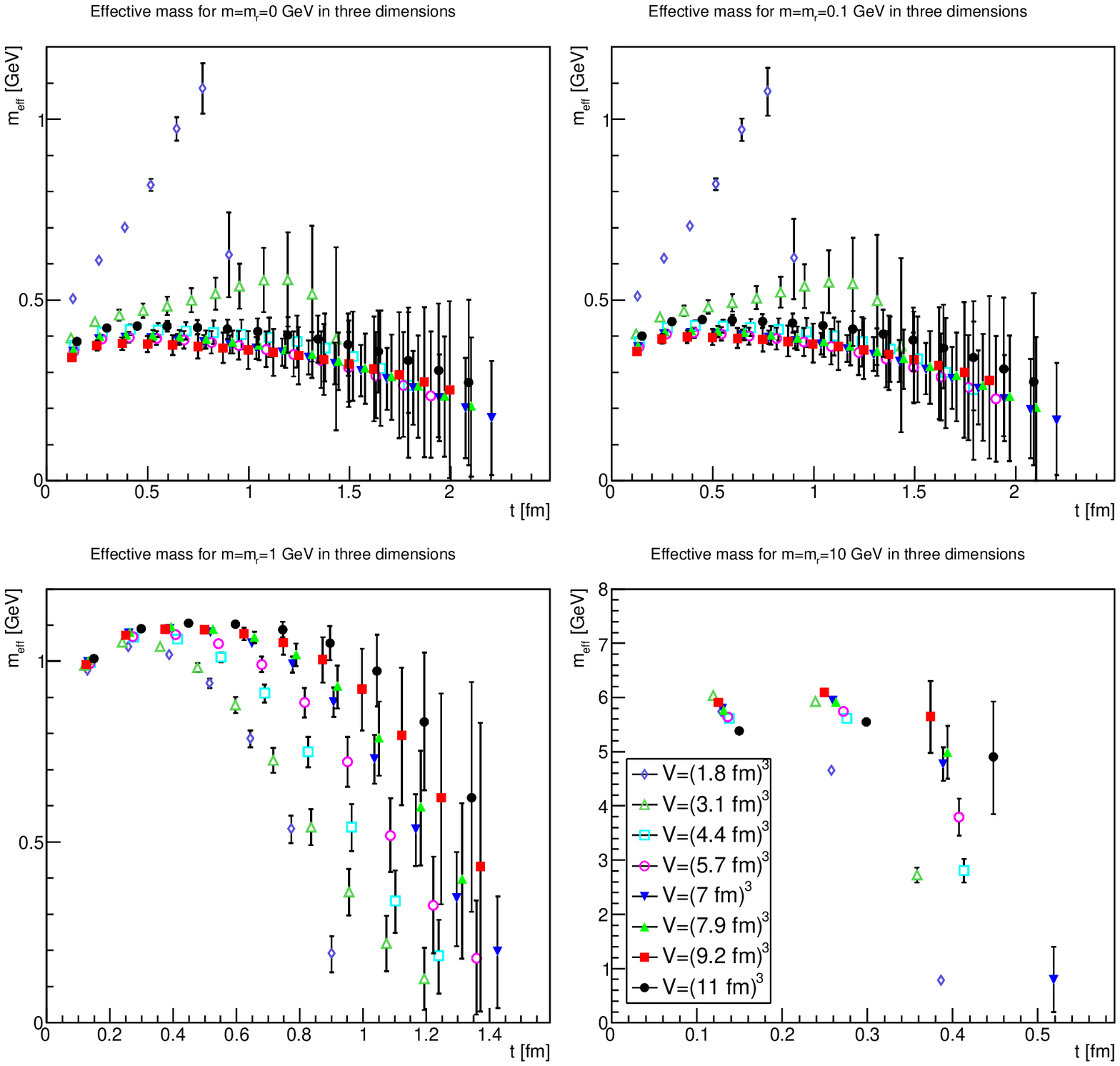}
\caption{\label{fig:m3d}The effective mass \pref{effmass} in three dimensions at lattice spacing $a^{-1}\approx 1.5$ GeV. The top-left panel shows the case of $m=m_r=0$ GeV, the top-right panel of $m=m_r=0.1$ GeV, the bottom-left panel of $m=m_r=1$ GeV, and the bottom-right panel of $m=m_r=10$ GeV. Points with more than 100\% relative error are suppressed.}
\end{figure}

\begin{figure}
\includegraphics[width=\linewidth]{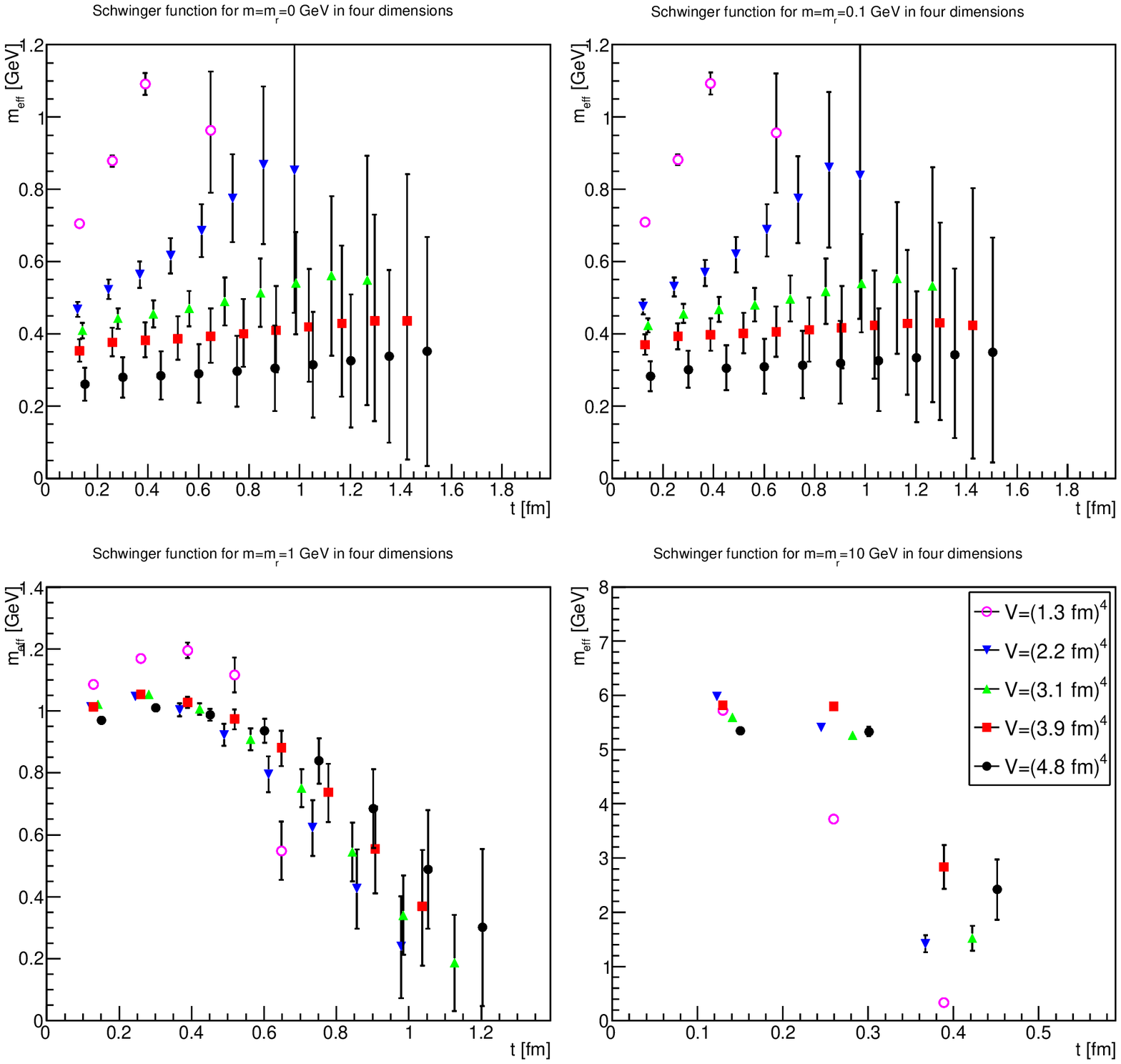}
\caption{\label{fig:m4d}The effective mass \pref{effmass} in four dimensions at lattice spacing $a^{-1}\approx 1.5$ GeV. The top-left panel shows the case of $m=m_r=0$ GeV, the top-right panel of $m=m_r=0.1$ GeV, the bottom-left panel of $m=m_r=1$ GeV, and the bottom-right panel of $m=m_r=10$ GeV. Points with more than 100\% relative error are suppressed.}
\end{figure}

Due to the same effects which already plagued the extraction of the screening mass in section \ref{ss:mom}, this is no longer the case. To actually observe a unique behavior requires to work at fixed lattice spacing. Even then, a substantial finite-volume effect remains. As an illustration, in figures \ref{fig:m2d}-\ref{fig:m4d} the effective mass \pref{effmass} is shown at fixed $a^{-1}\approx 1.5$ GeV in two, three, and four dimensions, respectively. As in the fundamental case \cite{Maas:2016edk}, the effective mass rises at short times, showing again that the particle is unphysical. The finite-volume then eventually makes it fall again. With increasing volume, the effective mass becomes flatter and flatter over a longer period of time, again similar to the fundamental case. 

\begin{figure}
\includegraphics[width=0.5\linewidth]{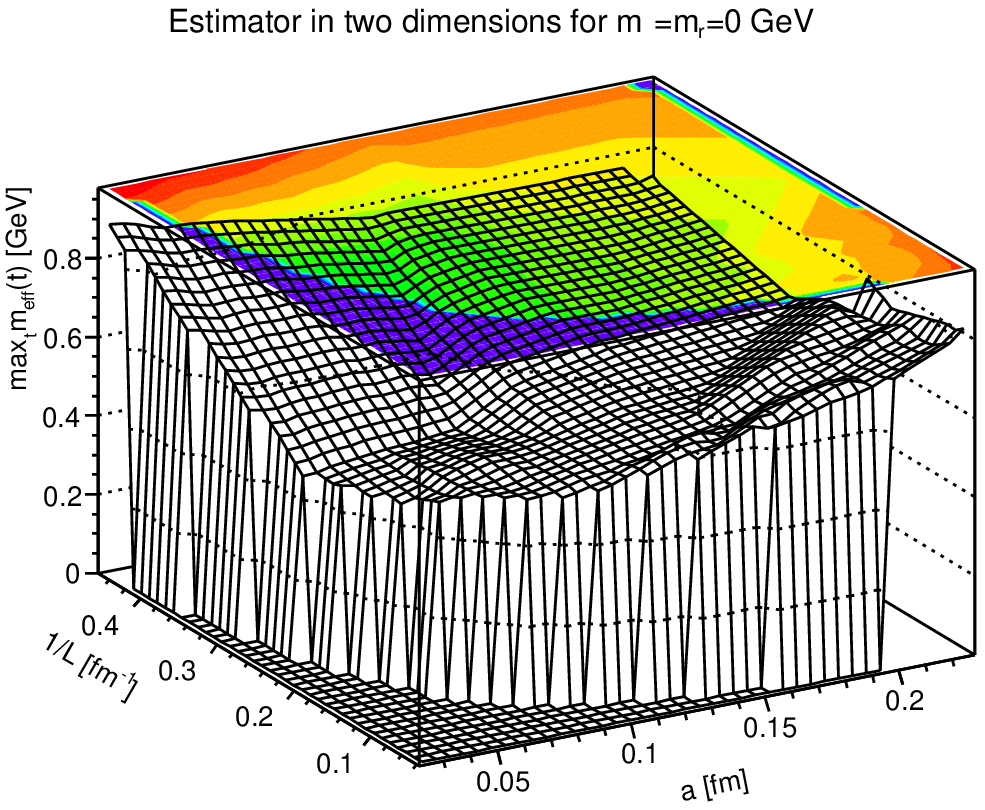}\includegraphics[width=0.5\linewidth]{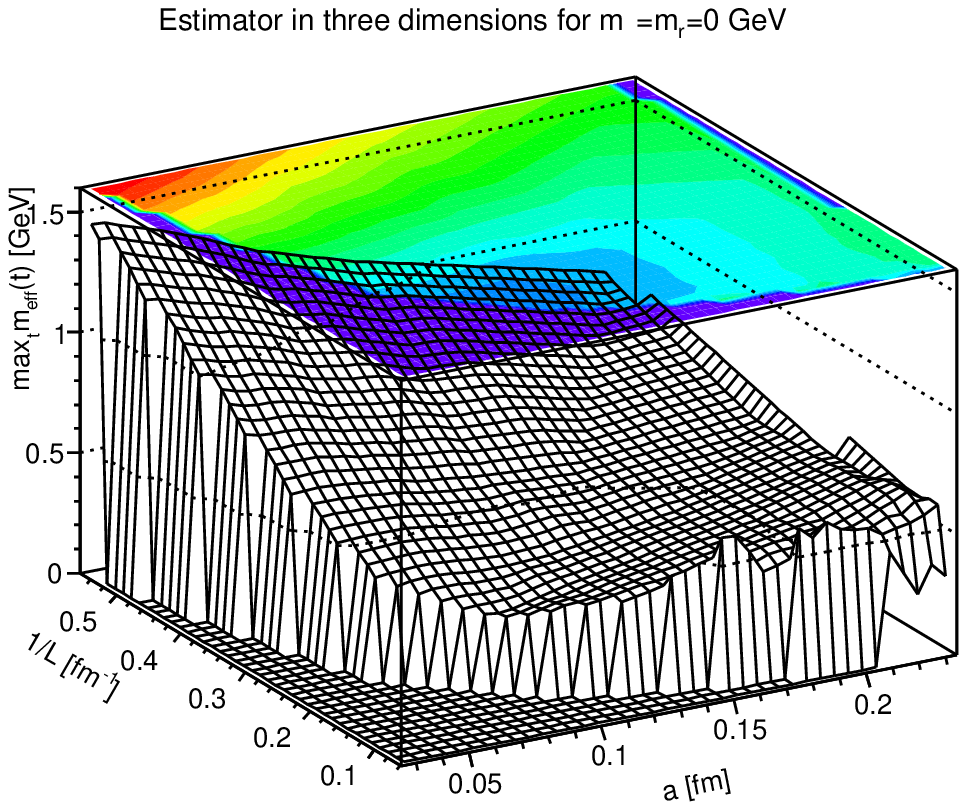}
\begin{center}
\includegraphics[width=0.5\linewidth]{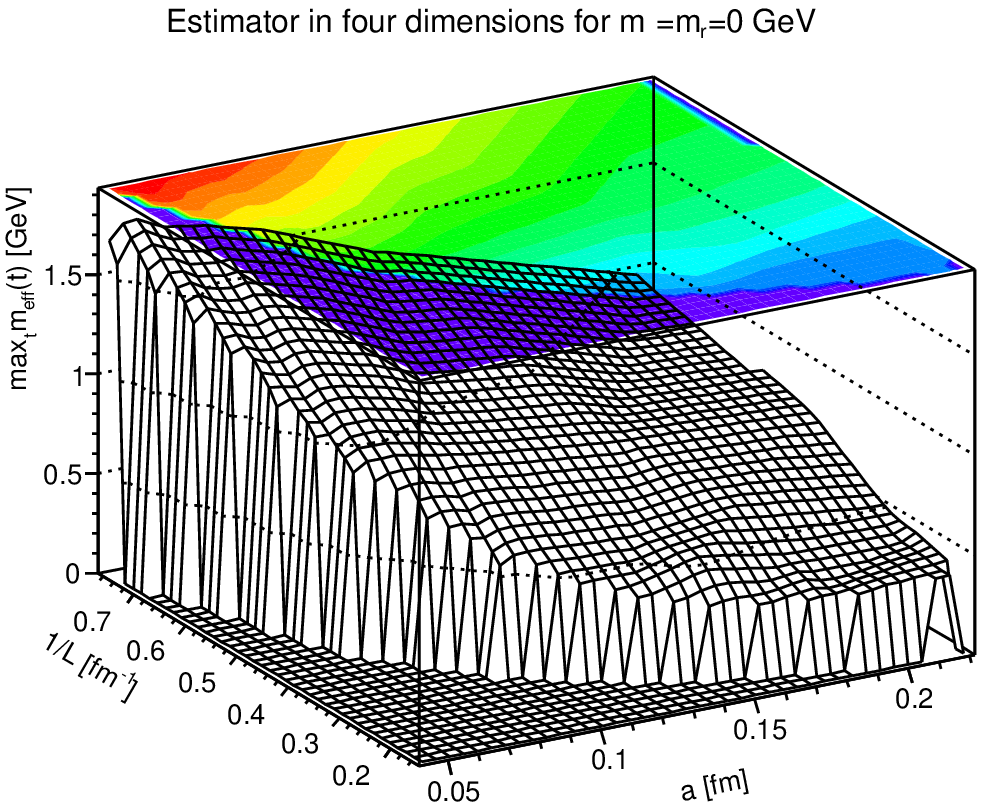}
\end{center}
\caption{\label{fig:pmest} The maximum of the effective mass \pref{effmass} in two dimensions (top-left panel), three dimensions (top-right panel), and four dimensions (bottom panel) as a function of lattice size and lattice spacing.}
\end{figure}

Thus, the maximum effective mass can be used as an estimator for an upper limit of such a would-be long-time pole mass. The result at $m=m_r=0$ GeV for this quantity is shown in figure \ref{fig:pmest}. In all cases, the estimator is strongly affected by the lattice parameters. In two dimensions, the estimator flattens out at about 500 MeV at sufficiently small lattice spacings and large volumes. In higher dimensions, it shows a complex combination of trends, dropping at least below 400 MeV closest to the thermodynamic limit. The results in three dimensions suggest that this process may come to a finite value in the thermodynamic limit. In four dimensions this is harder to judge. It is hence far less obvious if the adjoint scalar resembles at least for some distance regime a physical, massive particle as it was in the fundamental case \cite{Maas:2016edk}.

The situation is quite similar for $m=m_r=0.1$ GeV, and indeed leading to similar quantitative effects. Especially, the values for the estimator are essentially identical, and no trace of the differing renormalized masses remain. Quite contrary at $m=m_r=1$ GeV the same estimator is within some 10\% independent of the lattice parameters, and quickly converges to 1 GeV. Thus, no additional contribution to this mass estimator is observed. This is as in the fundamental case \cite{Maas:2016edk}. Hence, a large explicit mass completely overpowers any other contribution. For the largest mass a rise towards its value is also seen, but as the values of $a^{-1}$ are still below its value, this did not flatten out.

From this it can be concluded that there appears to be also in the adjoint case an additional mass generated. But because of much stronger lattice artifacts, this needs to be handled with much more care than in the fundamental case. Also, its value of about 400 MeV is not so much larger than the 250 MeV observed in the fundamental case as the difference between fundamental fermions and adjoint fermions suggest.

\begin{figure}
\includegraphics[width=\linewidth]{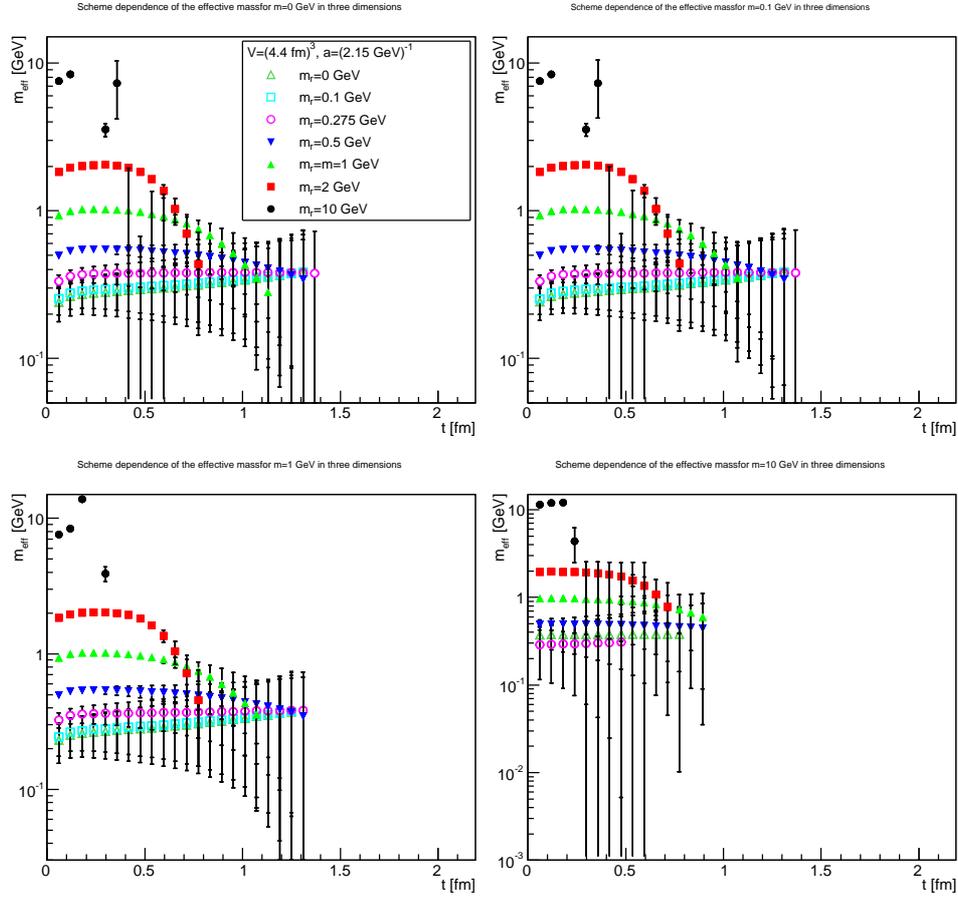}
\caption{\label{fig:scheme3d}The effective mass \pref{effmass} in three dimensions for different renormalization schemes. The top-left panel shows the case of $m=m_r=0$ GeV, the top-right panel of $m=m_r=0.1$ GeV, the bottom-left panel of $m=m_r=1$ GeV, and the bottom-right panel of $m=m_r=10$ GeV. Points with more than 100\% relative error are suppressed.}
\end{figure}

Finally, an investigations of the scheme-dependence along the lines of \cite{Maas:2016edk} reveals the same pattern: It is possible to shift the effective mass around, but it is not possible to push it below a certain limit, no matter the tree-level mass. This lower limit is similar for all tree-level masses, and of the same size as those shown  in figure \ref{fig:pmest}. I.\ e., just as in perturbation theory, this effective mass remains not a physical quantity, but remains scheme-dependent. Interestingly, as is shown in figure \ref{fig:scheme3d}, the flatness, of the effective mass curve is affected by the choice of renormalization scheme. There exists a 'sweet spot', in the example at about 275 MeV, where the effective mass becomes almost flat, except at very short times. Thus, there exists a scheme, which makes the adjoint scalar most 'particle-like'. Note that this value is actually independent from the tree-level mass. This mass value is also a value consistent with where the estimator in figure \ref{fig:pmest} are moving to in the thermodynamic limit.

As a flat curve would correspond to a physical particle of this mass, the best interpretation of this observation is that at this value of $m_r$ the renormalization scheme becomes the closest approximation to a pole scheme.

\section{Conclusions}\label{s:sum}

Summarizing, the propagator of an adjoint scalar in the quenched approximation has been studied in detail in two, three, and four dimensions. It shows both stronger modifications compared to its tree-level form as well as much stronger influences of lattice artifacts in comparison to a fundamental scalar \cite{Maas:2016edk}. Still, despite all quantitative differences, there appears to be little qualitative difference between both cases. Especially, in both cases a dynamical scale generation of about a few hundred MeV has been observed, independent of the tree-level mass. Moreover, neither particle exhibits a behavior which looks like a physical particle, though both approximate such a behavior at long distances. In the present adjoint case, this is much more sensitive to renormalization effects and lattice artifacts.

What is not seen is any indication, as was originally hoped for, which hint to an obvious connection to (Wilson) confinement. While the fundamental and adjoint Wilson string show, at the distance scales achieved here, qualitative different behavior in sufficiently high dimensions, this is not the case for the propagators. Especially the string breaking in the adjoint case seems to have no direct impact.

Also, there is no difference seen for the different dimensionality. Even though the physical picture, due to geometric Wilson confinement or the possibility of triviality in the dynamical case as allured to in the introduction, differs substantially between different dimensions, this seems to have next to no qualitative impact on the propagator. In fact, even quantitatively there is very little difference in the different dimensions.

While the present study cannot exclude some more subtle hint in the propagators, the lack of any obvious effect is unfortunate. Especially when searching for a possibility to obtain such information from low-order correlation functions. It may still be that such information is hidden in the vertices, as has been variously suggested \cite{Alkofer:2004it,Alkofer:2008tt,Fister:2010yw,Zwanziger:2010iz}. Thus, a study of the corresponding vertices is a logical next step \cite{Maas:unpublished,Maas:2011yx}.\\

\no{\bf Acknowledgments}\\

This work was supported by the DFG under grant numbers MA 3935/5-1, MA-3935/8-1 (Heisenberg program) and the FWF under grant number M1099-N16. Simulations were performed on the HPC clusters at the Universities of Jena and Graz. The author is grateful to the HPC teams for the very good performance of the clusters. The ROOT framework \cite{Brun:1997pa} has been used in this project.

\appendix

\section{Lattice setups}\label{a:ls}

The various lattice setups are listed in table \ref{tcgf}. The determination of the lattice spacings has been performed as in \cite{Maas:2014xma}.

\begin{longtable}{|c|c|c|c|c|c|c|c|}
\caption{\label{tcgf}Number and parameters of the configurations used, ordered by dimension, lattice spacing, and physical volume. In all cases $2(10N+100(d-1))$ thermalization sweeps and $2(N+10(d-1))$ decorrelation sweeps of mixed updates \cite{Cucchieri:2006tf} have been performed, and auto-correlation times of local observables have been monitored to be at or below one sweep. The number of configurations were selected such as to have a reasonable small statistical error for the renormalization constants determined in section \ref{s:ren}. The value $m_0$ denotes the value of the mass parameter in \pref{cov} to yield a tree-level mass of 1 GeV. The other tree-level masses are obtained by multiplying or dividing this number by 10, or setting it to zero for tree-level mass zero.}\\
\hline
$d$	& $N$	& $\beta$	& $a$ [fm] & $a^{-1}$ [GeV]	& L [fm]	&  $m_0$	& config.	\endfirsthead
\hline
\multicolumn{8}{|l|}{Table \ref{tcgf} continued}\\
\hline
$d$	& $N$	& $\beta$	& $a$ [fm] & $a^{-1}$ [GeV]	& L [fm]	&  $m_0$	& config.	\endhead
\hline
\multicolumn{8}{|r|}{Continued on next page}\\
\hline\endfoot
\endlastfoot
\hline
2	& 92	& 6.23	& 0.228		& 0.863	& 21	& 1.159		& 4994	\cr
\hline
2	& 106	& 6.33	& 0.226		& 0.870	& 24	& 1.149		& 5440	\cr
\hline
2	& 80	& 6.40	& 0.225		& 0.875	& 18	& 1.143		& 3957	\cr
\hline
2	& 58	& 6.45	& 0.224		& 0.879	& 13	& 1.138		& 3386	\cr
\hline
2	& 18	& 6.55	& 0.222		& 0.886	& 4.0	& 1.129		& 3661	\cr
\hline
2	& 122	& 6.60	& 0.221		& 0.890	& 27	& 1.124		& 4750	\cr
\hline
2	& 34	& 6.64	& 0.221		& 0.893	& 7.5	& 1.120		& 2970	\cr
\hline
2	& 68	& 6.64	& 0.221		& 0.893	& 15	& 1.120		& 3456	\cr
\hline
2	& 10	& 6.68	& 0.220		& 0.895	& 2.2	& 1.117		& 2192	\cr
\hline
2	& 50	& 6.68	& 0.220		& 0.895	& 11	& 1.117		& 3299	\cr
\hline
2	& 26	& 6.72	& 0.219		& 0.898	& 5.7	& 1.113		& 3410	\cr
\hline
2	& 42	& 6.73	& 0.219		& 0.900	& 9.2	& 1.112		& 3370	\cr
\hline
2	& 106	& 8.13	& 0.198		& 0.994 & 21	& 1.006		& 5440	\cr
\hline
2	& 122	& 8.24	& 0.197		& 1.00	& 24	& 0.9990	& 5614	\cr
\hline
2	& 92	& 8.33	& 0.196		& 1.01	& 18	& 0.9933	& 4104	\cr
\hline
2	& 68	& 8.70	& 0.191		& 1.03	& 13	& 0.9708	& 3456	\cr
\hline
2	& 58	& 8.83	& 0.190		& 1.04	& 11	& 0.9632	& 3386	\cr
\hline
2	& 80	& 9.03	& 0.188		& 1.05	& 15	& 0.9519	& 3597	\cr
\hline
2	& 50	& 9.36	& 0.184		& 1.07	& 9.2	& 0.9341	& 3174	\cr
\hline
2	& 42	& 9.91	& 0.179		& 1.10	& 7.5	& 0.9066	& 3433	\cr
\hline
2	& 122	& 10.6	& 0.172		& 1.14	& 21	& 0.8752	& 5248	\cr
\hline
2	& 106	& 10.9	& 0.170		& 1.16	& 18	& 0.8625	& 4768	\cr
\hline
2	& 34	& 11.1	& 0.168		& 1.17	& 5.7	& 0.8543	& 2950	\cr
\hline
2	& 92	& 11.7	& 0.164		& 1.20	& 15	& 0.8312	& 4994	\cr
\hline
2	& 80	& 11.8	& 0.163		& 1.21	& 13	& 0.8275	& 3498	\cr
\hline
2	& 68	& 11.9	& 0.162		& 1.21	& 11	& 0.8239	& 3456	\cr
\hline
2	& 58	& 12.4	& 0.159		& 1.24	& 9.2	& 0.8065	& 3304	\cr
\hline
2	& 26	& 13.1	& 0.154		& 1.28	& 4.0	& 0.7838	& 3410	\cr
\hline
2	& 50	& 13.8	& 0.150		& 1.31	& 7.5	& 0.7629	& 3174	\cr
\hline
2	& 122	& 14.3	& 0.148		& 1.34	& 18	& 0.7490	& 5248	\cr
\hline
2	& 92	& 15.5	& 0.142		& 1.39	& 13	& 0.7185	& 4930	\cr
\hline
2	& 106	& 15.5	& 0.142		& 1.39	& 15	& 0.7185	& 5872	\cr
\hline
2	& 80	& 16.3	& 0.138		& 1.43	& 11	& 0.7001	& 2279	\cr
\hline
2	& 42	& 16.8	& 0.136		& 1.45	& 5.7	& 0.6893	& 3350	\cr
\hline
2	& 68	& 16.9	& 0.135		& 1.46	& 9.2	& 0.6872	& 3420	\cr
\hline
2	& 58	& 18.4	& 0.130		& 1.52	& 7.5	& 0.6578	& 3304	\cr
\hline
2	& 122	& 20.3	& 0.123		& 1.60	& 15	& 0.6254	& 2106	\cr
\hline
2	& 106	& 20.4	& 0.123		& 1.60	& 13	& 0.6239	& 4032	\cr
\hline
2	& 18	& 20.6	& 0.122		& 1.61	& 2.2	& 0.6208	& 3660	\cr
\hline
2	& 92	& 21.5	& 0.120		& 1.65	& 11	& 0.6074	& 4420	\cr
\hline
2	& 34	& 22.2	& 0.118		& 1.67	& 4.0	& 0.5974	& 2970	\cr
\hline
2	& 80	& 23.2	& 0.115		& 1.71	& 9.2	& 0.5841	& 3498	\cr
\hline
2	& 50	& 23.6	& 0.114		& 1.73	& 5.7	& 0.5791	& 3351	\cr
\hline
2	& 68	& 25.2	& 0.110		& 1.79	& 7.5	& 0.5600	& 3420	\cr
\hline
2	& 122	& 26.9	& 0.107		& 1.85	& 13	& 0.5417	& 2106	\cr
\hline
2	& 106	& 28.4	& 0.104		& 1.90	& 11	& 0.5269	& 5200	\cr
\hline
2	& 92	& 30.5	& 0.100		& 1.97	& 9.2	& 0.5082	& 4234	\cr
\hline
2	& 58	& 31.6	& 0.0983	& 2.00	& 5.7	& 0.4991	& 3300	\cr
\hline
2	& 42	& 33.6	& 0.0953	& 2.07	& 4.0	& 0.4838	& 3680	\cr
\hline
2	& 80	& 34.7	& 0.0938	& 2.10	& 7.5	& 0.4759	& 3498	\cr
\hline
2	& 122	& 37.4	& 0.0903	& 2.18	& 11	& 0.4582	& 4372	\cr
\hline
2	& 106	& 40.4	& 0.0868	& 2.27	& 9.2	& 0.4406	& 4592	\cr
\hline
2	& 26	& 42.4	& 0.0847	& 2.33	& 2.2	& 0.4300	& 2720	\cr
\hline
2	& 68	& 43.2	& 0.0839	& 2.35	& 5.7	& 0.4260	& 3505	\cr
\hline
2	& 92	& 45.7	& 0.0816	& 2.42	& 7.5	& 0.4140	& 3848	\cr
\hline
2	& 50	& 47.4	& 0.0801	& 2.46	& 4.0	& 0.4064	& 3215	\cr
\hline
2	& 122	& 53.3	& 0.0755	& 2.61	& 9.2	& 0.3831	& 4698	\cr
\hline
2	& 80	& 59.7	& 0.0713	& 2.76	& 5.7	& 0.3618	& 3505	\cr
\hline
2	& 106	& 60.5	& 0.0708	& 2.78	& 7.5	& 0.3593	& 4240	\cr
\hline
2	& 58	& 63.7	& 0.0690	& 2.86	& 4.0	& 0.3501	& 3276	\cr
\hline
2	& 34	& 72.3	& 0.0647	& 3.04	& 2.2	& 0.3285	& 3549	\cr
\hline
2	& 92	& 78.8	& 0.0620	& 3.18	& 5.7	& 0.3146	& 4848	\cr
\hline
2	& 122	& 80	& 0.03122	& 3.20	& 7.5	& 0.3122	& 4896  \cr
\hline
2	& 68	& 87.3	& 0.0589	& 3.35	& 4.0	& 0.2988	& 3472	\cr
\hline
2	& 106	& 104	& 0.0539	& 3.65	& 5.7	& 0.2736	& 4474	\cr
\hline
2	& 42	& 110	& 0.0524	& 3.76	& 2.2	& 0.2660	& 3122	\cr
\hline
2	& 80	& 120	& 0.0502	& 3.93	& 4.0	& 0.02546	& 3631	\cr
\hline
2	& 50	& 155	& 0.0441	& 4.47	& 2.2	& 0.2239	& 3105	\cr
\hline
2	& 92	& 159	& 0.0436	& 4.52	& 4.0	& 0.2211	& 4110	\cr
\hline
2	& 58	& 209	& 0.0380	& 5.19	& 2.2	& 0.1928	& 3304	\cr
\hline
2	& 106	& 211	& 0.0378	& 5.21	& 4.0	& 0.1919	& 5184	\cr
\hline
2	& 68	& 287	& 0.0324	& 6.08	& 2.2	& 0.1644	& 3716	\cr
\hline
2	& 80	& 398	& 0.0275	& 7.16	& 2.2	& 0.1396	& 3509	\cr
\hline
2	& 92	& 526	& 0.0239	& 8.24	& 2.2	& 0.1214	& 5664	\cr
\hline
2	& 106	& 698	& 0.0208	& 9.49	& 2.2	& 0.1054	& 2784	\cr
\hline
\hline
3	& 60	& 3.30	& 0.234		& 0.841	& 14	& 1.189		& 2752	\cr
\hline
3	& 74	& 3.34	& 0.231		& 0.854 & 17	& 1.170		& 2225	\cr
\hline
3	& 48	& 3.35	& 0.230		& 0.858	& 11	& 1.166		& 3654	\cr
\hline
3	& 66	& 3.37	& 0.228		& 0.864	& 15	& 1.157		& 2816	\cr
\hline
3	& 8	    & 3.40	& 0.225		& 0.874	& 1.8	& 1.144		& 3000	\cr
\hline
3	& 54	& 3.43	& 0.223		& 0.884	& 12	& 1.131		& 5623	\cr
\hline
3	& 14	& 3.44	& 0.222		& 0.887	& 3.1	& 1.127		& 3600	\cr
\hline
3	& 20	& 3.46	& 0.220		& 0.894	& 4.4	& 1.119		& 3160	\cr
\hline
3	& 26	& 3.47	& 0.220		& 0.897	& 5.7	& 1.115		& 2840	\cr
\hline
3	& 36	& 3.47	& 0.220		& 0.897	& 7.9	& 1.115		& 3300	\cr
\hline
3	& 42	& 3.47	& 0.220		& 0.897	& 9.2	& 1.115		& 5021	\cr
\hline
3	& 32	& 3.48	& 0.219		& 0.900	& 7.0	& 1.111		& 2996	\cr
\hline
3	& 66	& 3.56	& 0.213		& 0.927	& 14	& 1.079		& 2200 	\cr
\hline
3	& 54	& 3.68	& 0.204		& 0.966	& 11	& 1.035		& 5538	\cr
\hline
3	& 74	& 3.69	& 0.203		& 0.969	& 15	& 1.032		& 2225	\cr
\hline
3	& 60	& 3.73	& 0.201		& 0.982	& 12	& 1.018		& 2752	\cr
\hline
3	& 36	& 3.82	& 0.195		& 1.01	& 7.0	& 0.9883	& 3462	\cr
\hline
3	& 48	& 3.86	& 0.192		& 1.03	& 9.2	& 0.9756	& 3654	\cr
\hline
3	& 74	& 3.90	& 0.190		& 1.04	& 14	& 0.9632	& 2225	\cr
\hline
3	& 42	& 3.92	& 0.189		& 1.04	& 7.9	& 0.9572	& 3450	\cr
\hline
3	& 60	& 4.01	& 0.183		& 1.07	& 11	& 0.9308	& 2752	\cr
\hline
3	& 66	& 4.03	& 0.182		& 1.08	& 12	& 0.9251	& 2816	\cr
\hline
3	& 32	& 4.10	& 0.178		& 1.10	& 5.7	& 0.9058	& 3006	\cr
\hline
3	& 54	& 4.25	& 0.171		& 1.15	& 9.2	& 0.8671	& 5304	\cr
\hline
3	& 26	& 4.28	& 0.169		& 1.16	& 4.4	& 0.8597	& 2840	\cr
\hline
3	& 42	& 4.33	& 0.167		& 1.18	& 7.0	& 0.8477	& 3277	\cr
\hline
3	& 66	& 4.33	& 0.167		& 1.18	& 11	& 0.8477	& 2200	\cr
\hline
3	& 48	& 4.38	& 0.165		& 1.20	& 7.9	& 0.8360	& 4776	\cr
\hline
3	& 74	& 4.43	& 0.162		& 1.21	& 12	& 0.8247	& 2225	\cr
\hline
3	& 54	& 4.83	& 0.147		& 1.34	& 7.9	& 0.7439	& 3744	\cr
\hline
3	& 48	& 4.84	& 0.146		& 1.35	& 7.0	& 0.7420	& 3600	\cr
\hline
3	& 36	& 4.52	& 0.159		& 1.24	& 5.7	& 0.8050	& 3300	\cr
\hline
3	& 20	& 4.60	& 0.155		& 1.27	& 3.1	& 0.7883	& 3160	\cr
\hline
3	& 60	& 4.64	& 0.154		& 1.28	& 9.2	& 0.7802	& 2496	\cr
\hline
3	& 74	& 4.77	& 0.149		& 1.32	& 11	& 0.7550	& 2848	\cr
\hline
3	& 66	& 5.03	& 0.140		& 1.41	& 9.2	& 0.7091	& 2160	\cr
\hline
3	& 32	& 5.09	& 0.138		& 1.43	& 4.4	& 0.6993	& 3070	\cr
\hline
3	& 42	& 5.15	& 0.136		& 1.45	& 5.7	& 0.6897	& 3400	\cr
\hline
3	& 60	& 5.29	& 0.132		& 1.50	& 7.9	& 0.6685	& 4602	\cr
\hline
3	& 54	& 5.36	& 0.130		& 1.52	& 7.0	& 0.6583	& 8312	\cr
\hline
3	& 14	& 5.39	& 0.129		& 1.53	& 1.8	& 0.6540	& 3600	\cr
\hline
3	& 74	& 5.55	& 0.125		& 1.58	& 9.2	& 0.6322	& 2560	\cr
\hline
3	& 36	& 5.64	& 0.122		& 1.61	& 4.4	& 0.6206	& 3300	\cr
\hline
3	& 66	& 5.74	& 0.120		& 1.64	& 7.9	& 0.6081	& 4338	\cr
\hline
3	& 26	& 5.76	& 0.119		& 1.65	& 3.1	& 0.6057	& 2768	\cr
\hline
3	& 48	& 5.78	& 0.119		& 1.66	& 5.7	& 0.6033	& 2485	\cr
\hline
3	& 60	& 5.87	& 0.117		& 1.69	& 7.0	& 0.5927	& 2015  \cr
\hline
3	& 74	& 6.34	& 0.107		& 1.84	& 7.9	& 0.5428	& 2560	\cr
\hline
3	& 66	& 6.38	& 0.106		& 1.86	& 7.0	& 0.5389	& 2112	\cr
\hline
3	& 54	& 6.41	& 0.106		& 1.87	& 5.7	& 0.5361	& 3893	\cr
\hline
3	& 42	& 6.45	& 0.105		& 1.88	& 4.4	& 0.5323	& 3450	\cr
\hline
3	& 32	& 6.91	& 0.0970	& 2.03	& 3.1	& 0.4925	& 3070	\cr
\hline
3	& 60	& 7.04	& 0.0950	& 2.07	& 5.7	& 0.4824	& 4050	\cr
\hline
3	& 74	& 7.06	& 0.0947	& 2.08	& 7.0	& 0.4808	& 2560	\cr
\hline
3	& 48	& 7.27	& 0.0917	& 2.15	& 4.4	& 0.4653	& 5166	\cr
\hline
3	& 20	& 7.39	& 0.0900	& 2.19	& 1.8	& 0.4569	& 3160	\cr
\hline
3	& 66	& 7.67	& 0.0864	& 2.28	& 5.7	& 0.4384	& 4400	\cr
\hline
3	& 36	& 7.69	& 0.0861	& 2.29	& 3.1	& 0.4371	& 3387	\cr
\hline
3	& 54	& 8.08	& 0.0815	& 2.42	& 4.4	& 0.4139	& 3931	\cr
\hline
3	& 42	& 8.84	& 0.0739	& 2.67	& 3.1	& 0.3750	& 5359	\cr
\hline
3	& 60	& 8.89	& 0.0734	& 2.68	& 4.4	& 0.3727	& 2496	\cr
\hline
3	& 26	& 9.38	& 0.0692	& 2.84	& 1.8	& 0.3515	& 2840	\cr
\hline
3	& 66	& 9.71	& 0.0667	& 2.95	& 4.4	& 0.3386	& 2560	\cr
\hline
3	& 48	& 10.0	& 0.0646	& 3.05	& 3.1	& 0.3280	& 3894	\cr
\hline
3	& 74	& 10.8	& 0.0595	& 3.31	& 4.4	& 0.3019	& 2560	\cr
\hline
3	& 54	& 11.1	& 0.0577	& 3.41	& 3.1	& 0.2931	& 4973	\cr
\hline
3	& 32	& 11.3	& 0.0566	& 3.48	& 1.8	& 0.2875	& 3208	\cr
\hline
3	& 60	& 12.3	& 0.0517	& 3.81	& 3.1	& 0.2627	& 1892	\cr
\hline
3	& 36	& 12.7	& 0.0500	& 3.94	& 1.8	& 0.2539	& 3410	\cr
\hline
3	& 66	& 13.4	& 0.0472	& 4.17	& 3.1	& 0.2398	& 2450	\cr
\hline
3	& 42	& 14.6	& 0.0432	& 4.57	& 1.8	& 0.2191	& 3357	\cr
\hline
3	& 48	& 16.6	& 0.0377	& 5.22	& 1.8	& 0.1914	& 3769	\cr
\hline
3	& 54	& 18.6	& 0.0335	& 5.88	& 1.8	& 0.1700	& 4548	\cr
\hline
3	& 60	& 20.6	& 0.0301	& 6.54	& 1.8	& 0.01529	& 4381	\cr
\hline
3	& 66	& 22.6	& 0.0274	& 7.20	& 1.8	& 0.01389	& 2464	\cr
\hline
3	& 74	& 25.3	& 0.0244	& 8.09	& 1.8	& 0.01236	& 2464	\cr
\hline
\hline
4	& 14	& 2.179	& 0.221		& 0.889	& 3.1	& 1.124		& 2930	\cr
\hline
4	& 10	& 2.181	& 0.220		& 0.894	& 2.2	& 1.119		& 3000	\cr
\hline
4	& 26	& 2.183	& 0.219		& 0.898	& 5.7	& 1.114		& 3152	\cr
\hline
4	& 22	& 2.185	& 0.218		& 0.902	& 4.8	& 1.109		& 3186	\cr
\hline
4	& 6	& 2.188	& 0.217		& 0.908	& 1.3	& 1.101		& 2220	\cr
\hline
4	& 18	& 2.188	& 0.217		& 0.908	& 3.9	& 1.101		& 3284	\cr
\hline
4	& 30	& 2.188	& 0.217		& 0.908	& 6.5	& 1.101		& 3000	\cr
\hline
4	& 32	& 2.190	& 0.216		& 0.912	& 6.9	& 1.096		& 1646	\cr
\hline
4	& 30	& 2.241	& 0.190		& 1.03	& 5.7	& 0.9667	& 3639	\cr
\hline
4	& 26	& 2.252	& 0.185		& 1.06	& 4.8	& 0.9396	& 3406	\cr
\hline
4	& 32	& 2.266 & 0.178		& 1.10	& 5.7	& 0.9055	& 3495	\cr
\hline
4	& 22	& 2.268	& 0.177		& 1.11	& 3.9	& 0.9007	& 3096	\cr
\hline
4	& 18	& 2.279	& 0.172		& 1.14	& 3.1	& 0.8743	& 3277	\cr
\hline
4	& 30	& 2.305	& 0.160		& 1.23	& 4.8	& 0.8136	& 4152	\cr
\hline
4	& 14	& 2.311	& 0.158		& 1.25	& 2.2	& 0.7999	& 2910	\cr
\hline
4	& 26	& 2.328	& 0.150		& 1.31	& 3.9	& 0.7618	& 3256	\cr
\hline
4	& 32	& 2.328 & 0.150		& 1.31	& 4.8	& 0.7618	& 2128	\cr
\hline
4	& 22	& 2.349 & 0.141		& 1.40	& 3.1	& 0.7162	& 2996	\cr
\hline
4	& 10	& 2.376 & 0.130		& 1.52	& 1.3	& 0.6600	& 3000	\cr
\hline
4	& 30	& 2.376 & 0.130		& 1.52	& 3.9	& 0.6600	& 3412	\cr
\hline
4	& 18	& 2.395	& 0.123		& 1.61	& 2.2	& 0.6222	& 3105	\cr
\hline
4	& 32	& 2.396	& 0.122		& 1.61	& 3.9	& 0.6203	& 2430	\cr
\hline
4	& 26	& 2.403	& 0.120		& 1.65	& 3.1	& 0.6067	& 3253	\cr
\hline
4	& 30	& 2.448 & 0.103		& 1.91	& 3.1	& 0.5246	& 3380	\cr
\hline
4	& 22	& 2.457	& 0.100		& 1.96	& 2.2	& 0.5092	& 3052	\cr
\hline
4	& 32	& 2.467	& 0.0970	& 2.03	& 3.1	& 0.4925	& 1797	\cr
\hline
4	& 14	& 2.480	& 0.0929	& 2.12	& 1.3	& 0.4714	& 2930	\cr
\hline
4	& 26	& 2.507	& 0.0847	& 2.33	& 2.2	& 0.4299	& 3239	\cr
\hline
4	& 30	& 2.548	& 0.0734	& 2.68	& 2.2	& 0.3726	& 3434	\cr
\hline
4	& 18	& 2.552	& 0.0724	& 2.72	& 1.3	& 0.3674	& 3280	\cr
\hline
4	& 32	& 2.566	& 0.0689	& 2.86	& 2.2	& 0.03496	& 4235  \cr
\hline
4	& 22	& 2.609	& 0.0591	& 3.33	& 1.3	& 0.3001	& 3252	\cr
\hline
4	& 26	& 2.656	& 0.0501	& 3.93	& 1.3	& 0.2543	& 3706	\cr
\hline
4	& 30	& 2.698	& 0.0434	& 4.54	& 1.3	& 0.2204	& 3593	\cr
\hline
4	& 32	& 2.718 & 0.0407	& 4.84	& 1.3	& 0.2065	& 2020	\cr
\hline
\end{longtable}

\bibliographystyle{bibstyle}
\bibliography{bib}


\end{document}